\documentclass[final]{siamltex}

\usepackage{amsmath,amsfonts,epsfig,amsbsy}

\newcommand{\boldn}{{\mathbf n}}
\newcommand{\boldX}{{\mathbf X}}
\newcommand{\boldx}{{\mathbf x}}
\newcommand{\boldnu}{{\boldsymbol \nu}}
\newcommand{\er}{{\mathbb R}}

\newcommand{\dx}{\mbox{d}x}
\newcommand{\dxi}{\mbox{d}\xi}
\newcommand{\deta}{\mbox{d}\eta}
\newcommand{\dy}{\mbox{d}y}
\newcommand{\dz}{\mbox{d}z}

\newcommand{\dv}{\mbox{d}v}
\newcommand{\dw}{\mbox{d}w}
\newcommand{\dt}{\mbox{d}t}
\newcommand{\dX}{\mbox{d}X}

\newcommand{\dW}{\mbox{d}W}
\newcommand{\por}{\overline{q}}

\newcommand{\picturesAB}[4]{
\centerline{
\hskip #4
\raise #3 \hbox{\raise 0.9mm \hbox{(a)}}
\hskip -8mm
\epsfig{file=#1,height=#3}
\!\!\!\!
\raise #3 \hbox{\raise 0.9mm \hbox{(b)}}
\hskip -8mm
\epsfig{file=#2,height=#3}
}}
\newcommand{\picturesABal}[5]{
\centerline{
\hskip #4
\raise #3 \hbox{\raise 0.9mm \hbox{(a)}}
\hskip -8mm
\epsfig{file=#1,height=#3}
\hskip #5
\raise #3 \hbox{\raise 0.9mm \hbox{(b)}}
\hskip -8mm
\epsfig{file=#2,height=#3}
}}

\title{Analysis of a stochastic chemical system close
to a SNIPER bifurcation of its mean-field model}

\author{\quad \quad \quad 
Radek Erban\thanks{University of Oxford, Mathematical Institute, 
24-29 St. Giles', Oxford, OX1 3LB, United Kingdom;
{\it e-mails:  erban@maths.ox.ac.uk; chapman@maths.ox.ac.uk}.} \quad
\and \quad  S. Jonathan Chapman\footnotemark[1]\quad
\and \quad Ioannis G. Kevrekidis\thanks{Princeton University,
Department of Chemical Engineering, PACM and Mathematics,
Engineering Quadrangle,
Olden Street, Princeton, NJ 08544, USA;
{\it e-mail: yannis@princeton.edu}.} \quad \quad \quad 
\and \quad Tom\'a\v{s} Vejchodsk\'y\thanks{
Institute of Mathematics, Czech Academy of Sciences,
\v{Z}itn\'a 25, 115 67 Praha 1, Czech Republic;
{\it e-mail:  vejchod@math.cas.cz}.}
}

\begin{document}

\maketitle

\begin{abstract}
A framework for the analysis of stochastic models of chemical 
systems for which the deterministic mean-field description is 
undergoing a saddle-node infinite period (SNIPER) bifurcation
is presented.
Such a bifurcation occurs for example in the modelling of 
cell-cycle regulation. 
It is shown that the stochastic system possesses 
oscillatory solutions even for parameter values for which the mean-field 
model does not oscillate. 
The dependence of the mean period of these oscillations on 
the parameters of the model (kinetic rate constants) and the size of the 
system (number of molecules present) is studied. 
Our approach is based on the
chemical Fokker-Planck equation. To get some insights into advantages 
and disadvantages of the method, a simple one-dimensional 
chemical switch is first analyzed, before
the chemical SNIPER problem is studied in detail. 
First, results obtained by solving the Fokker-Planck equation 
numerically are presented. 
Then an asymptotic analysis of the Fokker-Planck
equation is used to derive explicit formulae for the period of oscillation
as a function of the rate constants and as a function of the system size. 
\end{abstract}



\pagestyle{myheadings}
\thispagestyle{plain}
\markboth{RADEK ERBAN ET AL.}
{STOCHASTIC BIFURCATIONS}

\section{Introduction}
Bifurcation diagrams are often used for the analysis of deterministic
models of chemical systems. In recent years, they have also been
applied to models of biological (biochemical) systems. For example, 
the cell-cycle model of Tyson {\it et al.} \cite{Tyson:2002:DCC} is a system 
of ordinary differential equations (ODEs) which describes the time 
evolution of concentrations of biochemical species involved in
cell-cycle regulation. The dependence of the qualitative behaviour 
on the parameters of the model is studied using standard bifurcation 
analysis of ODEs \cite{Doedel:1991:NAC,Doedel:1991:NAC2}. Biological
systems are intrinsically noisy because of the low copy numbers
of some of the biochemical species involved. In such a case, one
has to use stochastic simulation algorithms (SSAs) to include the
intrinsic noise in the modelling \cite{Gillespie:1977:ESS}.
The SSA models use the same set of parameters as the ODE models.
However, even if we choose the same values of the rate constants
for both the SSA model and the ODE model, the behaviour of the
system can differ qualitatively \cite{DeVille:2006:NDL}.
For example, the transition from the $G_2$ phase to mitosis in the
model of Tyson {\it et al.} is governed by a SNIPER
(saddle-node infinite period) bifurcation\footnote{
A SNIPER bifurcation is sometimes
called a SNIC bifurcation (saddle-node bifurcation on invariant circle).}
\cite{Tyson:2002:DCC}. In the ODE setting, a SNIPER bifurcation
occurs whenever a saddle and a stable node collapse into
a single stationary point on a closed orbit (as a bifurcation
parameter is varied). In particular, the limit cycle is born with 
infinite period at the bifurcation point. If we use the SSA model 
instead of ODEs, the bifurcation behaviour will be altered. 
As we will see in Section \ref{secSSAsim}, intrinsic noise 
causes oscillations with finite average period at the deterministic 
bifurcation point. Moreover, the stochastic system oscillates even 
for the parameter
values for which the deterministic model does not have a periodic
solution. Clearly, there is a need to understand the changes
in the bifurcation behaviour if a modeller decides to use 
SSAs instead of ODEs. In this paper, we focus on the SNIPER bifurcation. 
However, the approach presented can be applied to more
general chemical systems. 

In Section \ref{secSSAsim},
we introduce a simple chemical system for which the ODE model
undergoes a SNIPER bifurcation.
We use this illustrative example to motivate the phenomena which 
we want to study. The study of stochastic 
chemical systems in the neighbourhood of determistic bifurcation 
points will be done using the chemical Fokker-Planck 
equation \cite{Gillespie:2000:CLE}. The analysis of the
Fokker-Planck equation is much easier in one dimension. Thus,
we start with an analysis of a simple one-dimensional chemical
switch in Section \ref{sec1Dswitch}. Using the one-dimensional
setting, we can show some advantages and disadvantages of our
approach without going into technicalities. In Section \ref{secnumerFPres}, 
we present a computer assisted analysis of the SSAs by
exploiting the chemical 
Fokker-Planck equation, using the illustrative model from 
Section \ref{secSSAsim}. In Section \ref{secanalFPres}, we provide 
core analytical results. We study the dependence of the period of 
oscillation on the parameters of the model, namely
kinetic rate constants 
and the size of the system (number of molecules present). We derive 
analytical formulae for the period of oscillation using the asymptotic 
expansion of the two-dimensional Fokker-Planck equation. 
Finally, in Section \ref{secconclusion}, we present our conclusions.

\section{A chemical system undergoing a SNIPER bifurcation}

\label{secSSAsim}

We consider two chemical species $X$ and $Y$ in a well-mixed reactor
of volume $V$ which are subject to the following set of seven chemical 
reactions
\begin{equation}
\mbox{ \raise 1mm
  \hbox{
   $\emptyset \;\displaystyle\mathop{\displaystyle\longrightarrow}^{k_{1d}}\;
   Y \;\displaystyle\mathop{\displaystyle\longrightarrow}^{k_{2d}}\; X
   \;\displaystyle\mathop{\displaystyle\longrightarrow}^{k_{3d}}\; \emptyset,$}
     }
\qquad\qquad\qquad
{\mbox{ \raise 0.851 mm \hbox{$2X$}}}
\;
\mathop{\stackrel{\displaystyle\longrightarrow}\longleftarrow}^{k_{4d}}_{k_{5d}}
\;
{\mbox{\raise 0.851 mm\hbox{$3X,$}}}
\qquad\qquad\label{model1}
\end{equation}
\begin{equation}
X + Y \;\displaystyle\mathop{\displaystyle\longrightarrow}^{k_{6d}}\; X + 2Y,
\qquad\qquad\qquad\quad
2X + Y \;\displaystyle\mathop{\displaystyle\longrightarrow}^{k_{7d}}\; 2 X.
\qquad\label{model2}
\end{equation}
The first reaction is the production of the chemical $Y$ from a source 
with constant rate $k_{1d}$ per unit of volume, i.e. the units
of $k_{1d}$ are $\mbox{sec}^{-1} \mbox{mm}^{-3}$, 
see (\ref{sniperparam}) below.  The second reaction
is the conversion of $Y$ to $X$ with rate constant $k_{2d}$ and 
the third
reaction is the degradation of $X$ with rate constant $k_{3d}$. The remaining
reactions are of second-order or third-order. They were chosen 
so that the mean-field description corresponding to 
(\ref{model1})--(\ref{model2})  
undergoes a SNIPER bifurcation as shown below. Clearly,
there exist many other chemical models with a SNIPER bifurcation, 
including the more realistic model of the cell-cycle regulation 
\cite{Tyson:2002:DCC}. The advantage of the model 
(\ref{model1})--(\ref{model2}) 
is that it involves only two chemical species $X$ and $Y$, making the 
visualization of our results clearer: two-dimensional phase planes
are much easier to plot than the phase spaces of high-dimensional 
models. The generalization of our results to models involving
more than two chemical species will be discussed 
in Section \ref{secconclusion}.  

\begin{figure}
\picturesAB{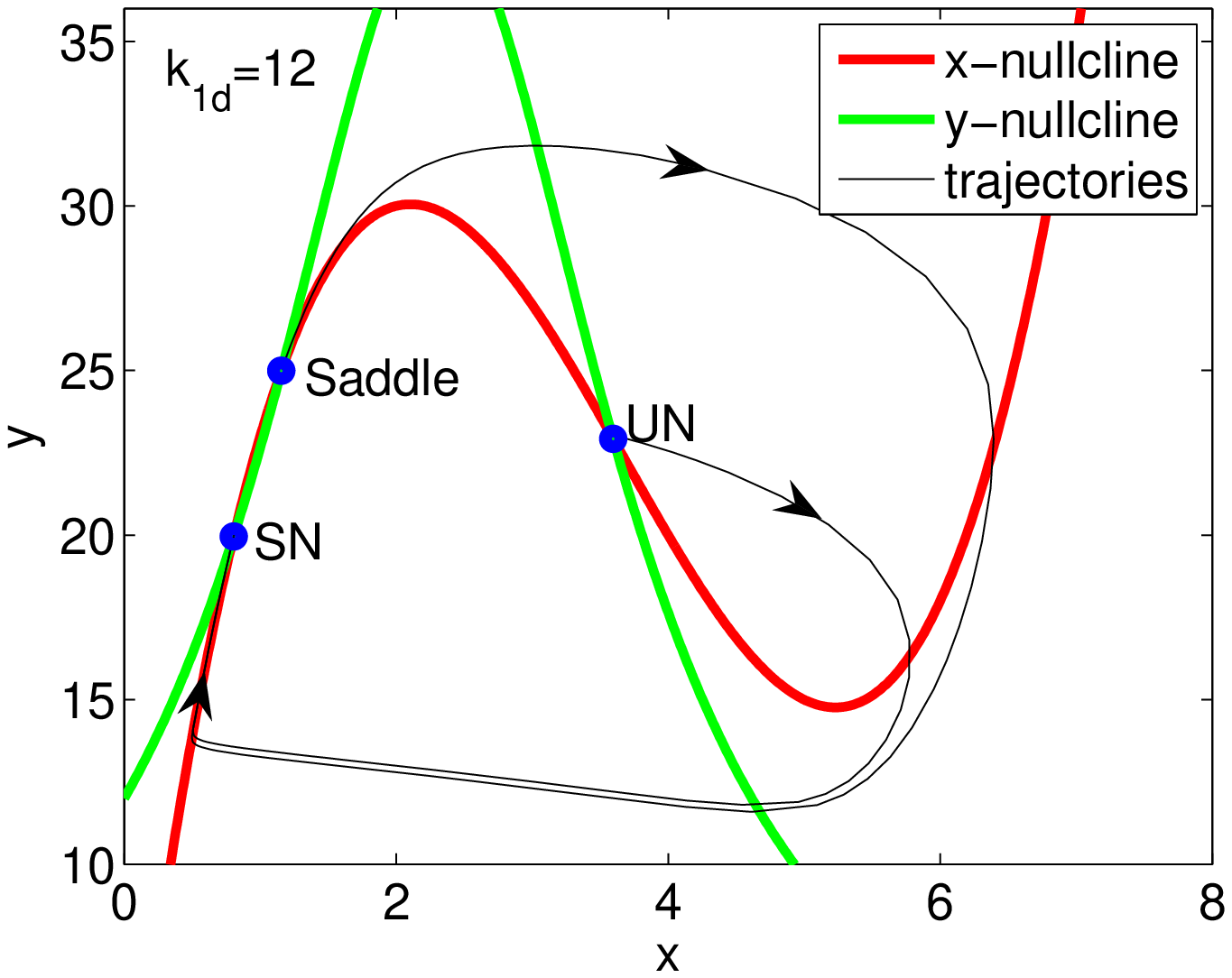}
{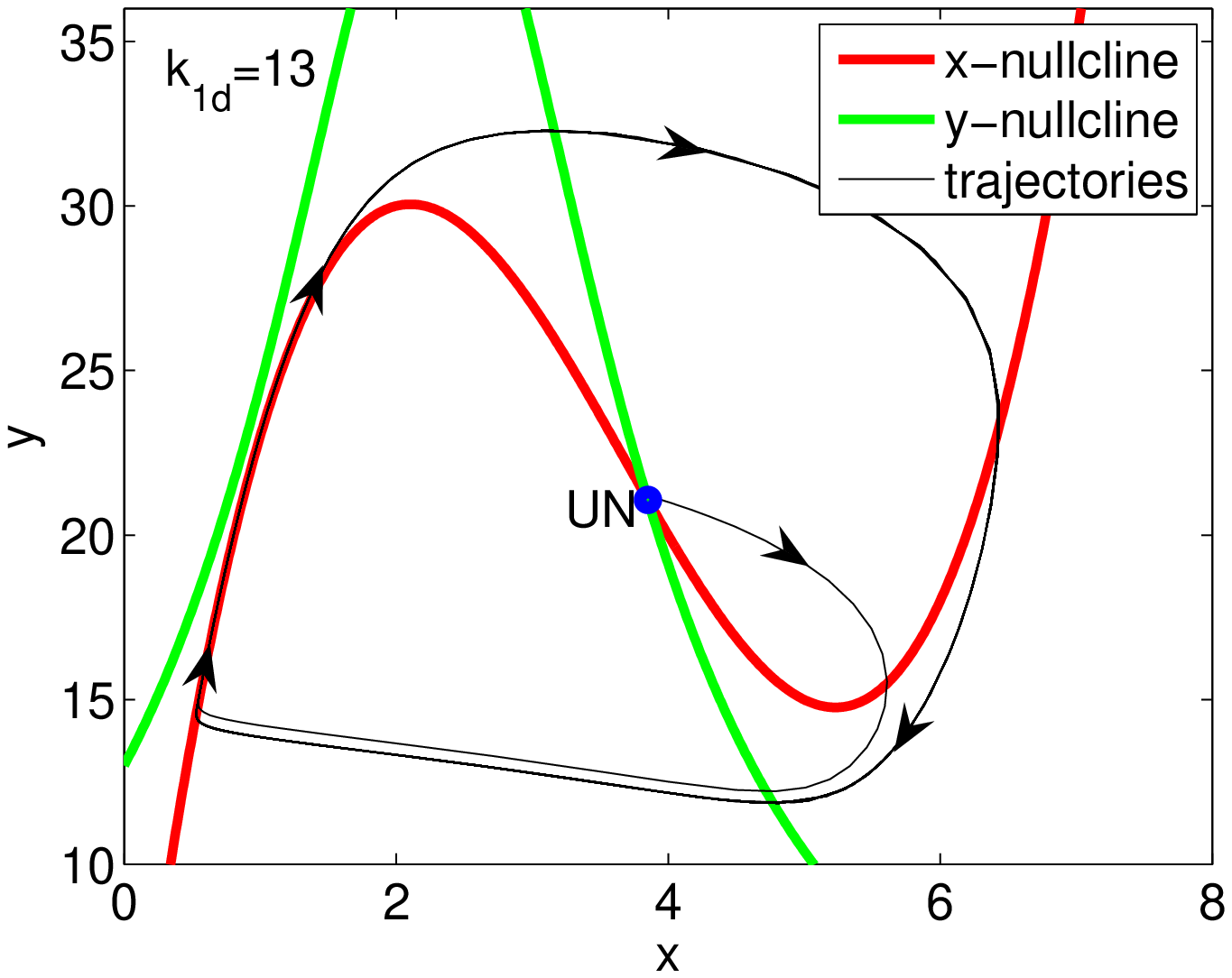}{2.1 in}{6 mm}
\caption{{\it Nullclines of the ODE system $(\ref{eq1})$--$(\ref{eq2})$
for $k_{1d}=12$ (left panel) and $k_{1d}=13$ (right panel).
The other parameter values are given by $(\ref{sniperparam})$.
The steady states are denoted by blue dots. 
Illustrative trajectories which start close to the steady 
states are plotted as thin black lines.
}}
\label{figsnipernullclines}
\end{figure}

Let $X \equiv X(t)$ and $Y \equiv Y(t)$ be the number of molecules of the 
chemical species $X$ and $Y$, respectively. The concentration of $X$ 
(resp. $Y$) will be denoted by $\widetilde{x} = X/V$ (resp. 
$\widetilde{y} = Y/V$). If we have 
enough molecules of $X$ and $Y$ in the system, we often describe
the time evolution of $\widetilde{x}$ and $\widetilde{y}$ 
by the mean-field ODE model. Using
(\ref{model1})--(\ref{model2}), this can be written as
\begin{eqnarray}
\frac{\mbox{d}\widetilde{x}}{\dt} 
& = & 
k_{2d} \, \widetilde{y} 
- 
k_{5d} \, \widetilde{x}^{\raise 0.5mm\hbox{{\scriptsize 3}}} 
+ 
k_{4d} \, \widetilde{x}^{\raise 0.5mm\hbox{{\scriptsize 2}}} 
- 
k_{3d} \, \widetilde{x},
\label{eq1} \\
\frac{\mbox{d}\widetilde{y}}{\dt} 
& = & 
- 
k_{7d} \, \widetilde{x}^{\raise 0.5mm\hbox{{\scriptsize 2}}} \, \widetilde{y}
+ 
k_{6d} \, \widetilde{x} \, \widetilde{y} 
- 
k_{2d} \, \widetilde{y} + k_{1d}.
\label{eq2}
\end{eqnarray}
We choose the values of the rate constants as follows:
\begin{eqnarray}
k_{1d} = 12 \; [\mbox{sec}^{-1} \, \mbox{mm}^{-3}], 
\quad 
&& k_{2d} = 1 \; [\mbox{sec}^{-1}],
\quad 
k_{3d} = 33 \; [\mbox{sec}^{-1}],
\quad 
k_{4d} = 11 \; [\mbox{sec}^{-1} \, \mbox{mm}^{3}],
\nonumber
\\
k_{5d} = 1 \; [\mbox{sec}^{-1} \, \mbox{mm}^{6}],
\quad 
&& k_{6d} = 0.6 \; [\mbox{sec}^{-1} \, \mbox{mm}^{3}],
\quad 
k_{7d} = 0.13 \; [\mbox{sec}^{-1} \, \mbox{mm}^{6}],
\label{sniperparam}
\end{eqnarray}
where we have included the units of each rate constant 
to emphasize the dependence of each rate constant on the 
volume. From now 
on, we will treat all rate constants as numbers, dropping the 
corresponding units, to simplify our notation. The nullclines
of the ODE system (\ref{eq1})--(\ref{eq2}) are plotted
in Figure~\ref{figsnipernullclines}(a). The $x$-nullcline
(resp. $y$-nullcline)
is given by $d\widetilde{x}/dt=0$ (resp. $d\widetilde{y}/dt=0$).
The nullclines intersect
at three steady states which are denoted as SN (stable node),
Saddle and UN (unstable node). We also plot illustrative
trajectories which start close to each steady state (thin black 
lines). 
As the parameter values change, the stable node
(SN) and the saddle can approach each other. 
If they coalesce into one point
(with the dominant eigenvalue equal to 0), a limit cycle
with infinite period appears. Shifting the nullclines
further apart, we obtain a dynamical system with periodic 
solutions. This is demonstrated in 
Figure~\ref{figsnipernullclines}(b) where we plot the nullclines 
of (\ref{eq1})--(\ref{eq2})
for $k_{1d}=13$, with the other parameter values given by 
$(\ref{sniperparam})$. An illustrative trajectory (thin black line)
converges to the limit cycle. We see that there is a SNIPER 
(saddle-node infinite period) bifurcation as the parameter 
$k_{1d}$ is increased from 12 to 13.

The main goal of this paper is to understand and analyse changes
in the behaviour of chemical systems when deterministic
ODE models are replaced by SSAs, i.e. when the intrinsic noise is 
taken into account. The stochastic model of the chemical system 
(\ref{model1})--(\ref{model2}) is given by the Gillespie SSA 
\cite{Gillespie:1977:ESS} which is equivalent 
to solving the corresponding chemical
master equation -- see Appendix \ref{appendixCFPE}. 
We denote the reaction with the rate constant $k_{id}$, $i=1, 2, \dots, 7$, 
as the reaction $R_i$. Note that the reversible reaction 
in (\ref{model1}) is considered as two separate chemical reactions. 
To use the Gillespie SSA, we have to specify the propensity
function $\alpha_i(x,y)$, $i=1, 2, \dots, 7$, of each chemical 
reaction in (\ref{model1})--(\ref{model2}). The propensity function
is defined so that $\alpha_i(x,y) \, \dt$ is the probability that, 
given $X(t)= x$ and $Y(t) = y$, one $R_i$ reaction will occur 
in the next infinitesimal time interval $[t,t+\dt)$. 
For example, $k_{1d}$ is the rate of production of $Y$ molecules per unit 
of volume. Thus, $k_1 = k_{1d} V$ is the rate of production of $Y$ molecules 
per the whole reactor of volume $V$. Consequently, the probability that one 
$Y$ molecule is produced in the infinitesimally small time interval 
$[t,t+\dt)$ is equal to $k_1 \dt$, i.e. the propensity function of
the reaction $R_1$ is $\alpha_1(x,y) \equiv k_1 = k_{1d} V.$ To specify other
propensity functions, we first scale the rate constant $k_{id}$ 
with the appropriate power of the volume $V$, namely we define
\begin{equation}
k_{1} = k_{1d} V, 
\; \;
k_{2} = k_{2d},
\; \;
k_{3} = k_{3d},
\; \;
k_{4} = \frac{k_{4d}}{V},
\;\;
k_{5} = \frac{k_{5d}}{V^2},
\;\;
k_{6} = \frac{k_{6d}}{V},
\;\;
k_{7} = \frac{k_{7d}}{V^2}.
\label{rateconstantsvolume}
\end{equation}
Then the propensity function of the reaction $R_i$, 
$i=1,2, \dots, 7$, is given as the product of the
scaled rate constant and numbers of available reactant 
molecules, namely
$$
\alpha_1(x,y) = k_1, 
\; \;
\alpha_2(x,y) = k_2 y,
\; \;
\alpha_3(x,y) = k_3 x,
\; \;
\alpha_4(x,y) = k_4 x (x-1),
$$
\begin{equation}
\alpha_5(x,y) = k_5 x (x-1) (x-2),
\;\;
\alpha_6(x,y) = k_6 x y,
\;\;
\alpha_7(x,y) = k_7 x (x-1) y.
\label{propfunctions}
\end{equation}
Note that the propensity functions of the reactions $R_4$ and $R_7$ 
are proportional to the number $x(x-1)/2$ of available pairs of $X$ 
molecules, and not to $x^2$; similarly, $\alpha_5$ is proportional
to the number $x(x-1)(x-2)/6$ of available triplets of $X$ molecules
and not to $x^3$. For a general discussion of the propensity functions
see \cite{Gillespie:1977:ESS}.

\begin{figure}
\picturesABal{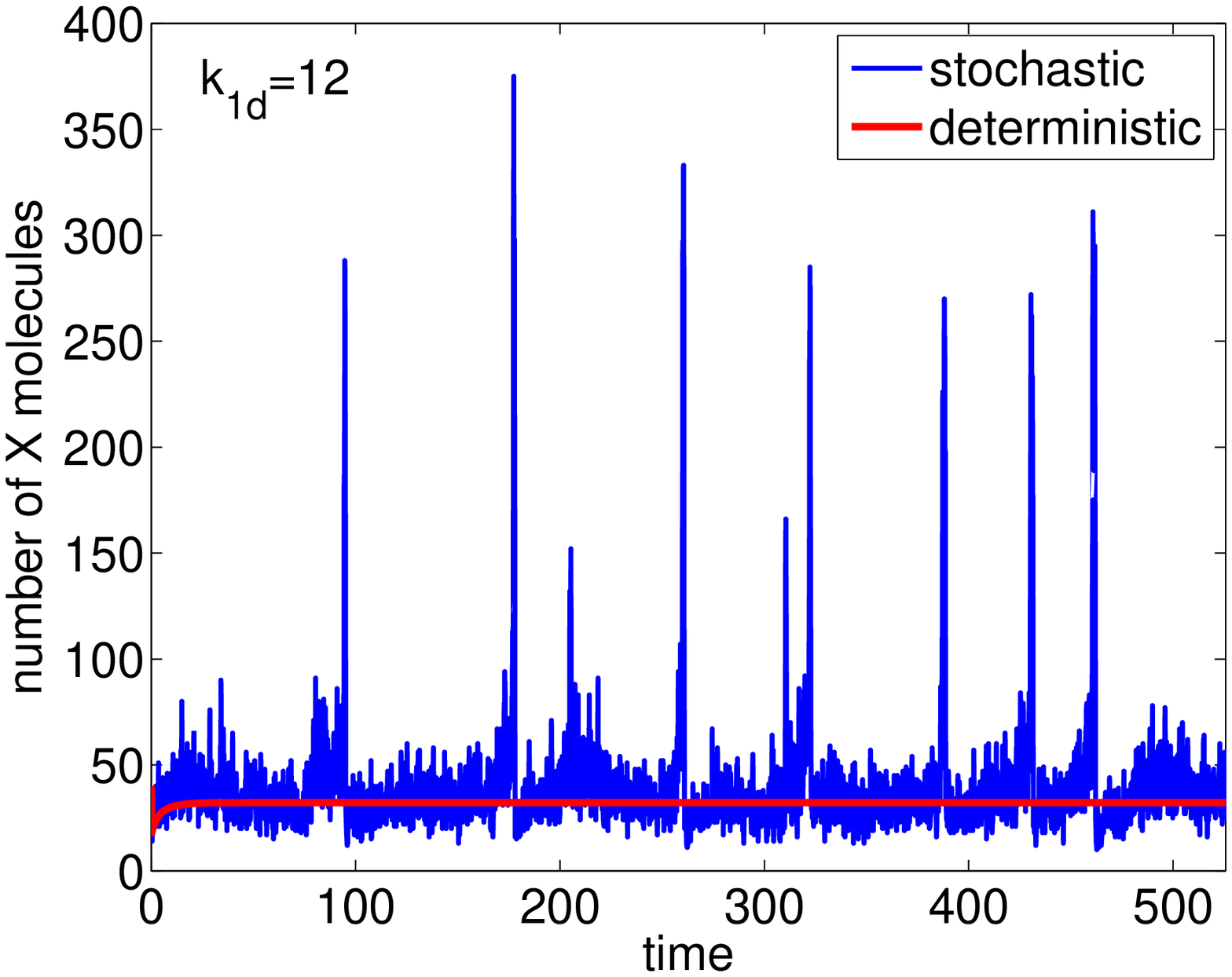}{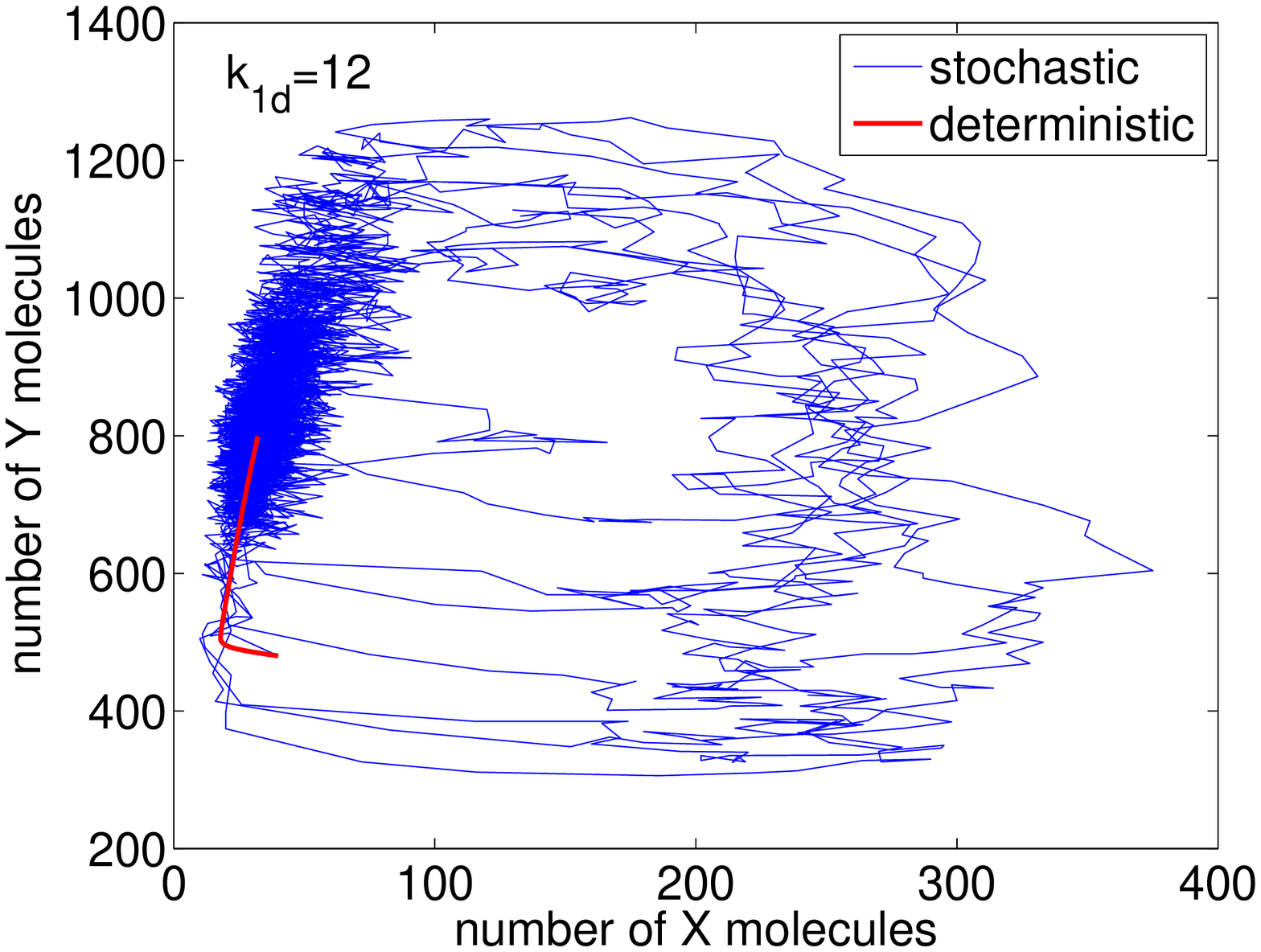}
{2.05in}{6mm}{2mm}
\caption{{\rm (a)} 
{\it The time evolution of $X$ given by the stochastic 
(blue line) and  deterministic (red line) models of the chemical 
system $(\ref{model1})$--$(\ref{model2})$. The rate constants are 
given by $(\ref{sniperparam})$ and $V=40.$}
{\rm (b)} {\it The same trajectory plotted in the $X$-$Y$ plane.}
}
\label{fig12timeevol}
\end{figure}

\begin{figure}
\picturesABal{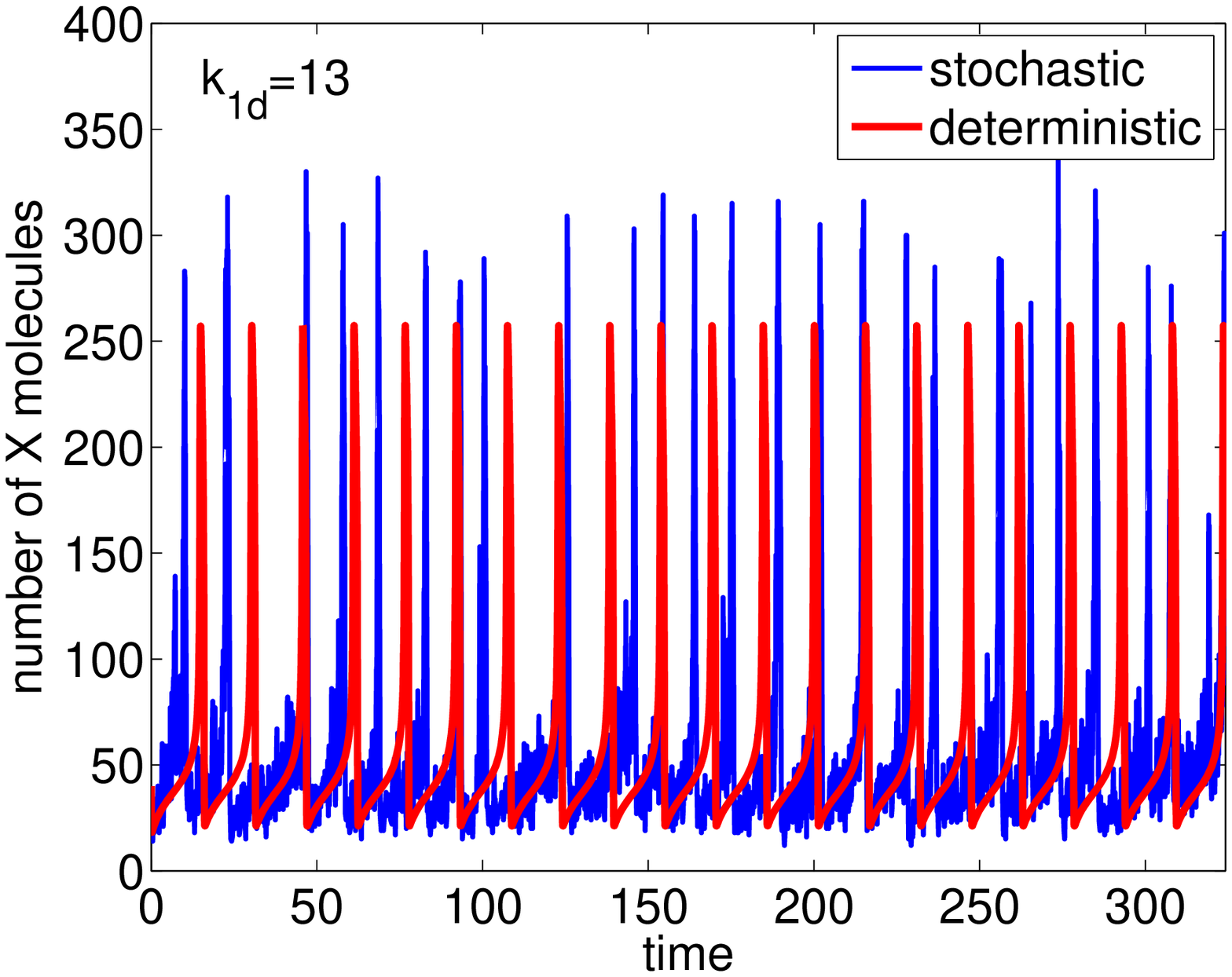}{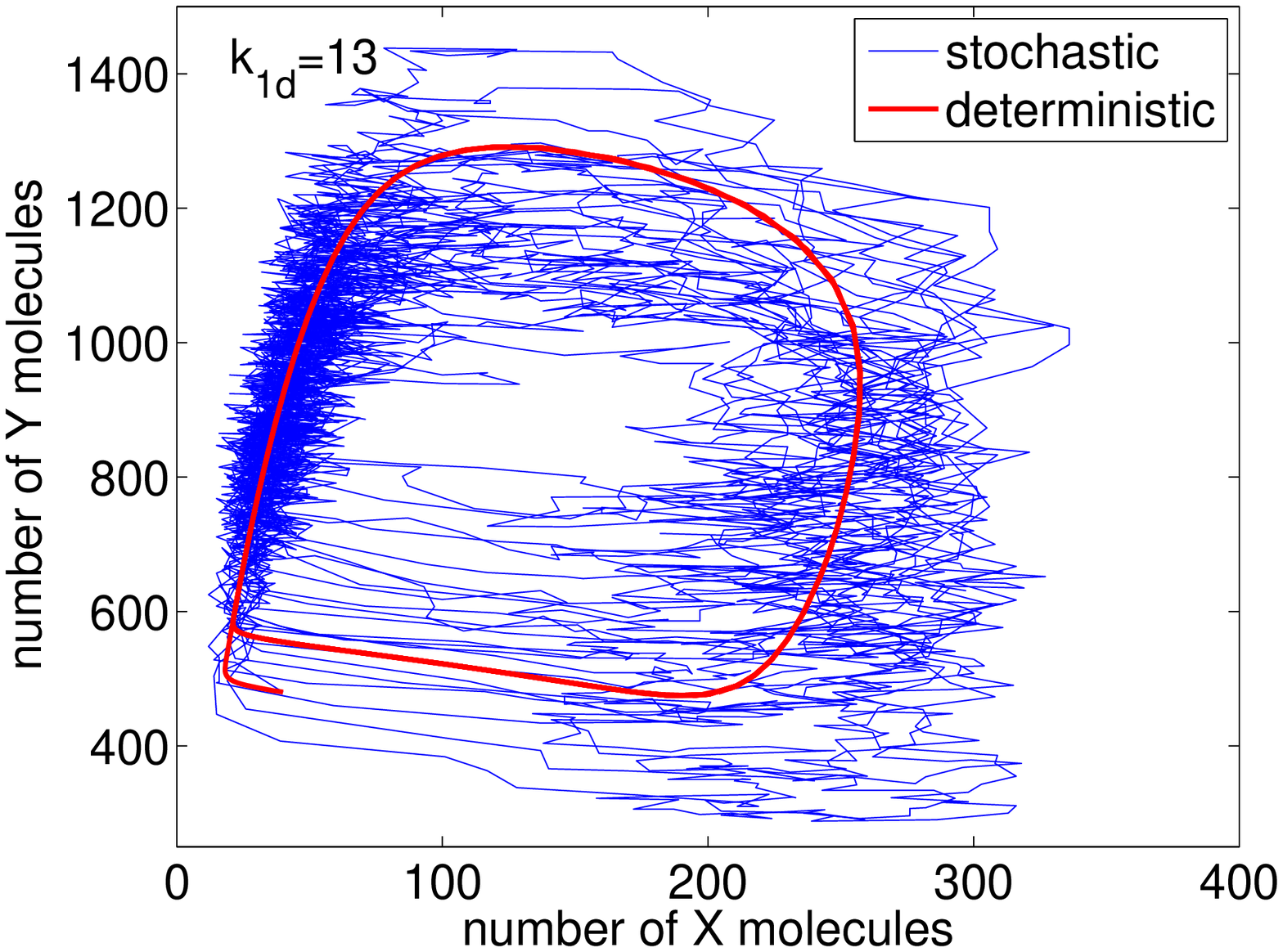}
{2.05in}{6mm}{2mm}
\caption{{\rm (a)} 
{\it The time evolution of $X$ given by the stochastic 
(blue line) and  deterministic (red line) models of the chemical 
system $(\ref{model1})$--$(\ref{model2})$ for $k_{1d} = 13$.
The other parameters are chosen as in Figure~$\ref{fig12timeevol}$.}
{\rm (b)}
{\it The same trajectory plotted in the $X$-$Y$ plane.}
}
\label{fig13timeevol}
\end{figure}

Using (\ref{propfunctions}) and the Gillespie SSA, we can simulate
the stochastic trajectories of (\ref{model1})--(\ref{model2}).
In Figure~\ref{fig12timeevol}(a), we compare the time evolution 
of $X$ given by the stochastic (Gillespie SSA) and deterministic
(the ODE system (\ref{eq1})--(\ref{eq2})) models. We use the same 
parameter values (\ref{sniperparam}) and the same initial 
condition $[X,Y] = [40,480]$ for both, stochastic and deterministic, 
simulations. The reactor volume is $V=40$. 
In Figure~\ref{fig12timeevol}(b), we plot both trajectories in the
$x$-$y$ plane. We see that the solution of the deterministic equations 
converges to a steady state while the stochastic model has oscillatory 
solutions. In Figure~\ref{fig13timeevol}, we present results obtained by 
the stochastic simulation of (\ref{model1})--(\ref{model2}) for 
$k_{1d}=13$. The other 
parameters are the same as in Figure~\ref{fig12timeevol}.
We see that both the stochastic and deterministic models oscillate for 
$k_{1d}=13$.

\begin{figure}
\picturesAB{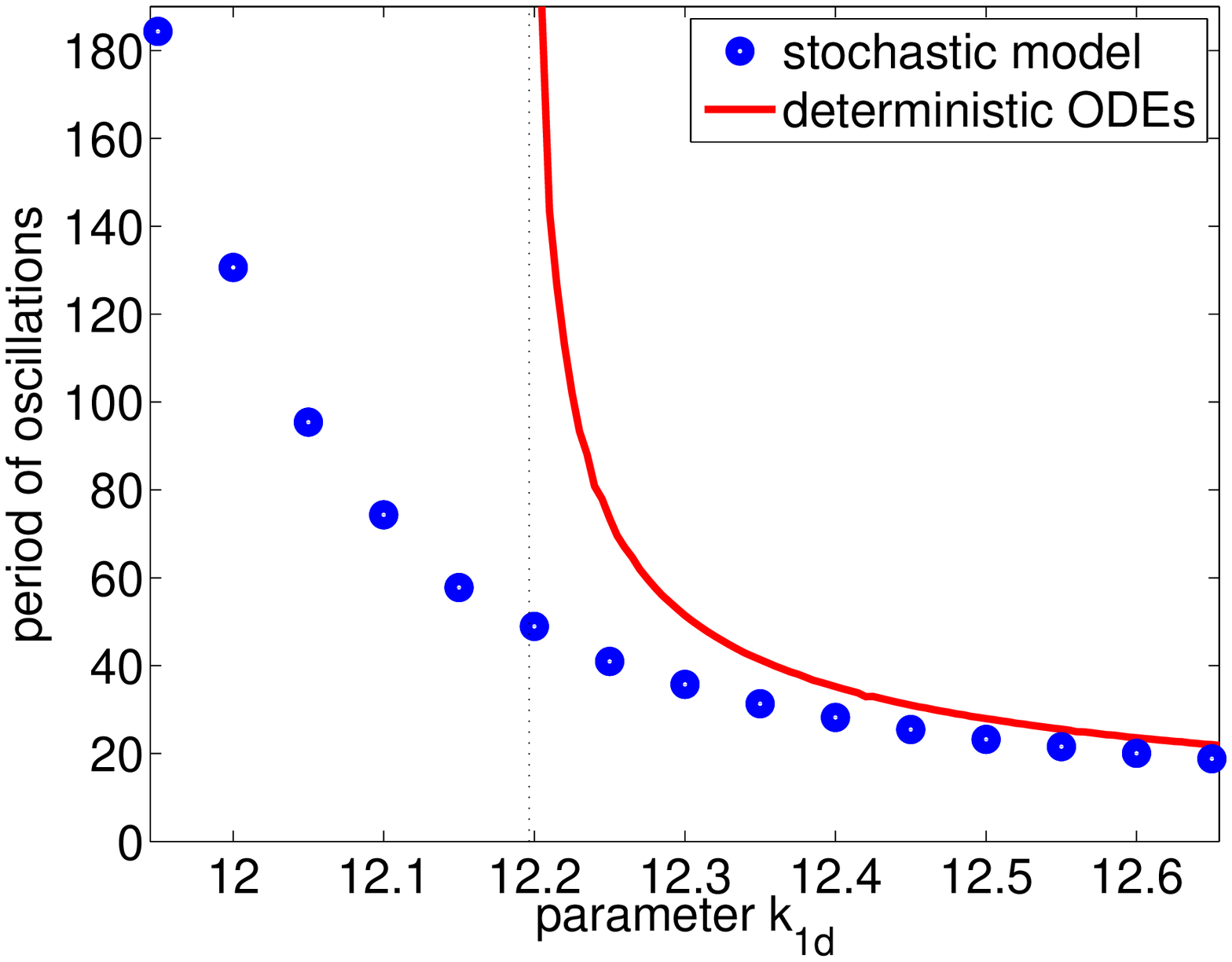}{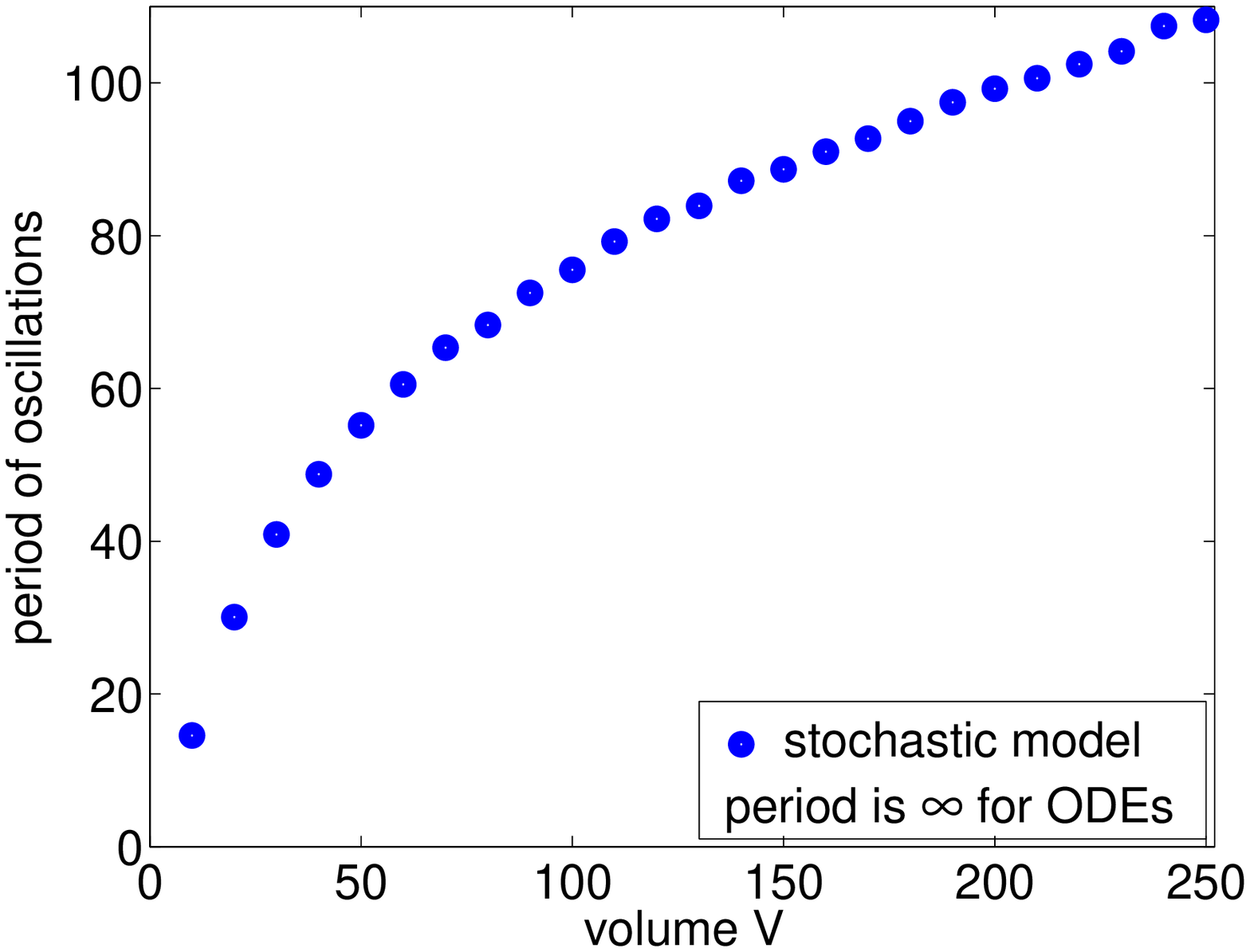}
{2.05 in}{6mm}
\caption{{\rm (a)}
 {\it The mean period of oscillation of the chemical
system $(\ref{model1})$--$(\ref{model2})$ as a function
of the parameter $k_{1d}$. The other parameters are given
by $(\ref{sniperparam})$ and $V=40.$ Each blue circle was obtained
as an average over 10,000 periods of the stochastic model. The
mean period of oscillation of the ODE model $(\ref{eq1})$--$(\ref{eq2})$
is plotted as the red line.} 
{\rm (b)} 
{\it The period of oscillation as a function of the volume $V$.
We use $k_{1d} = K_d \doteq 12.2$. Parameters $k_{2d}, \dots, k_{7d}$ 
are given by $(\ref{sniperparam})$.}
}
\label{figperosc}
\end{figure}

An important characteristic of oscillating systems is their period
of oscillation. This is a well-defined number for the deterministic
ODE model, but in the stochastic case the periods of 
individual oscillations vary. Thus, in the stochastic model
we are interested in the mean period of oscillation 
(averaged over many periods). The mean period of oscillation
is plotted in Figure~\ref{figperosc}(a) as a function of the 
rate constant $k_{1d}$ (blue circles). It was computed
as an average over 10,000 periods for each of the presented
values of $k_{1d}$. 
The starting/finishing point of every period was defined
as the time when the simulation enters the half-plane $X>200$.
More precisely, we start the computation of each period 
\label{pagedefcycle}
whenever the trajectory enters the area $X>200$. Then we 
wait until $X<60$ before we test the condition $X>200$ 
again. The extra condition $X<60$ guarantees that possible small random 
fluctuations around the point $X=200$ are not counted as two 
or more periods. The period of oscillation of the ODE model
(\ref{eq1})--(\ref{eq2})
is plotted in Figure~\ref{figperosc}(a) as the red line. It 
asymptotes to infinity as $(k_{1d} - K_d)^{-1/2}$ for
$k_{1d} \to K_d^+$ where $K_d \doteq 12.2$ is the bifurcation value 
of the parameter $k_{1d}$. 

In the limit $V \to \infty$ (which is the so-called
thermodynamic limit \cite{Gillespie:2000:CLE}), 
the stochastic description converges to 
the ODE model $(2.3)$-$(2.4)$, that is, the probability distributions 
become Dirac-like and their averages converge to the solution
of the ODEs $(2.3)$-$(2.4)$ for $V \to \infty$.
The dependence of the period of oscillation on the volume 
$V$ is shown in Figure~\ref{figperosc}(b) where we fix
$k_{1d}$ equal to the bifurcation value $K_d \doteq 12.2$
of the ODE model and we vary the volume $V$. The other parameter
values are given by $(\ref{sniperparam})$. Since $k_{1d}=K_d$,
the period of oscillation of the ODE model is infinity.
In Figure~\ref{figperosc}(b), we see that the period of oscillation 
of the stochastic model is an increasing function of $V$.
It approaches the period of oscillation of the ODE model
(infinity) as $V \to \infty$. 

The estimates of the period of oscillation (blue circles in 
Figure~\ref{figperosc}) were obtained as averages over 10,000 periods 
of the Gillespie SSA. Such an approach is computationally intensive.
The goal of this paper is to show that we can obtain the same 
information by solving and analyzing the chemical Fokker-Planck 
equation. In Section \ref{secnumerFPres}, we present results obtained 
by solving the Fokker-Planck equation using a suitable finite
element method. In Section \ref{secanalFPres}, we use the asymptotic analysis 
of the Fokker-Planck equation to derive explicit formulae for the 
period of oscillation, as a function of the rate constants and 
as a function of the volume $V$. The chemical Fokker-Planck equation
which corresponds to the chemical system (\ref{model1})--(\ref{model2})
is a two-dimensional partial differential equation. In particular, 
some parts
of its analysis are more technical than the analysis of one-dimensional
problems. To get some insights into our approach, we analyse
a one-dimensional chemical switch in the following section. We will present
only methods which are of potential use in the higher-dimensional
settings. A generalization of the analysis to the chemical SNIPER problem 
is shown in Section \ref{secanalFPres}.

\section{One-dimensional chemical switch} \label{sec1Dswitch}
\begin{figure}
\picturesAB{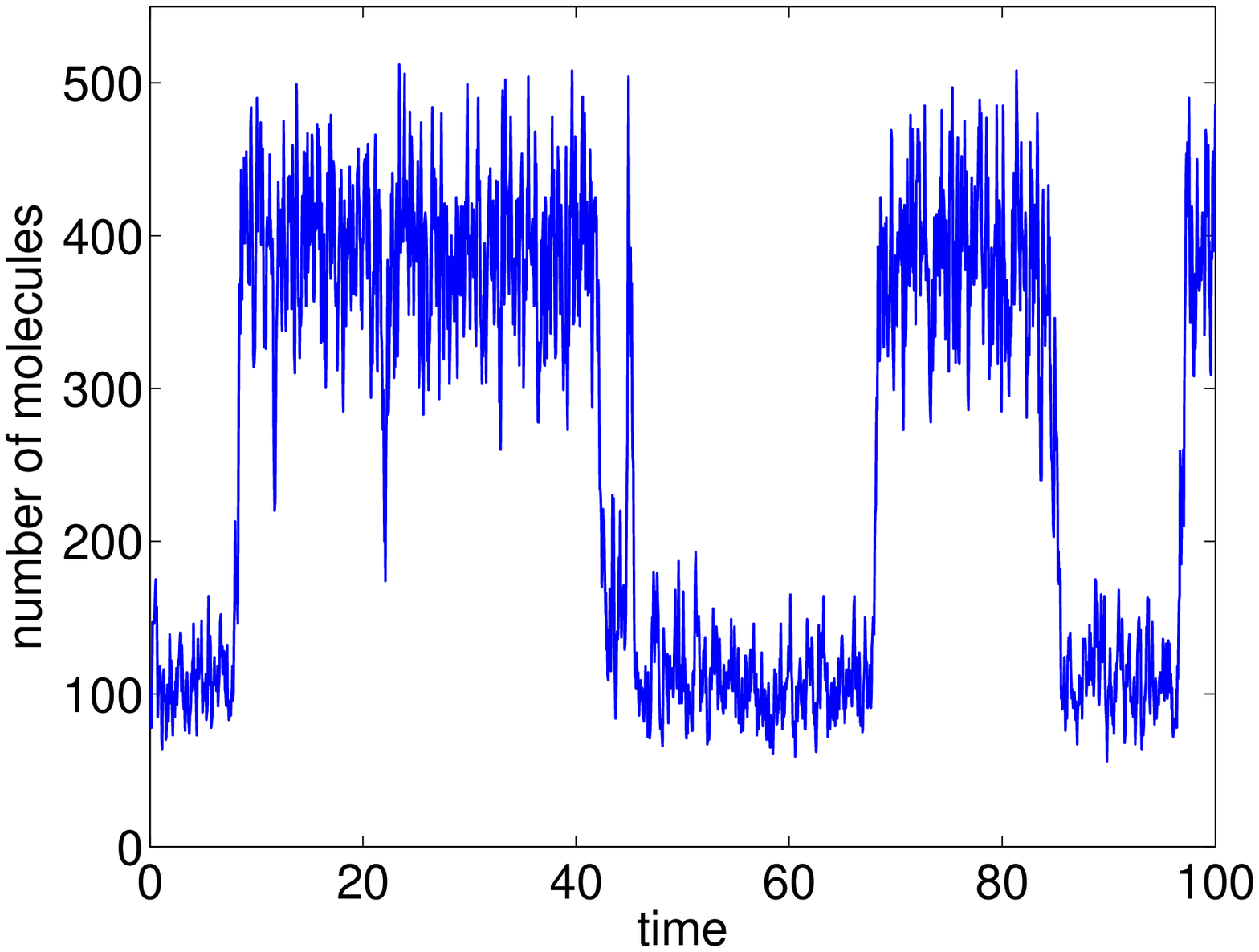}{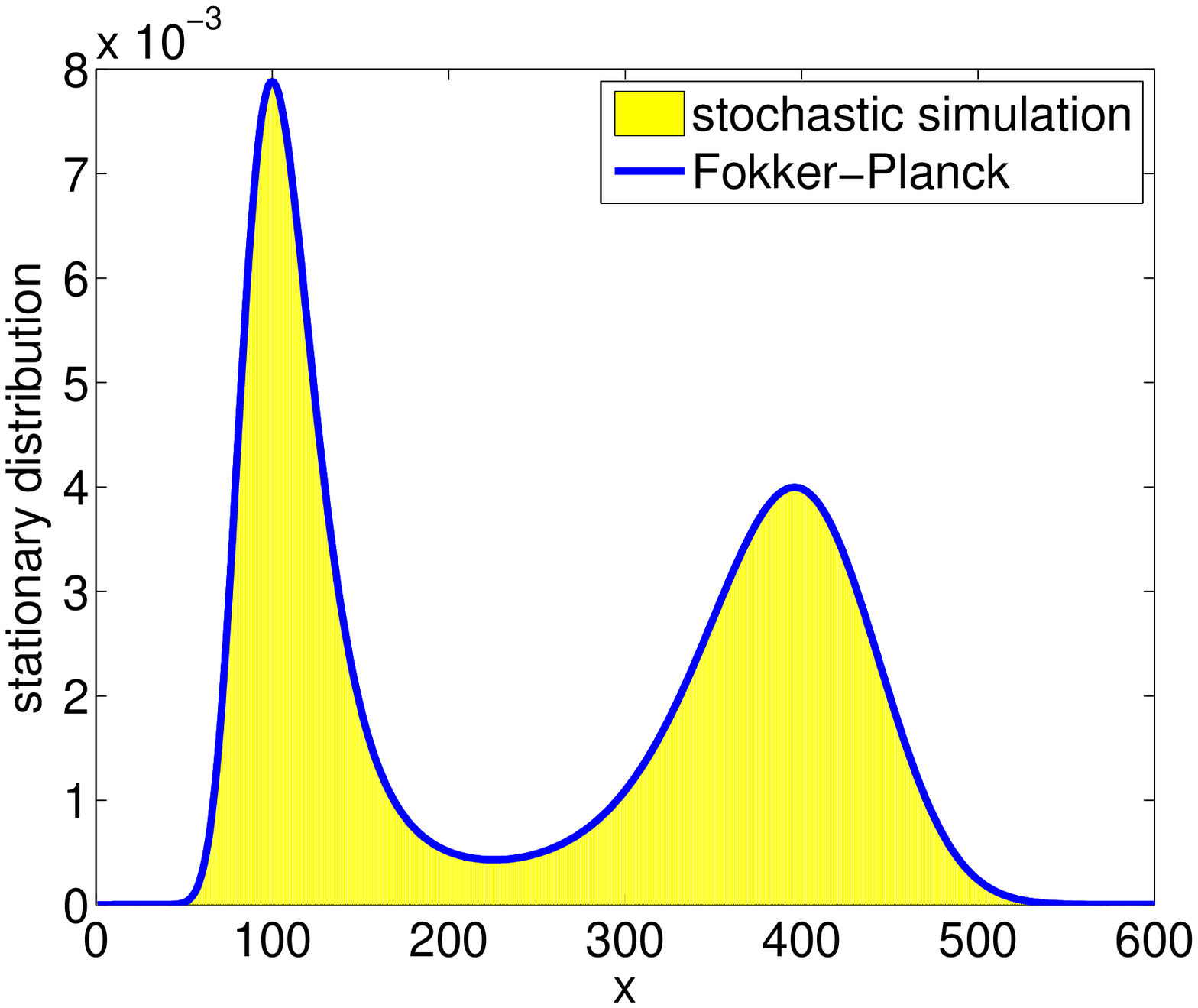}
{2.05 in}{6mm}
\caption{{\rm (a)} 
{\it Time evolution of $X$ obtained by 
the Gillespie SSA for the chemical system $(\ref{schlogl})$. 
The values of the rate constants are given by
$(\ref{paramschlogl})$.} 
{\rm (b)}
{\it Stationary distribution of $(\ref{schlogl})$ 
obtained by the Gillespie SSA (yellow histogram) for the parameters 
$(\ref{paramschlogl})$. The blue curve is the stationary
solution $(\ref{Psvd})$ of the chemical Fokker-Planck equation.}
}
\label{figtrajhistschlogl}
\end{figure}

We consider a chemical $X$ in a container of volume $V$ which is 
subject to the following set of chemical reactions 
(such a system was introduced by Schl\"ogl \cite{Schlogl:1972:CRM})
\begin{equation}
\emptyset
\;
\mbox{ 
\raise -0.6 mm 
\hbox{$\displaystyle
\mathop{
\stackrel{\displaystyle\longrightarrow}\longleftarrow}^{k_{1d}}_{k_{2d}}
$}}
\;\;
X,
\qquad\qquad\qquad
2 X
\;
\mbox{ 
\raise -0.6 mm 
\hbox{$\displaystyle
\mathop{\stackrel{\displaystyle\longrightarrow}\longleftarrow}^{k_{3d}}_{k_{4d}}
$}}
\;\;
3 X.
\label{schlogl}
\end{equation}
Let $X(t)$ be the number of molecules of the chemical $X$. 
The classical deterministic description of the chemical system
(\ref{schlogl}) is given by the following mean-field ODE for 
the concentration $\widetilde{x}=X/V$:
\begin{equation}
\frac{\mbox{d}\widetilde{x}}{\dt}
=
k_{1d} 
- 
k_{2d} \, \widetilde{x}
+ 
k_{3d} \, \widetilde{x}^{\raise 0.5mm\hbox{{\scriptsize 2}}} 
- 
k_{4d} \, \widetilde{x}^{\raise 0.5mm\hbox{{\scriptsize 3}}}.
\label{detschloglODE}
\end{equation} 
To obtain the stochastic description, we first scale the rate constants with 
the appropriate powers of the volume $V$, by defining
\begin{equation}
k_{1} = k_{1d} V, 
\qquad
k_{2} = k_{2d},
\qquad
k_{3} = \frac{k_{3d}}{V},
\qquad
k_{4} = \frac{k_{4d}}{V^2}.
\label{rateconstantsvolumeschlogl}
\end{equation}
Then the propensity functions of the chemical reactions (\ref{schlogl}) 
are given by
\begin{equation}
\alpha_1(x)= k_1,
\;\,
\alpha_2(x)= k_2 x,
\;\,
\alpha_3(x)= k_3 x (x-1),
\;\,
\alpha_4(x)= k_4 x (x-1) (x-2),
\label{propschlogl}
\end{equation}
i.e. the probability that, given $X(t)=x$, the $i$-th reaction occurs in 
the time interval
$[t, t + \dt)$ is equal to $\alpha_i(x) \; \dt$. Given the propensity
functions (\ref{propschlogl}), we can simulate the time evolution
of the system (\ref{schlogl}) by the Gillespie SSA \cite{Gillespie:1977:ESS}. 
We choose $V=1$ in what follows, i.e. $k_{i} = k_{id}$, $i = 1, 2, 3, 4.$
In Figure~\ref{figtrajhistschlogl}(a), we plot illustrative results 
obtained by the Gillespie SSA for the following set of rate constants
\begin{equation}
    k_1 = 2250, \qquad 
    k_2 = 37.5, \qquad
    k_3 = 0.18, \qquad 
    k_4 = 2.5 \times 10^{-4}. 
\label{paramschlogl}
\end{equation}%
We see that the system (\ref{schlogl}) has two favourable states 
for the parameter values (\ref{paramschlogl}). We also plot stationary 
distributions (yellow histograms) obtained by long time simulation of 
the Gillespie SSA in Figure~\ref{figtrajhistschlogl}(b).
The chemical master equation corresponding to (\ref{schlogl}) 
can be written as follows
\begin{equation}
\frac{\partial}{\partial t}
p(x,t)
=
\sum_{i=1}^4 
\Big[
\alpha_i(x+(-1)^i) \, p(x+(-1)^i,t)
- 
\alpha_i(x) p(x,t)
\Big]
\label{CMEschlogl}
\end{equation}
where $p(x,t)$ is the probability that $X(t)=x$, i.e. the probability
that there are $x$ molecules of the chemical species $X$ at time $t$
in the system. The stationary solution of this infinite set of ODEs
can be found in a closed form \cite{Hinch:2005:EST}, 
i.e. one can find an exact formula 
for the stationary distribution plotted in 
Figure~\ref{figtrajhistschlogl}(b). However, our goal is not to solve
the well-known Schl\"ogl system using its master equation. 
We want to motivate the approach which is used later for the 
chemical SNIPER problem, which is based on the approximate 
description of chemical systems given by the chemical
Fokker-Planck equation \cite{Gillespie:2000:CLE}, see
Appendix \ref{appendixCFPE}. To write this equation, we have to consider
$p(x,t)$ as a function of the real variable $x$, i.e. we
smoothly extend the function $p(x,t)$ to non-integer 
values of $x$. Using Appendix \ref{appendixCFPE}, the chemical Fokker-Planck
equation for the chemical system (\ref{schlogl}) is
\begin{equation}
\frac{\partial p}{\partial t}(x,t)
= 
\frac{\partial^2}{\partial x^2}
\Big(
d(x) p(x,t)
\Big)
-
\frac{\partial}{\partial x}
\Big(
v(x) p(x,t)
\Big)
\label{FPEschlogl}
\end{equation}
where the drift coefficient $v(x)$ and the diffusion coeficient $d(x)$ 
are given by
\begin{equation}
v(x) 
= 
\sum_{i=1}^4 (-1)^{i+1} \alpha_i(x)
=
k_1 - k_2 \, x + k_3 \, x (x-1) - k_4 \, x (x-1) (x-2),
\label{eqvx}
\end{equation}
\begin{equation}
d(x) 
= 
\frac{1}{2} \sum_{i=1}^4 \alpha_i(x)
=
\frac{k_1 + k_2 \, x + k_3 \, x (x-1) + k_4 \, x (x-1) (x-2)}{2}.
\label{eqdx}
\end{equation}
The stationary distribution $P_s(x) = \lim_{t \to \infty} p(x,t)$
is a solution of the stationary equation corresponding to
(\ref{FPEschlogl}), namely
\begin{equation}
\frac{\mbox{d}^2}{\dx^2}
\Big(
d(x) P_s(x)
\Big)
-
\frac{\mbox{d}}{\dx}
\Big(
v(x) P_s(x)
\Big)
=
0.
\label{statFPEschlogl}
\end{equation}
Integrating over $x$ and using the boundary conditions $P_s(x) \to 0$ as 
$x \to \pm \infty$, we obtain
\begin{equation} 
P_s(x)
=
\frac{c}{d(x)}
\exp \left[ \int_{0}^x \frac{v(z)}{d(z)} \,\dz \right]
\label{Psvd}
\end{equation}
where $c$ is the constant given by the normalization $\int_\er P_s(x) \dx = 1.$
The function $P_s$ is plotted in Figure~\ref{figtrajhistschlogl}(b)
for comparison as the blue line. We see that the chemical
Fokker-Planck equation gives a very good description of the 
chemical system (\ref{schlogl}).

The Fokker-Planck equation (\ref{FPEschlogl}) is equivalent to the 
Langevin equation (It\^{o} stochastic differential equation)
$$
\mbox{\dX}
=
v(X) \, \dt
+
\sqrt{2 d(X)} \, \dW
$$
where $dW$ represents white noise. In particular, the deterministic part 
of the dynamics is given by the ODE $\dX/dt = v(X).$ Substituting 
(\ref{eqvx}) for $v(X)$,
using (\ref{rateconstantsvolumeschlogl}) and dividing by $V$, 
we obtain the following ODE for the concentration
$\widetilde{x} = X/V$:
\begin{equation}
\frac{\mbox{d}\widetilde{x}}{\dt}
=
k_{1d} 
- 
k_{2d} \, \widetilde{x} 
+ 
k_{3d} \, \widetilde{x} \left( \widetilde{x} - \frac{1}{V} \right)
- 
k_{4d} \, \widetilde{x} \left( \widetilde{x} - \frac{1}{V} \right)
\left( \widetilde{x} - \frac{2}{V} \right).
\label{detschlogldriftODE}
\end{equation} 
This equation slightly differs from the classical 
mean-field deterministic description (\ref{detschloglODE}), but
the equations (\ref{detschloglODE}) and (\ref{detschlogldriftODE})
are equivalent in the limit of large $X$ in which 
$\widetilde{x} \gg 1/V$. This can also be thought of
as the limit of large $V$ after a suitable non-dimensionalization. 
Let us note that the chemical Fokker-Planck equation is actually 
derived from the Kramers-Moyal expansion in the limit of 
large $V$ by keeping only the first and the second derivatives 
in the expansion \cite{Gillespie:2000:CLE}.

\subsection{Mean switching time}
\begin{figure}[t]
\picturesAB{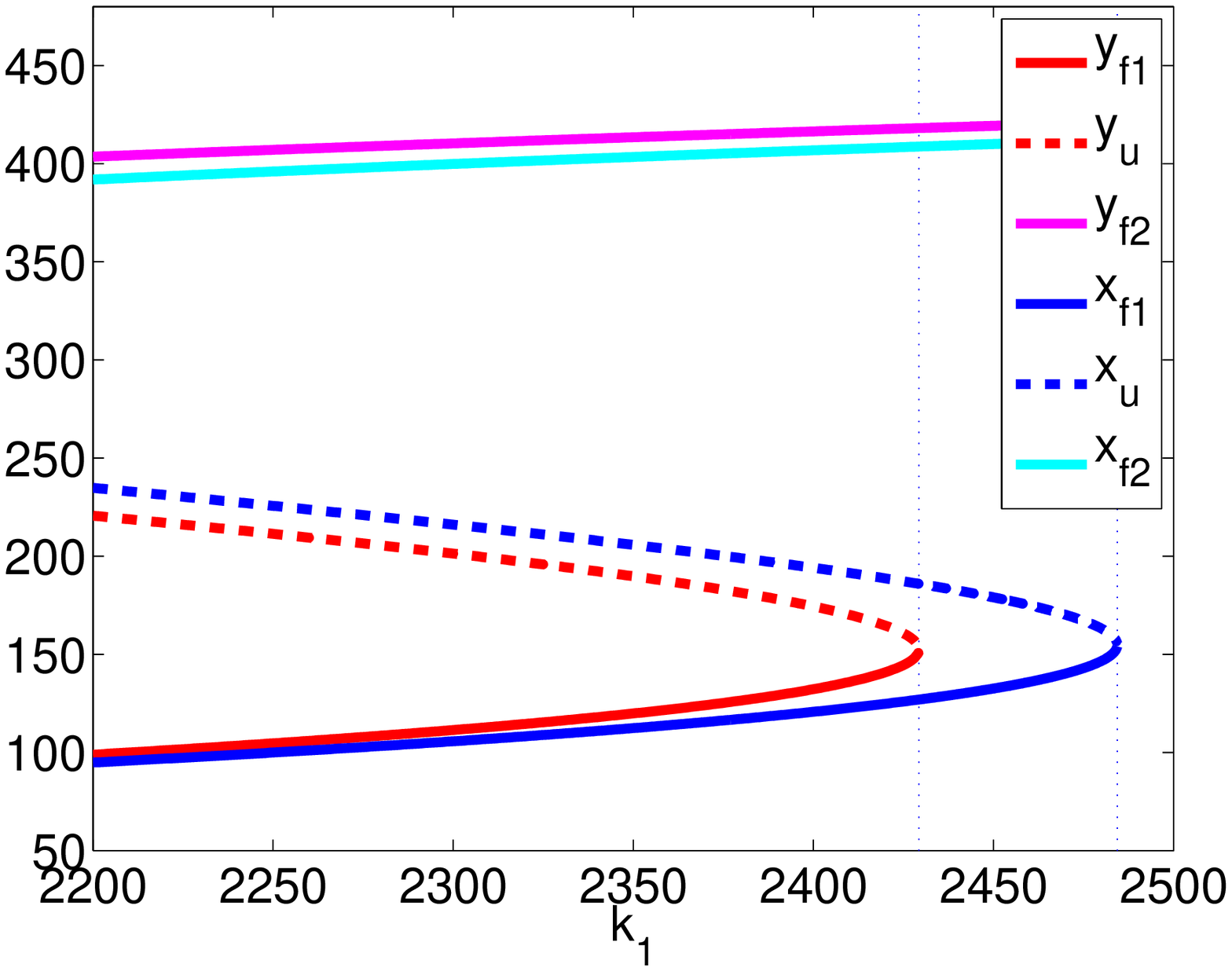}{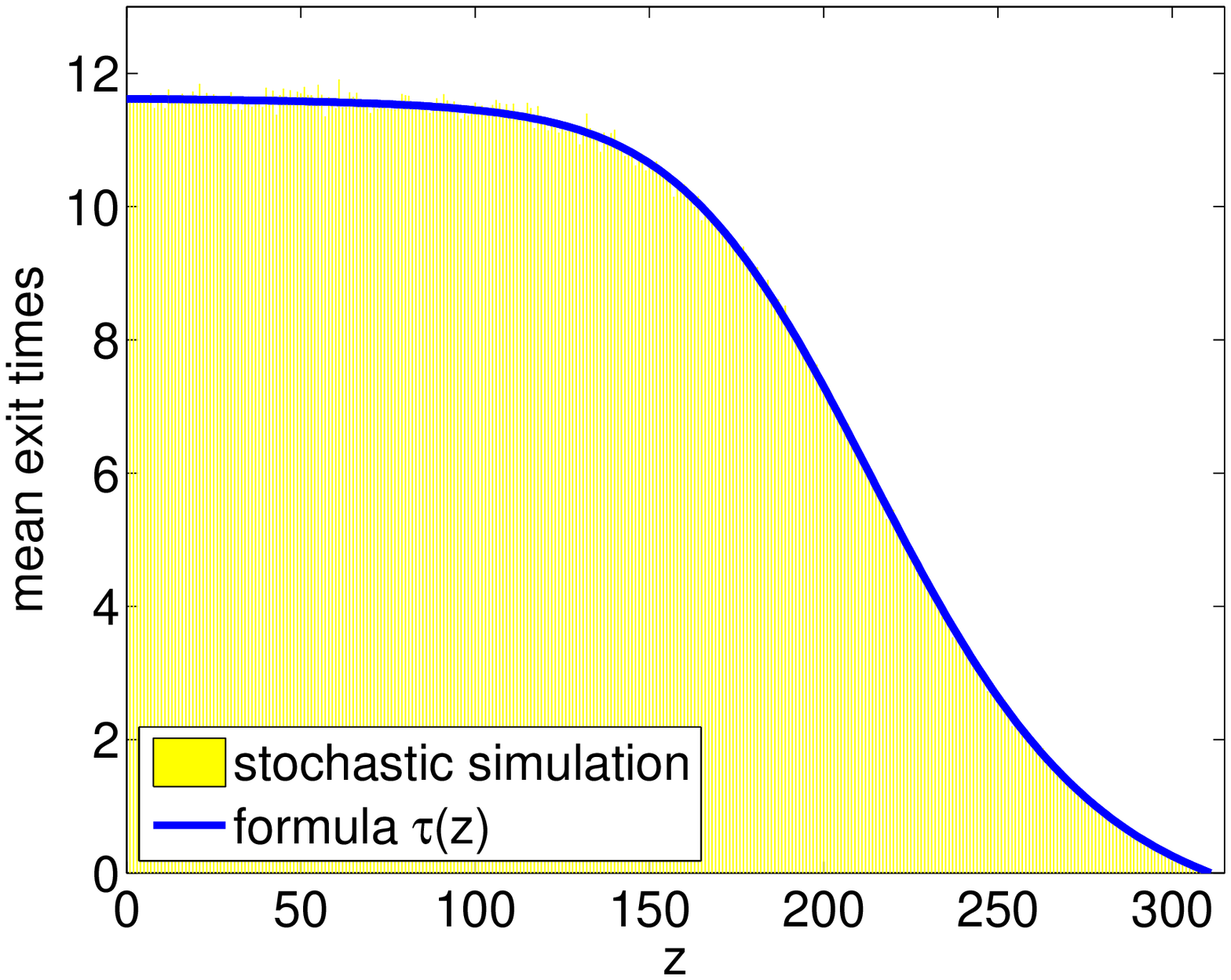}{2.05 in}{6mm}
\caption{{\rm (a)} 
{\it The local extrema of the potential
(i.e. the solutions $x_{f1} < x_{u} < x_{f2}$ of the
equation $(\ref{xfueq})$) and the solutions 
 $y_{f1} < y_{u} < y_{f2}$ of the equation
$v(y)=0$ as functions of the parameter $k_1$.
The critical values $K$ and $K_x$ are plotted as the blue
dotted lines.} 
{\rm (b)} 
{\it Mean exit times $\tau(z)$ to leave the 
interval $(-\infty, (x_u + x_{f2})/2)$,
given that initially $X(0)=z$. We compare the results obtained by
the Gillespie SSA (yellow histogram, averages over $10^4$ exits
for each $z$) 
with the results obtained by the formula $(\ref{tauzformula})$
(blue curve). Parameter values are given by $(\ref{paramschlogl})$.}
}
\label{figxyfuexittime}
\end{figure}

Let $x_{f1}$ and $x_{f2}$ be favourable states of the chemical system 
(\ref{schlogl}) which we define as the arguments of the maxima of $P_s$ 
(given by (\ref{Psvd})). 
Let $x_u$ be the local minimum of $P_s$ which lies between the two 
favourable states, so that $x_{f1} < x_u < x_{f2}$.
We can find the local extrema $x_{f1},$ $x_{u}$ and $x_{f2}$ of
$P_s$ as the solutions of $P_s^\prime(x)=0$, which, using (\ref{Psvd}), 
is equivalent to the cubic equation
\begin{equation}
v(x)-d^\prime(x) = 0.
\label{xfueq}
\end{equation}
Another way to define the favourable states of the system is by
considering the stationary points of the ODE (\ref{detschlogldriftODE}),
i.e. the points where the drift coefficient is zero:
\begin{equation}
v(x) 
=
k_1  - k_2 \, x  + k_3 \, x (x-1)- k_4 \, x (x-1) (x-2)
=
0.
\label{yfueq}
\end{equation}
This cubic equation has three roots which we denote
$y_{f1}$, $y_u$ and $y_{f2}$, where $y_{f1} < y_{u} < y_{f2}.$ 
Let us note that $x_{f1} \ne y_{f1}$,
$x_u \ne y_u$ and $x_{f2} \ne y_{f2}$ because the equation
(\ref{xfueq}) differs from the equation (\ref{yfueq}) by the
additional term $d^\prime(x)$. In Figure~\ref{figxyfuexittime}(a), 
we plot $y_{f1},$ $y_{u},$ $y_{f2}$ as functions of $k_1$.
The other parameter values are chosen as in (\ref{paramschlogl}).
We also plot $x_{f1},$ $x_{u},$ $x_{f2}$. We see that there are two critical 
values of the parameter $k_1$, namely
$K \doteq 2429.35$ and $K_x \doteq 2484.39$.
If $k_1=K$, then $y_{f1} = y_u$, while if $k_1=K_x$, then $x_{f1} = x_u$,
i.e. the number of local maxima of $P_s$ changes at $k_1 = K_x$. 
From our point of view, the more interesting critical value is $K$,
corresponding to the bifurcation value of the deterministic
description (\ref{detschlogldriftODE}), but it is important to note
that this value differs from the value $K_x$. To find the value of $K$, 
we solve the quadratic equation $v^\prime(x)=0$. The relevant root
is given by:
\begin{equation}
y_{k}
=
1
+
\frac{k_3}{3 k_4}
-
\frac{1}{6 k_4}
\sqrt{(6 k_4 + 2 k_3)^2 - 12 k_4 (2 k_4 + k_3 + k_2)}.
\label{valueofyk}
\end{equation}
The value of $K$ is determined by the equation $y_{f1} = y_u = y_k$,
giving
$$
K
=
k_4 \, y_{k} (y_{k}-1) (y_{k}-2) 
- 
k_3 \, y_{k} (y_{k}-1) 
+
k_2 \, y_{k}.
$$
For the parameter values (\ref{paramschlogl}), we obtain 
$y_k \doteq 152.45$ and $K \doteq 2429.35.$

Let $\tau(y)$ be the average time to leave the interval $(-\infty,b)$ 
given that we start at $X(0)=y$. Using the backward Kolmogorov equation
(adjoint equation to the Fokker-Planck equation (\ref{FPEschlogl})), 
one can derive a differential equation for $\tau(y)$ 
(see \cite{Fox:1986:FAS} and also Appendix \ref{appendixmtime}),
namely
\begin{equation}
- 1
=
v(y)
\frac{\mbox{d} \tau}{\dy} (y) 
+
d(y)
\frac{\mbox{d}^2 \tau}{\dy^2} (y)
\qquad
\mbox{for}
\;
y \in (-\infty,b),
\label{equationfortau}
\end{equation}
with boundary conditions
\begin{equation}
\frac{\mbox{d} \tau}{\dy} (-\infty) = 0,
\qquad
\tau(b) = 0.
\label{bctauschlog}
\end{equation}
Solving (\ref{equationfortau}), we obtain
\begin{eqnarray*}
\frac{\mbox{d} \tau}{\dy} (y)
& = &
- 
\exp \left[ - \int_{0}^y \frac{v(z)}{d(z)} \, \dz \right]
\int_{-\infty}^y
\frac{1}{d(x)}
\exp \left[ \int_{0}^x \frac{v(z)}{d(z)} \, \dz \right]
\dx
\\
& = &
- \frac{1}{d(y) P_s(y)} \int_{-\infty}^y P_s(x) \dx,
\end{eqnarray*}
where $P_s$ is the stationary distribution (\ref{Psvd}). 
Integrating over $y$ in the interval $(z,b)$ and using
(\ref{bctauschlog}), we obtain
\begin{equation}
\tau(z)
=
- \int_z^{b} \frac{\mbox{d} \tau}{\dy} (y) \dy
=
\int_z^{b}
\frac{1}{d(y) P_s(y)}  \int_{-\infty}^y P_s(x) 
\dx \dy.
\label{tauzformula}
\end{equation}
The important characteristic of the system is the mean time 
$T(k_1)$ for the system to switch from the favourable state 
$x_{f1}$ to the favourable state $x_{f2}.$ To determine
this we set
\begin{equation}
b 
=
\left\{
\begin{matrix}
(x_u + x_{f2})/2, & \qquad \mbox{for} \; k_1 < K_x, \\
284,              & \qquad \mbox{for} \; k_1 \ge K_x, 
\end{matrix}
\right.
\label{bdeffortau}
\end{equation}
and we define $T(k_1)$ precisely as the mean time to leave 
the interval $(-\infty,b)$ given that $X(0) = 0$.
This definition provides a natural extension
of the mean switching time for values $k_1 > K_x$. 
Note that the number $284$ in (\ref{bdeffortau})
is the value of $(x_u + x_{f2})/2$ at the critical
point $k_1 = K_x$. Using 
(\ref{tauzformula}), we can compute $T(k_1)$ as the
integral
\begin{equation}
T(k_1)
=
\int_0^{b}
\exp \left[ - \int_{0}^y \frac{v(z)}{d(z)} \,\dz \right]
\int_{-\infty}^y
\frac{1}{d(x)}
\exp \left[ \int_{0}^x \frac{v(z)}{d(z)} \,\dz \right]
\dx \dy
\label{taudef2}
\end{equation}
where $b$ is given by (\ref{bdeffortau}). Let us note that 
we could also put the right boundary $b$ at the points
$x_u$ or $y_u$. If the number of molecules reaches 
the value $x_u$ (resp. $y_u$), then there is, roughly
speaking, 50\% chance to return back to the favourable state 
$x_{f1}$ and 50\% chance to continue to the second favourable state
$x_{f2}$. Thus the mean switching time between the states could
be estimated by multiplying the time to reach the point
$x_u$ (resp. $y_u$) by the factor of 2. The problem with this
definition is that $x_u \ne y_u$. The difference between the
results obtained by choosing the escape boundary at $x_u$
or $y_u$ is not negligible because the drift is small 
between $x_u$ and $y_u$. On the other hand, the drift 
is large at $(x_u + x_{f2})/2$. Replacing $(x_u + x_{f2})/2$
by any number close to it, e.g. by $(y_u + y_{f2})/2$, leads
to negligible errors. In that sense, the boundary $b$  
given by (\ref{bdeffortau}) yields a more robust definition
of the mean switching time. The other reason for presenting
analysis for the definition (\ref{taudef2}) is that it is
naturally transferable to the chemical SNIPER problem
which is the main interest of this paper.

In Figure~\ref{figxyfuexittime}(b), we compare the results obtained
by the Gillespie SSA and by formula (\ref{tauzformula})
for the parameter values (\ref{paramschlogl}) and
the right boundary at $b = (x_u + x_{f2})/2 = 310.8$.
We have $x_{f1} = 99.9$, $x_u = 225.6$ and $x_{f2} = 396.0$ for the 
parameter values (\ref{paramschlogl}). In Figure~\ref{figxyfuexittime}(b), 
we plot mean exit times computed as averages over $10^4$ exits 
from the domain $[0,b]$, starting at $X(0)=z$ for every integer 
value of $z < b.$ The results are in excellent agreement
with the results computed by the formula (\ref{tauzformula}).

In Figure~\ref{figxyfuexittime}(b), we also see that there is no
significant difference between the exit times if we start at 
$X(0)=x_{f1}$ or $X(0)=0$ for $k_1 \ll K_x$. This justifies the 
choice $X(0)=0$ in the definition of the mean switching time 
$T(k_1)$. If $k_1 \approx K_x$, than $x_{f1}$ is close to $x_u$ 
and the results obtained by the starting point at $X(0)=0$ and 
at $X(0)=x_{f1}$ will differ. In the definition 
(\ref{taudef2}), we use $X(0)=0$ for any value of $k_1$ 
because (i) the results for $X(0)=0$ will provide
some insights into the estimation of the period of oscillation 
of the SNIPER problem studied later; (ii) the results 
are robust to small changes in the initial condition $X(0)=0$;
(iii) the starting point $X(0)=0$ is defined for any value 
of $k_1$ (note that $x_{f1}$ is not defined for $k_1 > K_x$). 

In the remainder of this section, we provide approximations
of $T(k_1)$ defined by the formula (\ref{taudef2}). We fix 
the parameters $k_2$, $k_3$ and $k_4$ as in (\ref{paramschlogl})
and we vary the parameter $k_1$. In Section \ref{secouter},
we provide the estimation of $T(k_1)$ for $k_1 \ll K$ (outer solution).
In Section \ref{secinner}, we provide the estimation of $T(k_1)$ for 
$k_1 \approx K$ (inner solution). Finally, in Section \ref{secmatching},
we match the inner and outer solutions to obtain the uniform approximation 
for any $k_1$. 

We note here that one could approximate
$T(k_1)$ by approximating the integral (\ref{taudef2}). Using
the method of steepest descent, we would obtain the generalization
of the well-known Kramers formula 
\begin{equation}
T(k_1)
\approx
T_k(k_1)
\equiv
\frac{2 \pi \exp[ \Phi(x_u) - \Phi(x_{f1})]}
{d(x_u) \sqrt{\Phi^{\prime\prime} (x_{f1}) |\Phi^{\prime\prime} (x_u)|}}
\label{tauKramers}
\end{equation}
where $\Phi(x)$ is defined by $\Phi(x) = - \int_0^x v(y)/d(y) \dy + \log(d(x)).$
This approximation is valid for $k_1 \ll K$ \cite{Risken:1989:FPE}. 
Haataja {\it et al.} \cite{Haataja:2004:AHD} suggest another generalization
of the Kramers formula replacing $d(x_u)$ in the denominator of
(\ref{tauKramers}) by $\{d(x_u) + d(x_{f1})\}/2$, but this approximation
is worse than (\ref{tauKramers}). 
Unfortunately, such an integral representation
is not available for higher-dimensional Fokker-Planck equations, including
the SNIPER problem studied later. Since our main goal is to present methods
which are applicable in higher-dimensions also, in Sections 
\ref{secouter}, \ref{secinner} and \ref{secmatching} we focus
on approximating the mean switching time $T(k_1)$ by analysing
the $\tau$-equation (\ref{equationfortau}). The chemical Fokker-Planck 
equation and the corresponding $\tau$-equation are available in any 
dimension -- see Appendix \ref{appendixCFPE} and Appendix \ref{appendixmtime}.

\subsection{Approximation of the mean switching time $T(k_1)$
for $k_1 \ll K$}

\label{secouter}

\begin{figure}[t]
\picturesAB{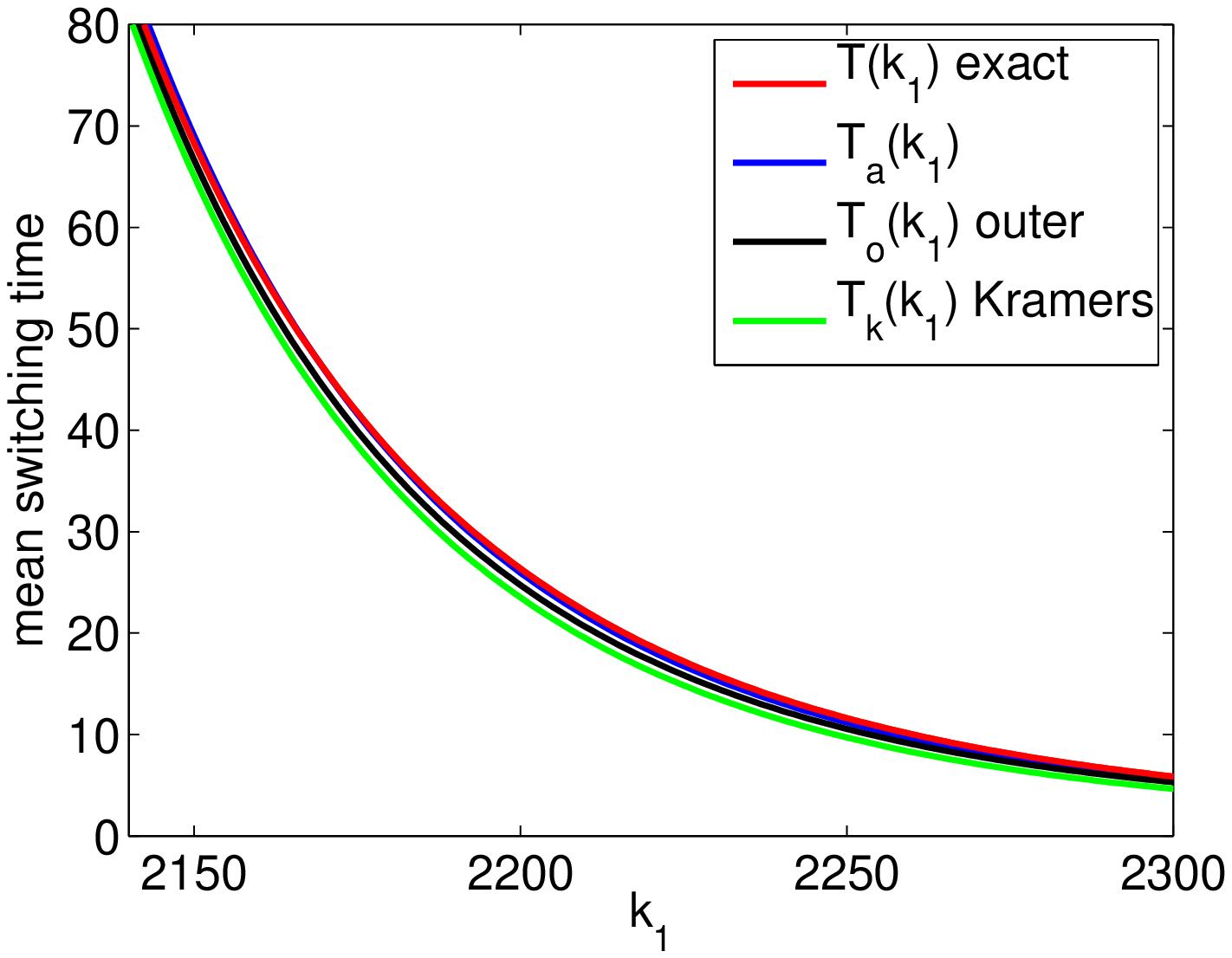}
{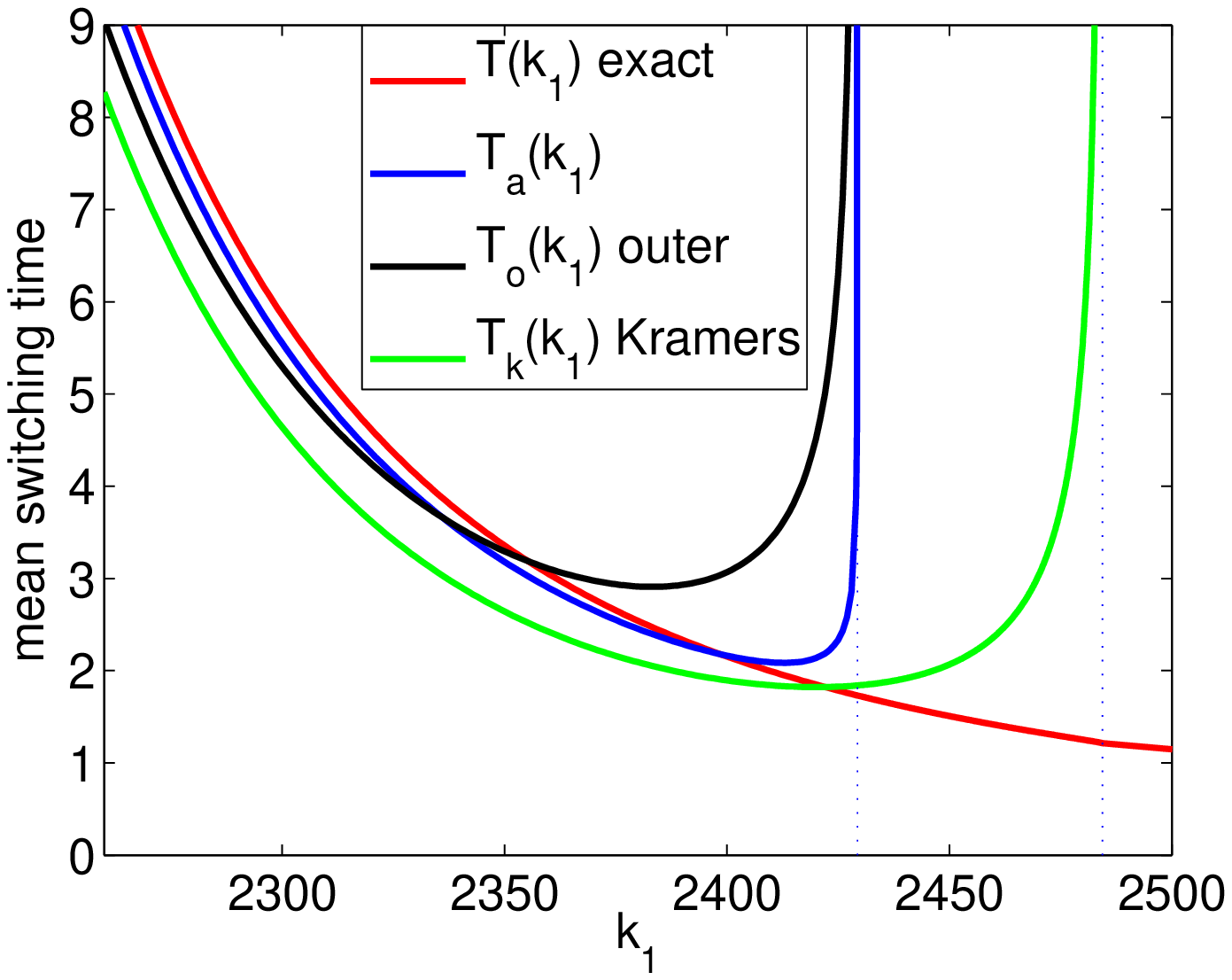}{2.05 in}{3mm}
\caption{{\it Approximations $(\ref{tauKramers})$, $(\ref{tauappj})$ and 
$(\ref{tauouterapp})$ of the mean switching time $T(k_1)$. The
red line is the exact value given by $(\ref{taudef2})$.}
{\rm (a)} 
{\it Region $k_1 \ll K$ for which the derivation was made.}
{\rm (b)} 
{\it Behaviour of approximations close to the critical points
$k_1 = K$ and $k_1 = K_x$. The critical values $K$ and $K_x$ 
are plotted as the dotted lines, $K < K_x$.}
}
\label{figcompoutapp}
\end{figure}

Formula (\ref{eqdx}) and the parameter values (\ref{paramschlogl}) 
imply that $d(x)$ is of the order $10^4$ while $x$ is of the order 
$10^{2}$. Considering small $\varepsilon \sim 10^{-2}$,
we use the following scaling
\begin{equation}
v = \frac{\overline{v}}{\varepsilon^2},
\qquad
d = \frac{\overline{d}}{\varepsilon^2},
\qquad
y = \frac{\overline{y}}{\varepsilon},
\qquad
\tau = \varepsilon \, \overline{\tau}.
\label{scalingone}
\end{equation}
Then we can rewrite the equation (\ref{equationfortau}) to
\begin{equation}
\overline{v}(\overline{y})
\frac{\mbox{d} \overline{\tau}}{\mbox{d} \overline{y}} 
(\overline{y}) 
+
\varepsilon
\overline{d}(\overline{y})
\frac{\mbox{d}^2 \overline{\tau}}{\mbox{d} \overline{y}^2}
(\overline{y})
=
-1.
\label{equationforovertau}
\end{equation}
Later we will see that in fact $\overline{\tau}$ is exponentially
large in $\varepsilon$, so that the left-hand side of this
equation dominates the right-hand side. Then $\overline{\tau}$
will be approximately constant except when $\overline{v} \approx 0$,
that is, the main variation in $\overline{\tau}$ occurs near
$\overline{y} = \overline{y}_u$.
For $\overline{y}$ close to $\overline{y}_u$, we use the transformation 
of variables $\overline{y} = \overline{y}_u + \sqrt{\varepsilon} \, \eta$ 
and approximate
\begin{equation}
\overline{v}(\overline{y})
\approx
\sqrt{\varepsilon} \, \eta
\frac{\mbox{d} \overline{v}}{\mbox{d} \overline{y}}(\overline{y}_u),
\qquad
\overline{d}(\overline{y})
\approx
\overline{d}(\overline{y}_u),
\qquad
\overline{\tau} \gg 1,
\label{dvapprox}
\end{equation}
to give
\begin{equation}
\eta
\frac{\mbox{d} \overline{v}}{\mbox{d} \overline{y}}(\overline{y}_u)
\frac{\mbox{d} \overline{\tau}}{\mbox{d} \eta} 
(\eta) 
+
\overline{d}(\overline{y}_u)
\frac{\mbox{d}^2 \overline{\tau}}{\mbox{d} \eta^2}
(\eta)
=
0.
\label{equationforovertaueta}
\end{equation}
Integrating over $\eta$, we get
\begin{equation}
\exp \left[ 
\frac{1}{2 \overline{d}(\overline{y}_u)} 
\frac{\mbox{d} \overline{v}}{\mbox{d} \overline{y}}(\overline{y}_u)
\eta^2
\right] 
\frac{\mbox{d} \overline{\tau}}{\mbox{d} \eta} 
= 
\frac{c_1}{
\overline{d}(\overline{y}_u) 
}
\label{taudereta}
\end{equation}
where $c_1$ is a real constant. To determine $c_1$
we cannot use the local expansion near $\overline{y}_u$
alone; $c_1$ is determined by the number $-1$ on the
right-hand side of (\ref{equationforovertau}), which
is the only place the scale of $\overline{\tau}$ is set.
To determine $c_1$ we use the projection method of Ward
\cite{Ward:1999:EAC}.
The stationary Fokker-Planck equation 
(\ref{statFPEschlogl}) can be rewritten
\begin{equation}
-
\frac{\mbox{d}}{\mbox{d} \overline{y}} 
\left[
\overline{v}(\overline{y})
P_s(\overline{y}) 
\right]
+
\varepsilon
\frac{\mbox{d}^2}{\mbox{d} \overline{y}^2}
\left[
\overline{d}(\overline{y})
P_s(\overline{y})
\right]
=
0,
\label{eqPsovervariables}
\end{equation}
which is the adjoint equation to the homogeneous version of
(\ref{equationforovertau}).
Multiplying the equation (\ref{equationforovertau}) by $P_s$, 
the equation (\ref{eqPsovervariables}) by $\overline{\tau}$,
integrating over $\overline{y}$ in the interval 
$[-\infty,\overline{y}_u]$ and subtracting the resulting equations, 
we obtain
\begin{eqnarray*}
&&
\int_{-\infty}^{\overline{y_u}}
\frac{\mbox{d}}{\mbox{d} \overline{y}} 
\left[
P_s(\overline{y})
\overline{v}(\overline{y})
\overline{\tau}(\overline{y})
\right]
\mbox{d}\overline{y}
\\
+\,
\varepsilon&&
\int_{-\infty}^{\overline{y}_u}
\left(
\frac{\mbox{d}}{\mbox{d} \overline{y}} 
\left[
P_s(\overline{y})
\overline{d}(\overline{y})
\frac{\mbox{d} \overline{\tau}}{\mbox{d} \overline{y}}
(\overline{y})
\right]
-
\frac{\mbox{d}}{\mbox{d} \overline{y}} 
\left[
\overline{\tau}(\overline{y})
\frac{\mbox{d}}{\mbox{d} \overline{y}}
\left[
\overline{d}(\overline{y})
P_s(\overline{y})
\right]
\right]
\right)
\mbox{d}\overline{y}
=
-
\int_{-\infty}^{\overline{y}_u}
P_s(\overline{y})
\mbox{d}\overline{y}.
\end{eqnarray*}
Evaluating the integrals on the left hand side with the help 
of (\ref{eqPsovervariables}) and the boundary condition (\ref{bctauschlog}), 
we find
\begin{equation}
\frac{\mbox{d} \overline{\tau}}{\mbox{d} \overline{y}}
(\overline{y}_u)
=
-
\frac{1}{\varepsilon
P_s(\overline{y}_u)
\overline{d}(\overline{y}_u)}
\int_{-\infty}^{\overline{y}_u}
P_s(\overline{y})
\mbox{d}\overline{y}.
\label{dertauxu}
\end{equation}
Since $P_s$ is exponentially localized near $\overline{y}_{f1}$
and $\overline{y}_{f2}$, $P_s(\overline{y}_{u})$ is exponentially
small in $\varepsilon.$ Using (\ref{taudereta}) in (\ref{dertauxu}), 
we find
$$
c_1
= 
\overline{d}(\overline{y}_u)
\frac{\mbox{d} \overline{\tau}}{\mbox{d} \eta}(0) 
=
\sqrt{\varepsilon} \;
\overline{d}(\overline{y}_u)
\frac{\mbox{d} \overline{\tau}}{\mbox{d} \overline{y}}(\overline{y}_u)
=
-
\frac{1}{\sqrt{\varepsilon} \, P_s(\overline{y}_u)}
\int_{-\infty}^{\overline{y}_u}
P_s(\overline{y})
\mbox{d}\overline{y},
$$
so that
$$
\frac{\mbox{d} \overline{\tau}}{\mbox{d} \eta} 
\approx
-
\frac{1}{\sqrt{\varepsilon} \,
P_s(\overline{y}_u)
\overline{d}(\overline{y}_u)}
\,
\exp \left[
- 
\frac{1}{2 \overline{d}(\overline{y}_u)} 
\frac{\mbox{d} \overline{v}}{\mbox{d} \overline{y}}(\overline{y}_u)
\eta^2
\right]
\int_{-\infty}^{\overline{y}_u}
P_s(\overline{y})
\mbox{d}\overline{y}.
$$
Integrating over $\eta$ in $[-\infty,\infty]$, we obtain
\begin{equation}
\lim_{\eta \to -\infty}
\overline{\tau}(\eta)
 = 
\frac{\sqrt{2 \pi}}
{\sqrt{\varepsilon \overline{d}(\overline{y}_u)} \,
P_s(\overline{y}_u)}
\left(
\frac{\mbox{d} \overline{v}}{\mbox{d} \overline{y}}(\overline{y}_u)
\right)^{\!\!\!-1/2}
\int_{-\infty}^{\overline{y}_u}
P_s(\overline{y})
\mbox{d}\overline{y},
\label{overtaueta}
\end{equation}
where we used that $\overline{\tau}(\eta) \to 0$ as $\eta \to \infty$, 
i.e. the switching time is going to zero if the starting point approaches
the boundary $b$, see Figure~\ref{figxyfuexittime}(b). The limit 
$\eta \to -\infty$ on the left hand side of (\ref{overtaueta}) is an
approximation of the mean switching time $T(k_1)$. It is the plateau value
of $\tau(z)$ in Figure~\ref{figxyfuexittime}(b). Transforming 
(\ref{overtaueta}) to the original variables, we obtain the following
approximation for the mean switching time
\begin{equation}
T(k_1)
\approx
T_{\mbox{\rm \scriptsize a}}(k_1)
\equiv 
\frac{\sqrt{2 \pi}}{\sqrt{d(y_u)} \,
P_s(y_u)}
\left(
\frac{\mbox{d} v}{\mbox{d}y}(y_u)
\right)^{\!\!\!-1/2}
\int_{-\infty}^{y_u}
P_s(y)
\mbox{d}y.
\label{tauappj}
\end{equation}
Let us define the potential $\Psi$ by
$$
\Psi(x) 
=
-
\int_{0}^x \frac{v(z)}{d(z)} \,\dz.
$$
Then (\ref{tauappj}) can be rewritten as
$$
T_{\mbox{\rm \scriptsize a}}(k_1)
=
\sqrt{2 \pi d(y_u)} \, \exp[\Psi(y_u)]
\left(
\frac{\mbox{d} v}{\mbox{d}y}(y_u)
\right)^{\!\!\!-1/2}
\int_{-\infty}^{y_u}
\frac{\exp[-\Psi(y)]}{d(y)}
\mbox{d}y.
$$
The local extrema of the potential $\Psi$ are equal
to $y_{f1}$, $y_u$ and $y_{f2}$. Using the Taylor expansion
$\Psi(y) \approx \Psi(y_{f1}) + (y-y_{f1})^2 \Psi^{\prime\prime}(y_{f1})/2$
in the integral on the right hand side, we approximate
$$
T(k_1)
\,
\approx
\,
T_{\mbox{\rm \scriptsize a}}(k_1)
\,
\approx 
\,
\exp[\Psi(y_u)-\Psi(y_{f1})]
\left(
\frac{\mbox{d} v}{\mbox{d}y}(y_u)
\right)^{\!\!\!-1/2}
\frac{2 \pi \sqrt{d(y_u)}}{d(y_{f1}) \sqrt{\Psi^{\prime\prime}(y_{f1})}}.
$$
Using the definition of $\Psi$, we obtain the following approximation of 
the mean switching time
\begin{equation}
T(k_1)
\approx
T_{o}(k_1)
\equiv 
2 \pi \sqrt{d(y_u)}
\, \exp \!
\left[
- \int_{y_{f1}}^{y_u} \frac{v(z)}{d(z)} \, \dz
\right]
\left(
\frac{\mbox{d} v}{\mbox{d}y}(y_u)
\left|
\frac{\mbox{d} v}{\mbox{d}y}(y_{f1})
\right| 
d(y_{f1}) 
\right)^{\!\!\!-1/2}.
\label{tauouterapp}
\end{equation}
We will call $T_{o}(k_1)$ the outer solution. In 
Figure~\ref{figcompoutapp}(a), we compare the approximations 
(\ref{tauappj}) and (\ref{tauouterapp})  with the exact value given by 
$(\ref{taudef2})$. We see that $T_{\mbox{\rm \scriptsize a}}$ 
and $T_o$ provide good approximation
of the mean switching time. In Figure~\ref{figcompoutapp}(b), we present
the behaviour of approximations close to the critical point $k_1 = K$.
They both blow up at the point $k_1 = K$. We also plot the results
obtained by the Kramers approximation $T_k(k_1)$ given by (\ref{tauKramers}).
We again confirm that it provides a good approximation for $k_1 \ll K$,
see Figure~\ref{figcompoutapp}(a), but it blows up at the point
$k_1 = K_x$, see Figure~\ref{figcompoutapp}(b). In the next section, 
we present the inner approximation which is valid close to the 
critical point $k_1 = K.$

\subsection{Approximation of the mean switching time $T(k_1)$
for $k_1 \approx K$} \label{secinner}
\begin{figure}
\picturesAB{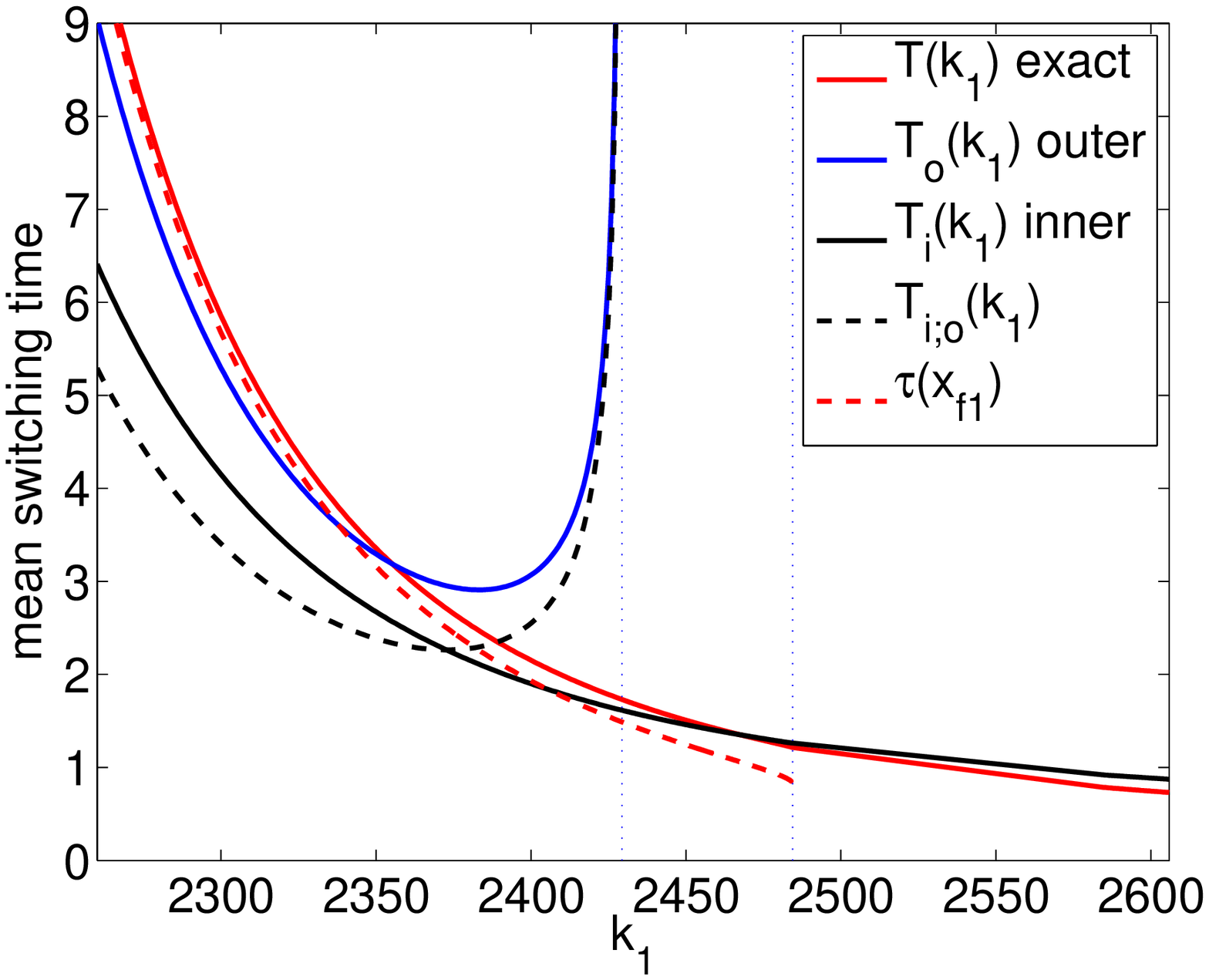}
{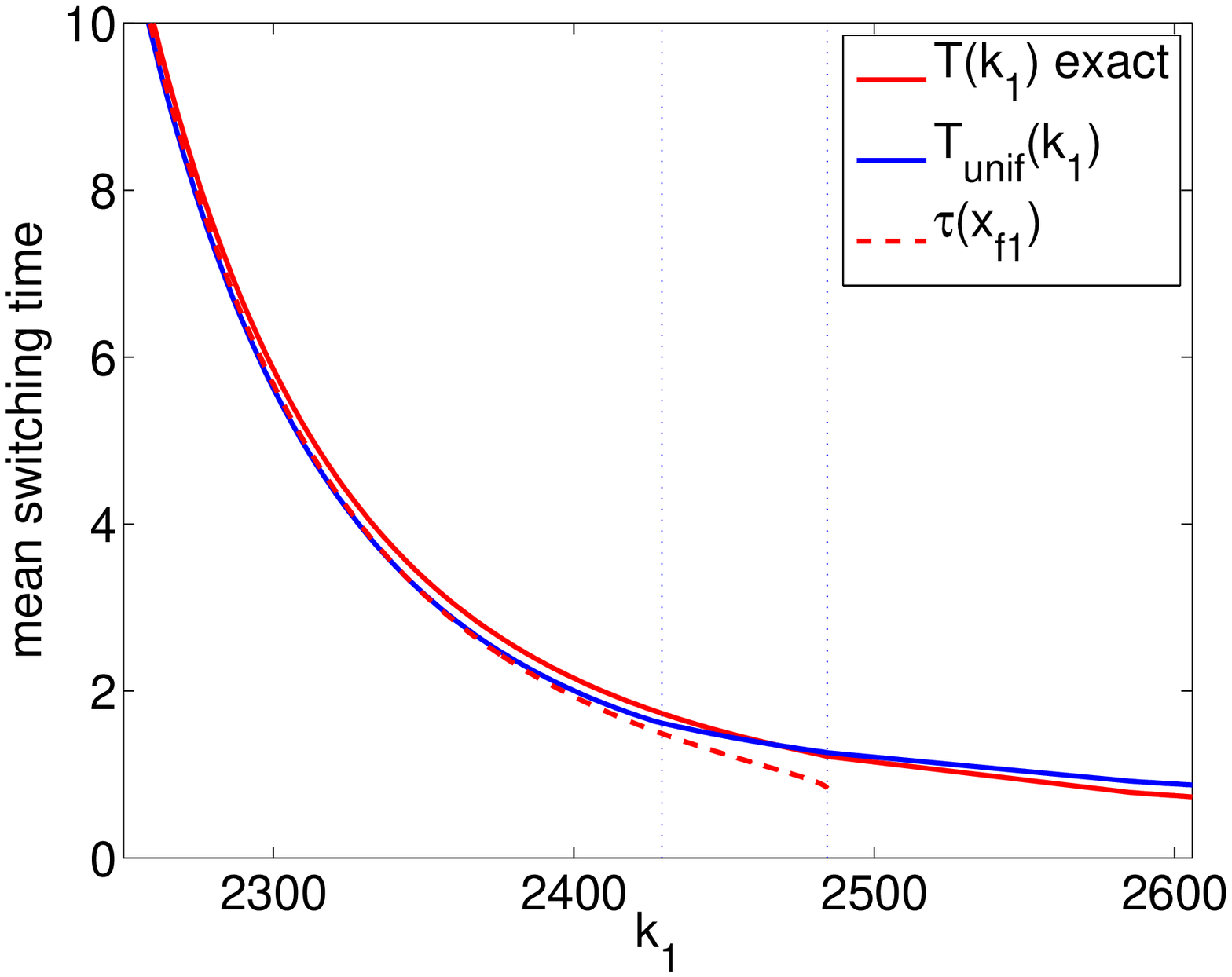}{2.05 in}{6mm}
\caption{{\rm (a)} 
{\it The inner solution $(\ref{tauinnerapp})$ as a function
of $k_1$ (black line). The red line is the exact value $T(k_1)$ 
given by $(\ref{taudef2})$. We also plot the outer solution 
$(\ref{tauouterapp})$ (blue line), the outer limit of the inner solution 
$(\ref{tauioform})$ (black dashed line) and $\tau(x_{f1})$ given by 
$(\ref{tauzformula})$ for $b=(x_u+x_{f2})/2$ (red dashed line).}
{\rm (b)}
{\it The uniform approximation $T_{unif}(k_1)$ 
which is given by $(\ref{tauunifform})$ (blue line). We also plot
the exact value $T(k_1)$ given by 
$(\ref{taudef2})$ (red line)
and $\tau(x_{f1})$ given by $(\ref{tauzformula})$ 
for $b=(x_u+x_{f2})/2$ (red dashed line).}
}
\label{figcompinnapp}
\end{figure}

If $k_1 = K$, then $\overline{y}_u = \overline{y}_{f1} = \overline{y}_k$
where $y_k$ is given by (\ref{valueofyk}). 
Consider the drift coefficient given as a function of $\overline{y}$ and
the parameter $k_1$, namely
\begin{equation}
\overline{v}(\overline{y},k_1) 
= 
k_1 
- 
k_2 \, \overline{y}
+ 
k_3 \, \overline{y} (\overline{y}-1)
- 
k_4 \, \overline{y} (\overline{y}-1) (\overline{y}-2).
\label{driftvk1}
\end{equation}
If $\overline{y}$ is close to $\overline{y}_k$ and $k_1$ is close to 
$K$, we can use the Taylor expansion to approximate
\begin{equation}
\overline{v}(\overline{y},k_1)  
\approx
\frac{1}{2}
\frac{\partial^2 \overline{v}}{\partial \overline{y}^2}(\overline{y}_k,K)
\left(\overline{y}-\overline{y}_k \right)^2
+
(k_1 - K),
\label{approxclosetoK}
\end{equation}
where we used
$$
\frac{\partial \overline{v}}{\partial k_1}(\overline{y}_k,K)
=
1
\;\;\;
\mbox{and}
\;\;\;
\overline{v}(\overline{y}_k, K) 
=
\frac{\partial \overline{v}}{\partial \overline{y}}
(\overline{y}_k, K) 
=
\frac{\partial^2 \overline{v}}{\partial \overline{y} \partial k_1}
(\overline{y}_k, K) 
=
\frac{\partial^2 \overline{v}}{\partial k_1^2}
(\overline{y}_k, K) 
= 
0.
$$
We use the transformation of variables 
$\overline{y} = \overline{y}_k + \varepsilon^{1/3} \, \eta$ and 
$k_1 = K + \varepsilon^{2/3} \, \kappa$. Then (\ref{approxclosetoK}) 
implies
\begin{equation}
\overline{v}(\overline{y},k_1)  
\approx
\varepsilon^{2/3}
\left(
\frac{1}{2}
\frac{\partial^2 \overline{v}}{\partial \overline{y}^2}(\overline{y}_k,K)
\eta^2
+
\kappa
\right).
\label{approxclosetoK3}
\end{equation}
Now in (\ref{scalingone}) we need to replace
$\tau = \varepsilon \, \overline{\tau}$
by $\tau = \varepsilon^{2/3} \, \overline{\tau}$. Then 
(\ref{equationforovertau}) reads as follows
\begin{equation}
\varepsilon^{-1/3}
\overline{v}(\overline{y})
\frac{\mbox{d} \overline{\tau}}{\mbox{d} \overline{y}} 
(\overline{y}) 
+
\varepsilon^{2/3}
\overline{d}(\overline{y})
\frac{\mbox{d}^2 \overline{\tau}}{\mbox{d} \overline{y}^2}
(\overline{y})
=
-1.
\label{equationforovertau2}
\end{equation}
Using (\ref{approxclosetoK3}) and approximating
$\overline{d}(\overline{y}) \approx \overline{d}(\overline{y}_k)$, 
we get
\begin{equation}
\left(
\frac{1}{2}
\frac{\partial^2 \overline{v}}{\partial \overline{y}^2}(\overline{y}_k,K)
\eta^2
+
\kappa
\right)
\frac{\mbox{d} \overline{\tau}}{\mbox{d} \eta} 
(\eta) 
+
\overline{d}(\overline{y}_k)
\frac{\mbox{d}^2 \overline{\tau}}{\mbox{d} \eta^2}
(\eta)
=
-1.
\label{equationforovertauetayfyu}
\end{equation}
This time, since we are close to the bifurcation point, the right-hand
side is not negligible. Equation (\ref{equationforovertauetayfyu})
can be rewritten as
$$
\frac{\mbox{d}}{\mbox{d} \eta}
\left(
\overline{d}(\overline{y}_k) 
\exp \left[ 
\frac{1}{\overline{d}(\overline{y}_k)} 
\left(
\frac{1}{2}
\frac{\partial^2 \overline{v}}{\partial \overline{y}^2}(\overline{y}_k,K)
\frac{\eta^3}{3}
+
\kappa
\eta
\right)
\right] 
\frac{\mbox{d} \overline{\tau}}{\mbox{d} \eta} 
\right)
$$
$$
=
-
\exp \left[ 
\frac{1}{\overline{d}(\overline{y}_k)} 
\left(
\frac{1}{2}
\frac{\partial^2 \overline{v}}{\partial \overline{y}^2}(\overline{y}_k,K)
\frac{\eta^3}{3}
+
\kappa
\eta
\right)
\right]. 
$$
Integrating over $\eta$, we get
\begin{eqnarray}
\frac{\mbox{d} \overline{\tau}}{\mbox{d} \eta} 
& = &
-
\frac{1}{\overline{d}(\overline{y}_k)} 
\exp \left[ 
-
\frac{1}{\overline{d}(\overline{y}_k)} 
\left(
\frac{1}{2}
\frac{\partial^2 \overline{v}}{\partial \overline{y}^2}(\overline{y}_k,K)
\frac{\eta^3}{3}
+
\kappa
\eta
\right)
\right] 
\nonumber
\\
& \times & 
\int_{-\infty}^\eta
\exp \left[ 
\frac{1}{\overline{d}(\overline{y}_k)} 
\left(
\frac{1}{2}
\frac{\partial^2 \overline{v}}{\partial \overline{y}^2}(\overline{y}_k,K)
\frac{\xi^3}{3}
+
\kappa
\xi
\right)
\right] 
\dxi. 
\label{pomicb}
\end{eqnarray}
We chose the exiting boundary so that $\tau(\eta) \to 0$ at $\eta \to \infty$.
Consequently, integrating (\ref{pomicb}) over $\eta$ in $[-\infty,\infty]$,
we obtain 
\begin{equation*}
\lim_{\eta \to -\infty}
\overline{\tau}(\eta) 
=
\frac{1}{\overline{d}(\overline{y}_k)}
\int_{-\infty}^{\infty}
\int_{-\infty}^\eta 
\exp \! \left[ 
-
\frac{1}{\overline{d}(\overline{y}_k)} 
\left(
\frac{\partial^2 \overline{v}}{\partial \overline{y}^2}(\overline{y}_k,K)
\, \frac{\eta^3 - \xi^3}{6}
+
\kappa (\eta - \xi)
\right)
\right] 
\dxi
\, 
\deta.
\end{equation*}
Substituting
$$
\eta = 
\left(
\frac{1}{3\overline{d}(\overline{y}_k)} 
\frac{\partial^2 \overline{v}}{\partial \overline{y}^2}(\overline{y}_k,K)
\right)^{\!-1/3}
(u + v),
\qquad
\xi = 
\left(
\frac{1}{3\overline{d}(\overline{y}_k)} 
\frac{\partial^2 \overline{v}}{\partial \overline{y}^2}(\overline{y}_k,K)
\right)^{-\!1/3}
(u - v),
$$
we obtain
\begin{equation}
\lim_{\eta \to -\infty}
\overline{\tau}(\eta) 
=
\frac{\overline{d}(\overline{y}_k) \beta^2}{2} 
\sqrt{\frac{\pi}{3}} 
\,
H \! \left(
\beta \kappa
\right).
\label{limemi}
\end{equation}
where
\begin{equation}
\beta
=
-
24^{1/3} 
\;
\overline{d}(\overline{y}_k)^{-2/3}
\left(
\frac{\partial^2 \overline{v}}{\partial \overline{y}^2}(\overline{y}_k,K)
\right)^{\!-1/3},
\label{definitionbeta}
\end{equation}
and the function $H: \er \to (0,\infty)$ is given by
\begin{equation}
H(z)
=
\int_{0}^\infty
\frac{\exp \left[ 
- v^3
+ z v
\right]}{\sqrt{v}}
\, \dv.
\label{defH}
\end{equation}
The limit of $\overline{\tau}(\eta)$ as $\eta \to -\infty$ is the 
approximation of 
the mean switching time $T(k_1)$ for $k_1$ close to $K$.
Transforming back to the original variables and 
using (\ref{definitionbeta}), we obtain the inner solution $T_i(k_1)$
in the following form
\begin{eqnarray}
T(k_1)
\approx
T_i(k_1)
& \equiv &
2  \sqrt{\pi} \; 3^{1/6} 
\left(
d(y_k)
\right)^{-1/3}
\left(
\frac{\partial^2 v}{\partial y^2}(y_k,K)
\right)^{\!-2/3}
\nonumber
\\
& \times &
H \! \left(
-
[k_1 - K]
\,
24^{1/3}
\,
\left(
d(y_k)
\right)^{-2/3}
\left(
\frac{\partial^2 v}{\partial y^2}(y_k,K)
\right)^{\!-1/3}
\right). \qquad \qquad
\label{tauinnerapp}
\end{eqnarray}
In Figure~\ref{figcompinnapp}(a), we compare the approximation 
(\ref{tauinnerapp}) with the exact value $T(k_1)$ given 
by $(\ref{taudef2})$. We also plot $\tau(x_{f1})$ given 
by (\ref{tauzformula}) for $b=(x_u+x_{f2})/2.$ As discussed 
before, $\tau(x_{f1})$ could be considered as another 
possible definition of the mean switching time. We see that 
the value of  the inner approximation $T_i(k_1)$ at the critical 
point $k_1 = K$ lies between the exact value $T(k_1)$ and 
$\tau(x_{f1})$. Thus we confirm that $T_i(k_1)$ is a good 
approximation of $T(k_1)$ for $k_1$ close the the critical 
point $K$.

\subsection{Uniform approximation of the mean switching time 
$T(k_1)$}

\label{secmatching}

In Section \ref{secouter}, we obtained the outer approximation
(\ref{tauouterapp}) of the mean switching time which is valid
for $k_1 \ll K$. In Section \ref{secinner}, we obtained the inner 
approximation (\ref{tauinnerapp}) of the mean switching time which 
is valid for $k_1 \approx K$. In this section, we match the inner
and the outer solutions to derive the uniform approximation 
of the mean switching time. First, we investigate the asymptotic 
behaviour of $H(z)$ as $z \to \infty$. Let $z > 0$. Using
the substitution $w = z^{-1/2} v$ in the definition (\ref{defH}),
we obtain
$$
H(z)
=
z^{1/4}
\int_{0}^\infty
\frac{1}{\sqrt{w}}
\,
\exp \left[ 
z^{3/2}
\left(
w - w^3
\right)
\right]
\, \dw
\;
\sim 
\;
\sqrt{\frac{\pi}{z}}
\exp \left[ 
\frac{2 z^{3/2}}{ 3\sqrt{3}}
\right]
\qquad
\mbox{as}
\;
z \to \infty.
$$
Consequently, the outer limit of the inner solution (\ref{tauinnerapp})
is
\begin{equation}
T_{i;o}
= 
\frac{2^{1/2} \pi}{\sqrt{|k_1 - K|}}
\left( 
\frac{\partial^2 v}{\partial y^2}(y_k,K)
\right)^{\!-1/2}
\exp \! \left[ 
\frac{4 \sqrt{2}}{3 d(y_k)}
|k_1 - K|^{3/2}
\left(
\frac{\partial^2 v}{\partial y^2}(y_k,K)
\right)^{\!-1/2}
\right].
\label{tauioform}
\end{equation}
In Figure~\ref{figcompinnapp}(a), we plot $T_{i;o}(k_1)$ as the black
dashed line. It is easy to check that the inner limit of the outer 
solution $T_o(k_1)$ is equal to the outer limit of the inner solution 
$T_i(k_1)$, 
i.e. it is also given by $T_{i;o}(k_1)$. Thus we can define the 
continuous uniform approximation of the mean switching time 
$T(k_1)$ by the formula
\begin{equation}
T_{unif}(k_1)
\equiv
\begin{cases}
T_{o}(k_1)
+
T_{i}(k_1)
- 
T_{i;o}(k_1)
-
C,
& \mbox{for} \; k_1 < K; \\
T_{i}(k_1), & \mbox{for} \; k_1 \ge K; \\
\end{cases}
\label{tauunifform}
\end{equation}
where the constant $C$ is defined by
$$
C
=
\lim_{\;\,k_1 \to K_{-}}
\left(
T_{o}(k_1)
-
T_{i;o}(k_1) 
\right).
$$
Thus, for $k_1 < K$, we add the inner and the outer solutions and subtract 
the ``overlap" solution (the inner limit of the outer solution) which has
been double-counted. In order to make the
approximation continuous at $k_1 = K$, we also subtract the constant
$C$ (which is in fact a higher order term). 
We approximate $T(k_1)$ by the inner solution
for $k_1 \ge K$. In Figure~\ref{figcompinnapp}(b), we plot the uniform 
approximation $T_{unif}(k_1)$  together with the exact 
value $T(k_1)$ given by $(\ref{taudef2})$. We also plot $\tau(x_{f1})$ 
given by (\ref{tauzformula}) for $b=(x_u+x_{f2})/2.$ The comparison
is excellent.

\section{Numerical results obtained by solving the chemical
Fokker-Planck equation for the chemical SNIPER problem}
\label{secnumerFPres}
\begin{figure}
\picturesAB{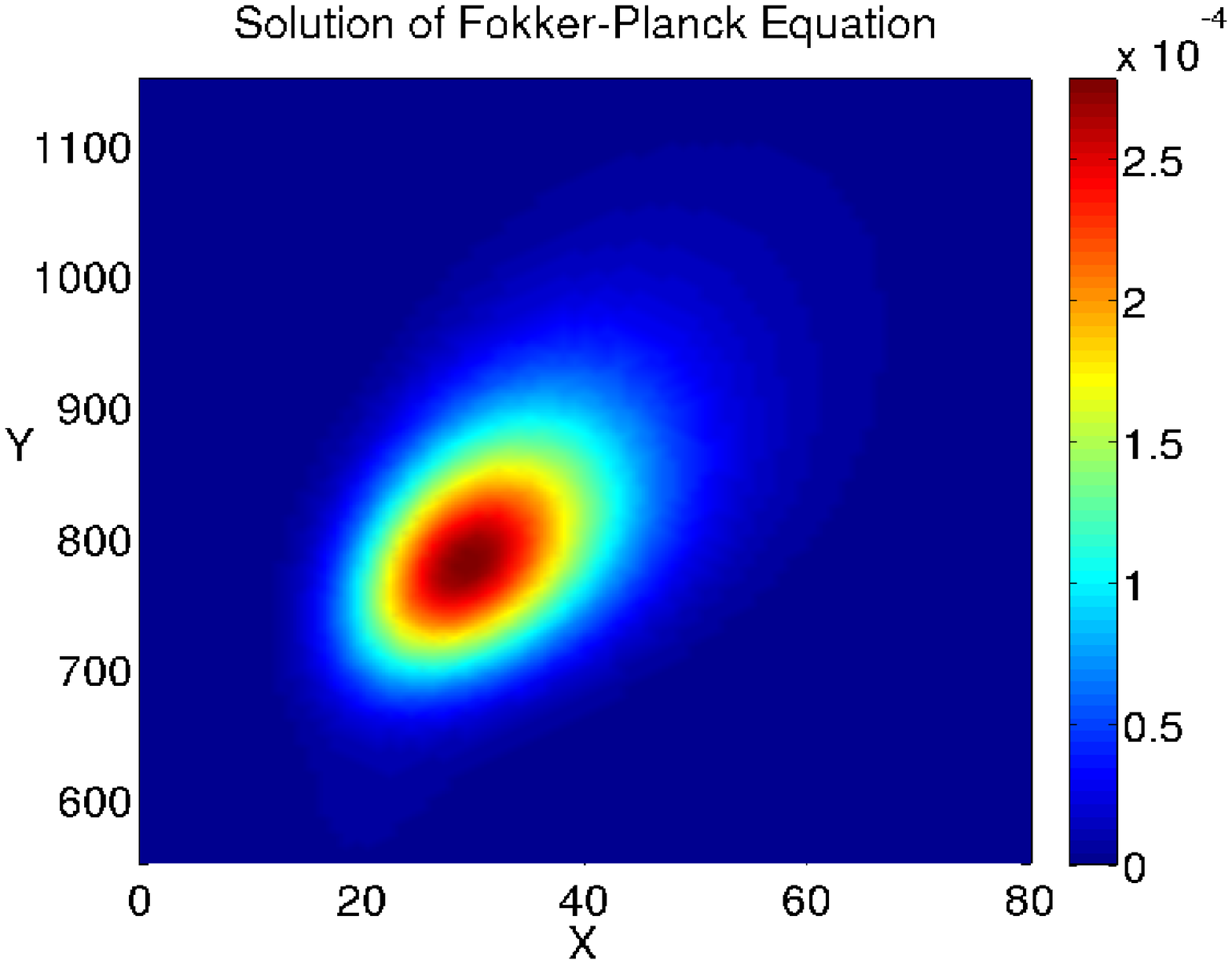}{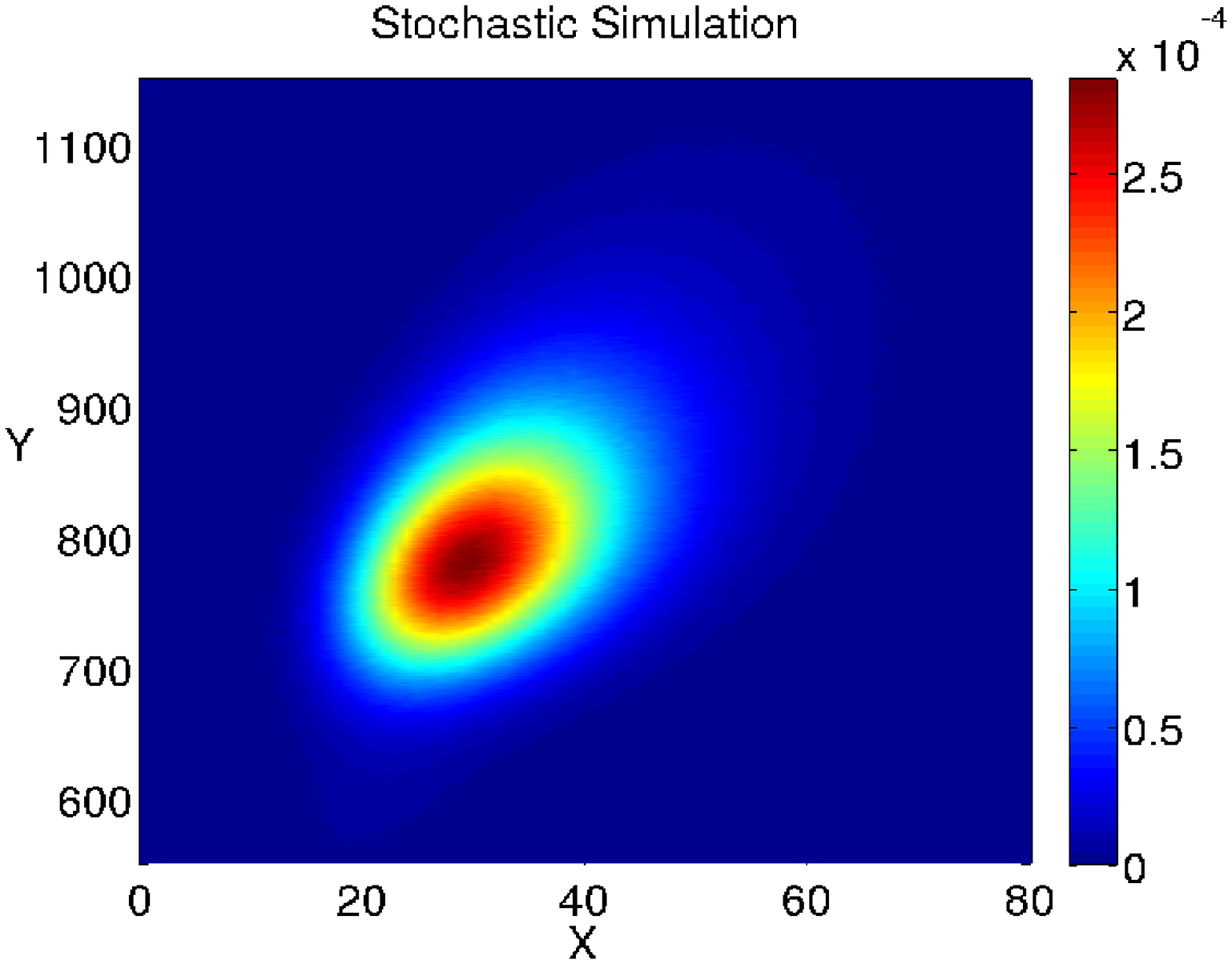}{2.05in}{6mm}
\caption{{\rm (a)} 
{\it
The FEM solution of the Fokker-Planck equation $(\ref{problemAF})$
restricted to the subdomain $(0,80)\times(550,1150)$.}
{\rm (b)} 
{\it $P_s(x,y)$ obtained by the Gillespie SSA.
We use the parameter values $(\ref{sniperparam})$ and $V=40$.}
}
\label{figstatdistrsniper}
\end{figure}

\begin{figure}
\picturesAB{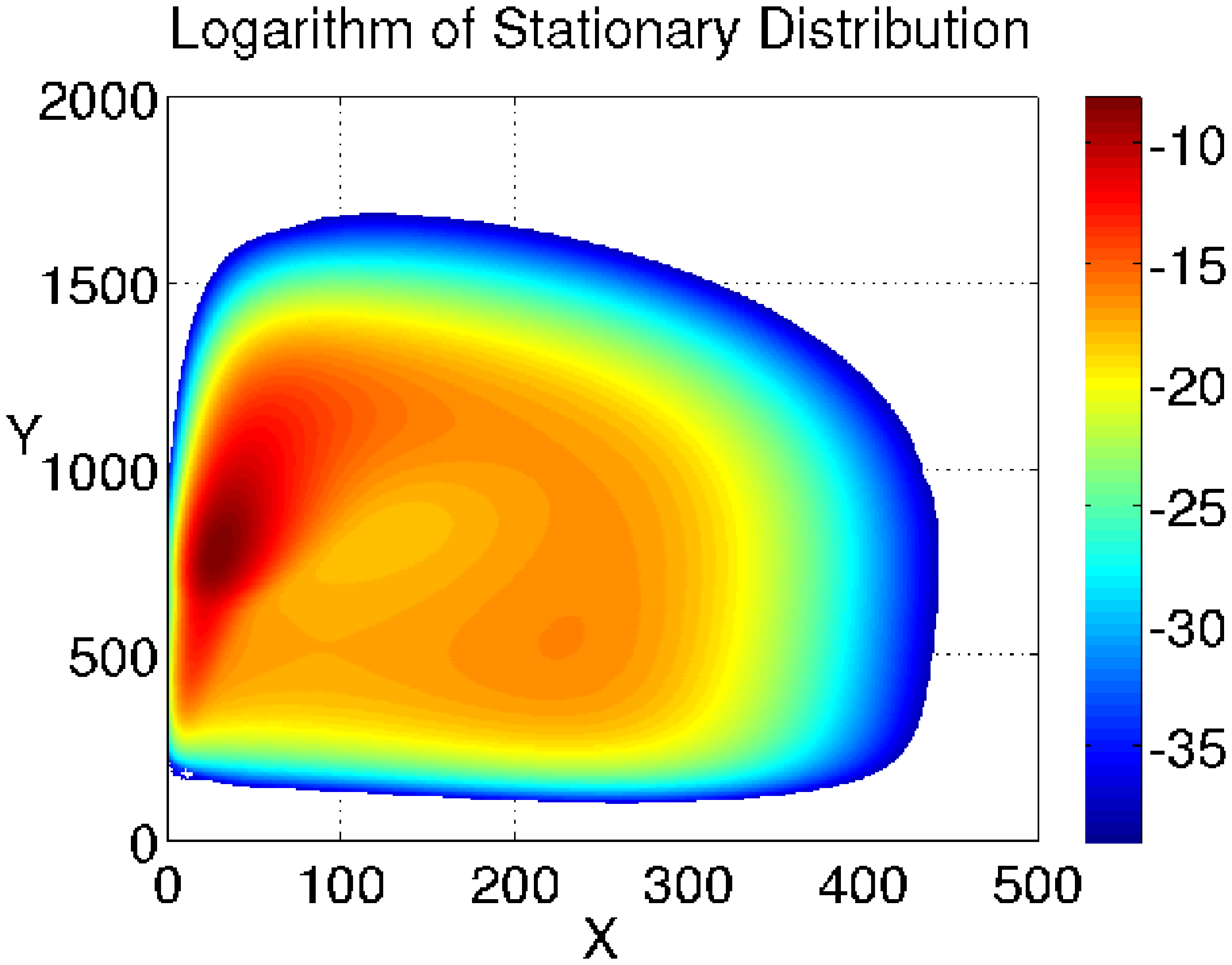}{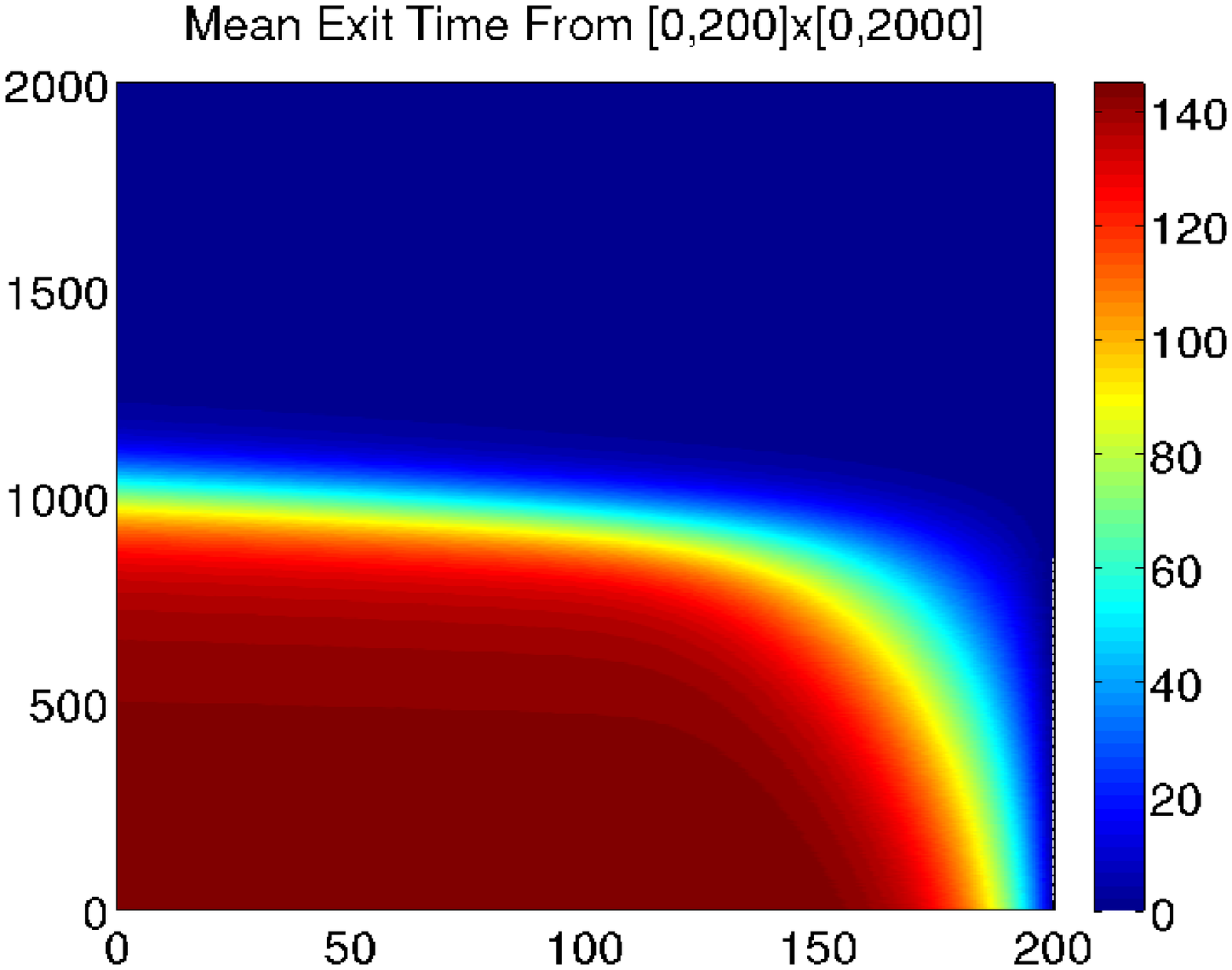}
{2.05in}{6mm}
\caption{
{\rm (a)} 
{\it 
Logarithm of the FEM solution of the Fokker-Planck equation 
$(\ref{problemAF})$ in the domain $S = (0,500)\times(0,2000)$.} 
{\rm (b)} 
{\it The solution $\tau(x,y)$ of $(\ref{finproblemtau})$ computed by the
FEM. We use the parameter values $(\ref{sniperparam})$ and $V=40$.}}
\label{figstatdistrsniper2}
\end{figure}

We consider the chemical system (\ref{model1})--(\ref{model2}).
Substituting the propensity functions (\ref{propfunctions})
into the equation (\ref{FPEgengen}) in Appendix \ref{appendixCFPE}, 
we obtain the chemical Fokker-Planck equation
\begin{equation}
\frac{\partial P}{\partial t}
=
\frac{\partial^2}{\partial x^2} \big[ d_x \, P \big]
+ \frac{\partial^2}{\partial x \partial y} \big[ d_{xy} \, P \big]
+ \frac{\partial^2}{\partial y^2} \big[ d_y \, P \big] 
- \frac{\partial}{\partial x} \big[ v_x \, P \big]
- \frac{\partial}{\partial y} \big[ v_y \, P \big]
\label{FokkerPlanck2D}
\end{equation}
where $P(x,y,t)$ is the joint probability distribution that
$X(t)=x$ and $Y(t)=y$, $x \in [0,\infty)$, $y \in [0,\infty)$, 
and the drift and diffusion coefficients are given by
\begin{eqnarray}
v_x(x,y) 
& = &
k_2 y - k_5 x (x-1) (x-2) + k_4 x (x-1) - k_3 x, 
\label{begcovx}
\\
v_y(x,y)
& = &
- k_7 x (x-1) y + k_6 x y - k_2 y + k_1, 
\label{midcovy}
\\
d_x(x,y) 
& = & 
[k_2 y + k_5 x (x-1) (x-2)  + k_4 x (x-1) + k_3 x]/2, \\
d_y(x,y)
& = & 
[k_7 x (x-1) y + k_6 x y + k_2 y + k_1]/2, \\
d_{xy}(x,y)
& = &
 - k_2 y.
\label{endcodxy}
\end{eqnarray}
The stationary distribution $P_s (x,y) = \lim_{t \to \infty} P(x,y,t)$
can be obtained as a solution of the corresponding elliptic problem:
\begin{equation}
0 =
\frac{\partial^2}{\partial x^2} \big[ d_x \, P_s \big]
+ \frac{\partial^2}{\partial x \partial y} \big[ d_{xy} \, P_s \big]
+ \frac{\partial^2}{\partial y^2} \big[ d_y \, P_s \big] 
- \frac{\partial}{\partial x} \big[ v_x \, P_s \big]
- \frac{\partial}{\partial y} \big[ v_y \, P_s \big],
\label{statFP2Dsniper1}
\end{equation}
\begin{equation}
\int_0^\infty \int_0^\infty P_s(x,y) \dx \dy = 1,
\label{statFP2Dsniper2}
\end{equation}
\begin{equation}
P_s (x,y) \ge 0, \qquad \qquad \mbox{for all} 
\; (x,y) \in [0,\infty) \times [0,\infty).
\label{statFP2Dsniper3}
\end{equation}
To solve the problem (\ref{statFP2Dsniper1})--(\ref{statFP2Dsniper3})
approximately by the finite element method (FEM) we have to reformulate 
it as a problem in a finite domain. Therefore, we first restrict 
the domain $[0,\infty) \times [0,\infty)$ to a rectangle $S$
which has to be sufficiently large to contain most of the trajectories.
For example, Figure~\ref{fig12timeevol}(b) shows the illustrative 
trajectory for the parameter values (\ref{sniperparam}) and $V=40$.
In this case, the rectangle $S$ can be chosen as 
$S = (0,500) \times (0,2000)$. On the boundary $\partial S$ we prescribe 
homogeneous Neumann boundary conditions. We reformulate 
(\ref{statFP2Dsniper1})--(\ref{statFP2Dsniper3})
as the following Neumann problem in the domain $S$
\begin{eqnarray}
\label{problemAF}
  -\operatorname{div} ( \mathcal{A} \nabla P_s + P_s \mathbf{b} ) &=& 0
\quad\text{in } S,
\\
\nonumber
  (\mathcal{A} \nabla P_s + P_s \mathbf{b} ) \cdot \mathbf{n} &=& 0
\quad\text{on }\partial S,
\end{eqnarray}
where $\nabla$ stands for the gradient,
$\mathbf{n}$ denotes the outward unit normal vector to $\partial S$,
and the $2\times2$ symmetric positive definite matrix
$\mathcal{A}$ and the vector $\mathbf{b}$ are given by
\begin{equation}
\mathcal{A}
=
\left(
\begin{matrix}
-d_x & -d_{xy}/2
\\
-d_{xy}/2 & -d_y
\end{matrix}
\right),
\quad  
\mathbf{b} = \left( 
          v_x - \frac{\partial d_x}{\partial x}
          - \frac12 \frac{\partial d_{xy}}{\partial y},
            v_y - \frac{\partial d_y}{\partial y}
          - \frac12 \frac{\partial d_{xy}}{\partial x}
	  \right)^T.
\label{calAb}	  
\end{equation}
To define the FEM solution, we need the weak 
formulation of (\ref{problemAF}).
The weak solution $P_s \in H^1(S)$ is uniquely determined by 
the requirement
$$
  a(P_s, \varphi) = 0 \quad\forall\varphi\in H^1(S),
$$
where $H^1(S)$ stands for the Sobolev space $W^{1,2}(S)$ 
and the bilinear form $a(\cdot,\cdot)$ is given by
$$  a(P_s, \varphi) = \int_S (\mathcal{A}\nabla P_s + P_s \mathbf{b})
    \cdot \nabla \varphi \,\dx\dy.
$$
We use first-order triangular elements.
First we define a triangulation
$\mathcal{T}_h$ of the domain $S$ and the corresponding
finite element subspace $W_h$ of the piecewise linear functions:
\begin{equation}
  W_h = \{ \varphi_h \in H^1(S) : \varphi_h|_K \in P^1(K),\ K\in\mathcal{T}_h \},
\label{defVh}
\end{equation}
where $P^1(K)$ stands for the three-dimensional space of linear functions 
on the triangle $K\in\mathcal{T}_h$.
The finite element problem then reads: find $P_{s,h} \in W_h$ such that
\begin{equation}
 \label{problemAFh}
 a(P_{s,h}, \varphi_h) = 0 \quad\forall\varphi_h\in W_h.
\end{equation}
Both problems (\ref{problemAF}) and (\ref{problemAFh}) posses
trivial solutions $P_s \equiv 0$ and $P_{s,h} \equiv 0$.
To get a non-zero solution we use appropriate numerical methods
for finding eigenvectors corresponding to the zero eigenvalue.
The computed nontrivial solution $P_{s,h}$ is then normalized to 
have $\int_S P_{s,h}(x,y) \dx \dy = 1$.
In Figure~\ref{figstatdistrsniper}(a), we present the FEM 
solution $P_{s,h}$ to the problem (\ref{problemAFh}) obtained 
on a uniform mesh with $2^{18}$ triangular elements. We use the
parameter values (\ref{sniperparam}) and $V=40$.
We plot only the stationary distribution in the subdomain
$[0,80] \times [550,1150],$ i.e. the part of the $X$-$Y$ phase
space where the system spends most of time. 
Running long time stochastic 
simulations, we can find the stationary distribution
by the Gillespie SSA, which is plotted 
in Figure~\ref{figstatdistrsniper}(b) for the same
parameter values $(\ref{sniperparam})$ and $V=40$.
The comparison with the results obtained by the chemical 
Fokker-Planck equation is visually excellent.
Plotting the numerical results in the whole computational domain 
$S = (0,500)\times(0,2000)$, we do not see any additional information
because most of $P_s$ is localized in the subdomain shown
in Figure~\ref{figstatdistrsniper}. 
We have to plot $\log(P_s)$ to see the underlying ``volcano-shaped" 
probability distribution -- see Figure~\ref{figstatdistrsniper2} 
(a).
Let us note that there is no bifurcation in the features of $P_s$,
i.e. the probability distribution is ``volcano-shaped" 
both before and after the ODE critical value.

\subsection{Computation of the period of oscillation}
Let us consider the SNIPER chemical system
(\ref{model1})--(\ref{model2}) with the parameter
values (\ref{sniperparam}) and $V=40$. An illustrative
stochastic trajectory is shown in Figure~\ref{fig12timeevol}. Let the domain $\Omega$ be defined
by $\Omega = \{ [x,y] \; | \; x < 200 \}$.
In each cycle (as we defined it on page \pageref{pagedefcycle}), 
the trajectory leaves the domain $\Omega$.
However, the time which the trajectory spends outside
the domain $\Omega$ is much smaller than the
time which the trajectory spends inside $\Omega$.
This observation is confirmed by Figure~\ref{figstatdistrsniper2}(a).
The probability that the system state is outside $\Omega$
is several orders of magnitude smaller than the probability
that it lies inside $\Omega$. 
Thus the period
of oscillation can be estimated as the mean time to leave
the domain $\Omega$ provided that the trajectory ``has
just entered it". Let $\tau(x,y)$ be the average time to 
leave the domain $\Omega$ provided that the trajectory
starts at $X(0)=x$ and $Y(0)=y$. As shown in Appendix 
\ref{appendixmtime}, $\tau(x,y)$ evolves according to the equation 
(\ref{tau2Dequation}):
\begin{equation}
d_x \frac{\partial^2 \tau}{\partial x^2}
+ 
d_{xy} \frac{\partial^2 \tau}{\partial x \partial y}
+ 
d_y \frac{\partial^2 \tau}{\partial y^2} 
+
v_x 
\frac{\partial \tau}{\partial x}
+
v_y
\frac{\partial \tau}{\partial y}
=
- 1,
\quad
\mbox{for} \; [x,y] \in \Omega,
\label{escapeequation}
\end{equation}
together with the boundary condition $\tau(200,y)=0$ 
for $y \in \er$. To solve this problem approximately, we truncate 
the domain $\Omega$ to get the finite domain 
$\widetilde{S} = (0,200) \times (0,2000)$. We consider homogeneous
Neumann boundary conditions on the parts of the boundary which are
not on the line $x=200$ and we rewrite the problem (\ref{escapeequation}) 
in the form 
\begin{eqnarray}
\label{finproblemtau}
  -\operatorname{div} ( \mathcal{A} \nabla \tau ) 
  + \mathbf{b}\cdot\nabla \tau &=& -1
\quad\text{in }\widetilde{S},
\\
\nonumber
  \tau & = & 0
   \quad\text{on the line } x=200, \label{bo1}
   \\
  (\mathcal{A}\nabla\tau) \cdot \boldn & = & 0
   \quad\text{on the lines } y=0,\ y=2000,\ x=0,
   \nonumber
\end{eqnarray}
where the $2\times 2$ matrix $\mathcal{A}$ and the vector 
$\mathbf{b}$ are given by (\ref{calAb}) and the coefficients 
$d_x$, $d_{xy}$, $d_y$, $v_x$, and $v_y$ are given by
(\ref{begcovx})--(\ref{endcodxy}). Notice the difference between
(\ref{finproblemtau}) and (\ref{problemAF}) and
that $\widetilde{S} \subset S.$
As in the previous section, we define the weak solution as 
$\tau \in \widetilde{W}$ satisfying
$$
  \widetilde{a}(\tau,\varphi) 
  = \int_{\widetilde{S}} -1 \cdot \varphi \,\dx\dy \qquad 
  \forall \varphi\in \widetilde{W},
$$
where $\widetilde{W} 
= \{ v \in H^1(\widetilde{S}) : v = 0$ on the line $x = 200\}$
and
$$
  \widetilde{a}(\tau,\varphi) = \int_{\widetilde{S}} 
  \mathcal{A} \nabla \tau \cdot \nabla\varphi \,\dx\dy
  + \int_{\widetilde{S}} \mathbf{b}\cdot\nabla \tau \varphi \,\dx\dy.
$$
The FEM solution $\tau_h \in \widetilde{W}_h$ is defined by the
requirement
$$
  \widetilde{a}(\tau_h,\varphi_h) 
  = \int_{\widetilde{S}} -1 \cdot \varphi_h \,\dx\dy, 
  \qquad \forall \varphi_h\in \widetilde{W}_h,
$$
where $\widetilde{W}_h \subset \widetilde{W}$ is the space of continuous 
and piecewise linear functions based
on a suitable triangulation $\mathcal{T}_h$ of $\widetilde{S}$
(compare with (\ref{defVh})).
The triangulation used does not follow the anisotropy of $\widetilde{S}$
and it consits of elements close to the equilateral triangle.
It has about $500,000$ triangles and the number of degrees of freedom
is about $250,000$. The FEM solution $\tau_h$ is shown in
Figure~\ref{figstatdistrsniper2}(b). Using the computed profile,
we can estimate the period of oscillation. One possibility
is to find the maximum of the stationary distribution $P_s(x,y)$ which
is attained at the point $[x_f,y_f] = [29.30, 781.25]$, and estimate the 
period of oscillation as $\tau_h(x_f,y_f) = \tau_h(29.30, 781.25) 
\doteq 134.1$. This compares well with the period of
oscillation estimated by long stochastic simulation
for the same parameter values. We obtained $130.4$ as an average 
over $10^5$ periods. 

\begin{figure}
\picturesAB{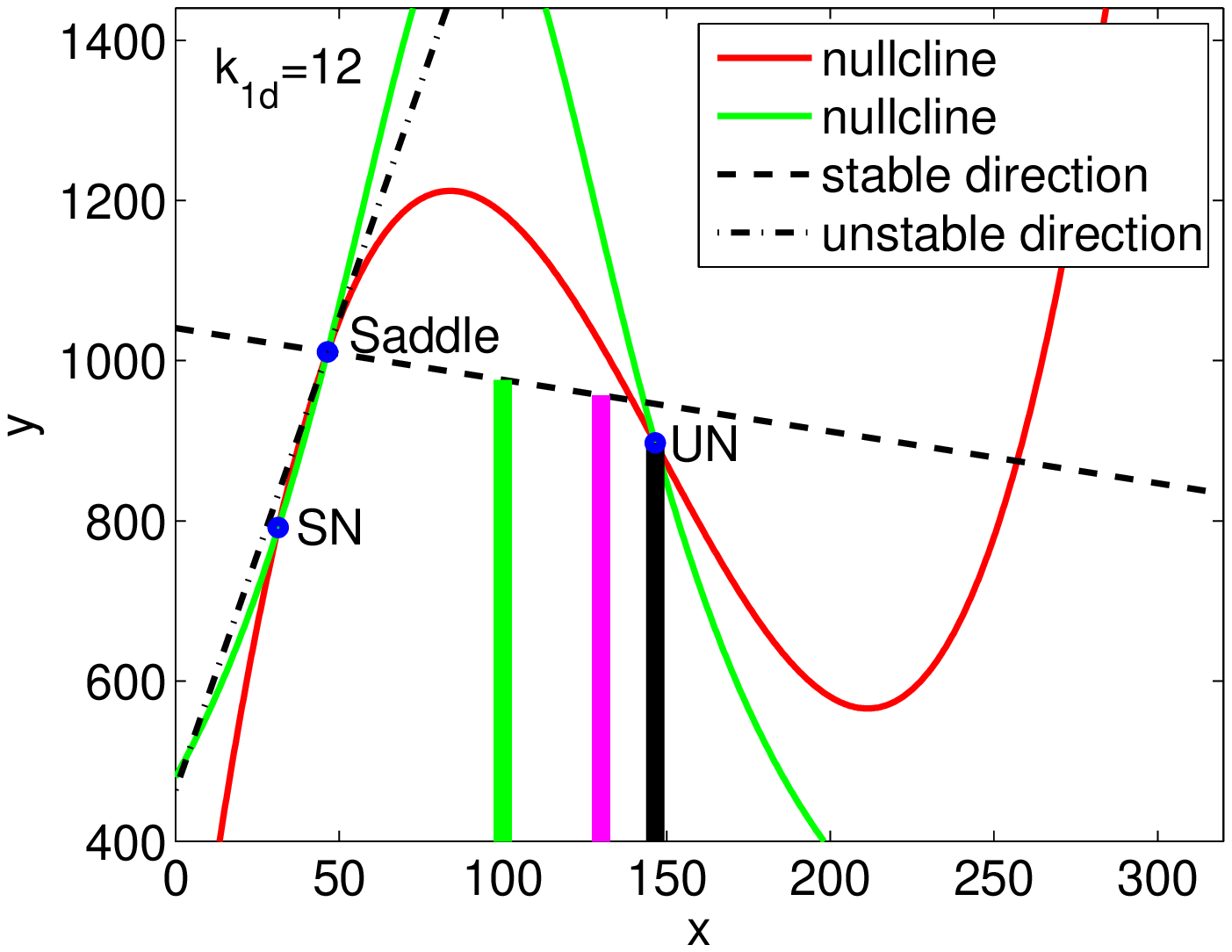}
{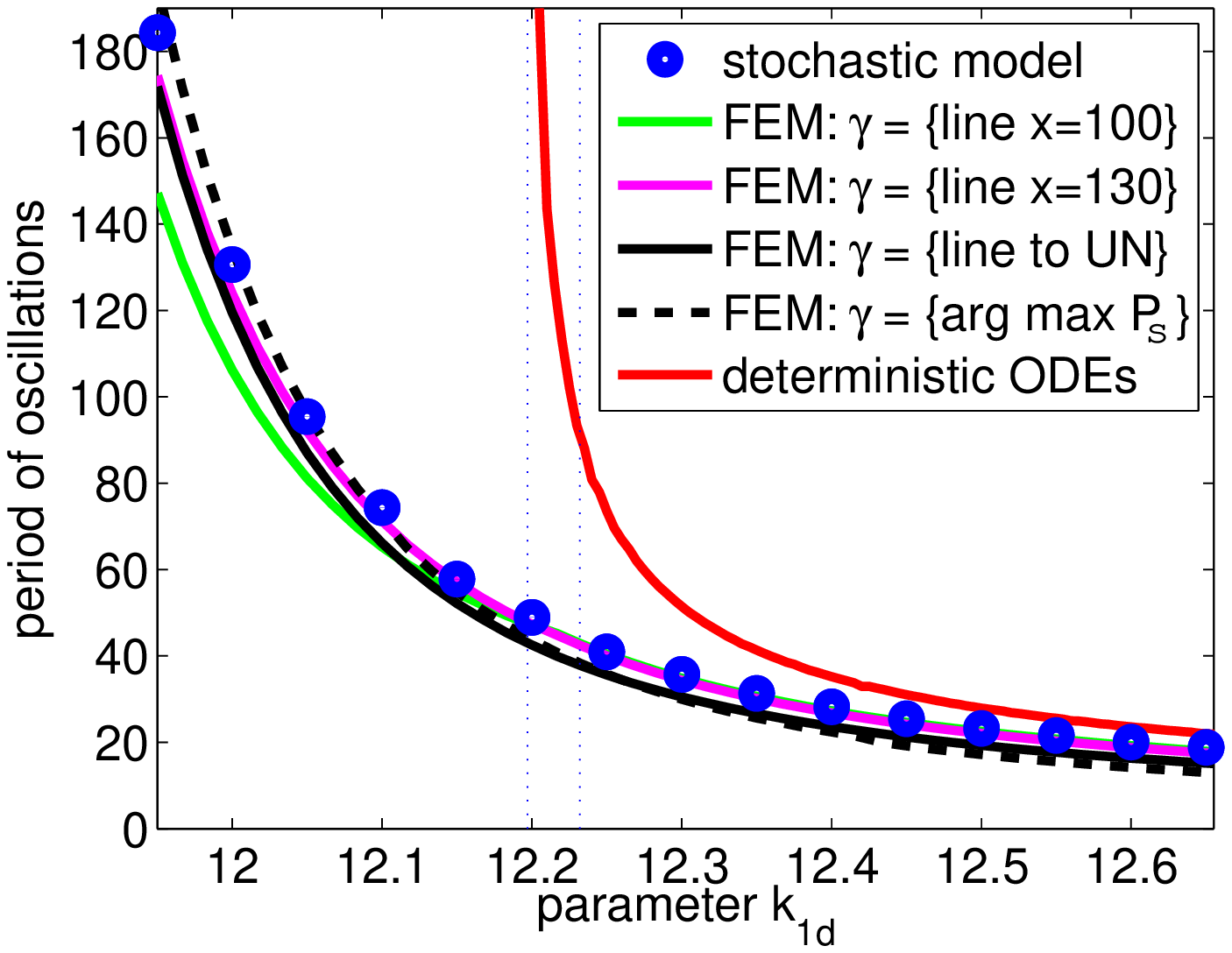}{2.05 in}{4mm}
\caption{{\rm (a)} 
{\it Nullclines and steady states of the ODE system
$(\ref{detSNIPERode})$ together with the lines $\gamma$ which 
are used in the formula $(\ref{apposc})$. The lines
$\gamma$ are plotted as the thick
green line (for $x=100$), the magenta line (for $x=130$)
and the black line (for the line to the unstable node).}
{\rm (b)} 
{\it The period of oscillation as a function
of the parameter $k_{1d}$. The results obtained by the Gillespie
SSA (blue circles) are compared with the results 
obtained by $(\ref{apposc})$ for the lines $\gamma$ shown
in the panel (a).}
}
\label{figsniperperiod2}
\end{figure}

The estimation of the period of oscillation is analogous 
to the estimation of the mean switching time of the Schl\"ogl
problem (\ref{schlogl}) which was studied in Section \ref{sec1Dswitch}.
We introduced the boundary $b$ given by (\ref{bdeffortau}) and
we asked what the mean time to leave the domain $(-\infty,b)$ is.
The biggest contribution to the mean exit time was given
by the behaviour close to the point $x_u$. The SNIPER equivalent of 
the boundary $b$ (resp. point $x_u$) is the line $x=200$
(resp. the saddle). In the case of the Schl\"ogl problem 
(\ref{schlogl}), we saw that the difference between the
escape times from $X(0)=0$ and $X(0)=x_{f1}$ increased as
we approached the bifurcation value. Similarly,
the estimation of the period of oscillation of the SNIPER
problem by $\tau_h(x_f,y_f)$ is fine for $k_{1d} \ll K_d$,
but it might provide less accurate results if the saddle 
and the stable node of the SNIPER problem are too close,
i.e. if $k_{1d}$ is close to the bifurcation value $K_d$. Thus it is better
to estimate the mean period of oscillation as the mean time
to leave the domain $\Omega$, starting from the subdomain 
$\gamma \subset \Omega$ where stochastic trajectories enter 
the domain $\Omega$. We approximate the mean period of oscillation
as a weighted average of $\tau(x,y)$ over a suitable subdomain 
$\gamma$, namely by
\begin{equation}
T(\gamma)
=
\frac{\displaystyle
\int_{\gamma}
\tau(x,y) \, P_s(x,y)
\, \dx \dy}{\displaystyle
\int_{\gamma}
P_s(x,y)
\,
\dx \dy}.
\label{apposc}
\end{equation}
Since the trajectories are entering the domain $\widetilde{S}$
at the smaller values of $Y$ and leaving the domain at the larger
values of $Y$, it is reasonable to choose the subdomain $\gamma$  
as the line segment parallel with the $Y$ axis
       between the $X$ axis and a suitable treshold value.
Three choices of $\gamma$ are shown in Figure
\ref{figsniperperiod2}(a) as thick black, magenta and green lines.
The black line is the line from $y=0$ to the unstable node.
The magenta line is a segment of $x=130$ and the thick green line
is a segment of $x=100$. In the case of $x=100$ (resp. $x=130$),
we define the threshold value as the intersection of the line
$x=100$ (resp. $x=130$) with the stable direction at the 
saddle point. Note that we use the nullclines and
the saddle point of the system of ODEs 
\begin{equation}
\frac{\mbox{d}x}{\dt} 
= 
v_x(x,y),
\qquad
\frac{\mbox{d}y}{\dt} 
= 
v_y(x,y)
\label{detSNIPERode}
\end{equation}
which slightly differs from the classical mean-field ODE 
description (\ref{eq1})--(\ref{eq2}). The ODE system 
(\ref{detSNIPERode}) is equivalent to the ODE system
(\ref{eq1})--(\ref{eq2}) in the limit $V \to \infty$.
See also the discussion after the equation (\ref{detschlogldriftODE})
about the differences between (\ref{detschloglODE}) and 
(\ref{detschlogldriftODE}).

\begin{figure}
\picturesAB{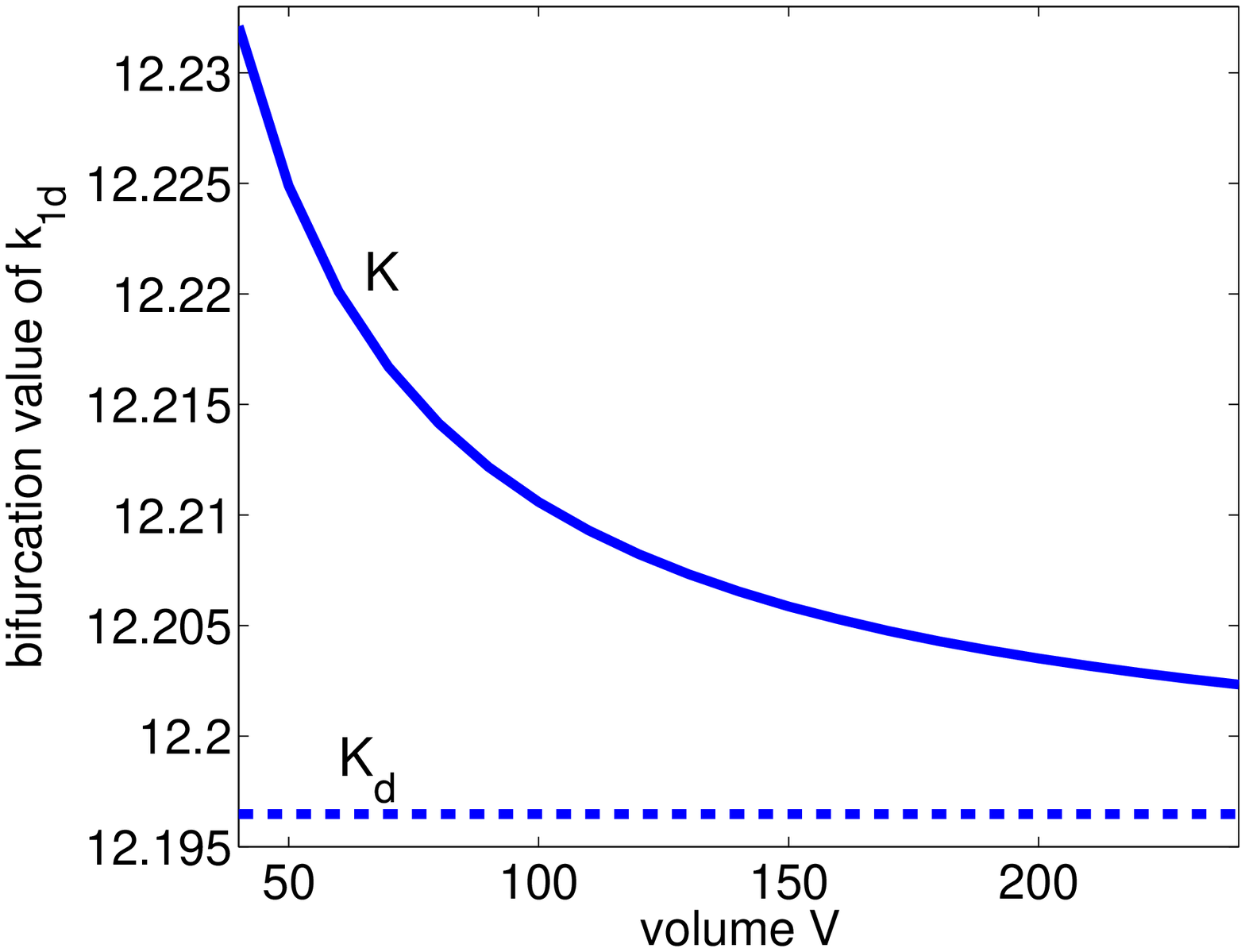}
{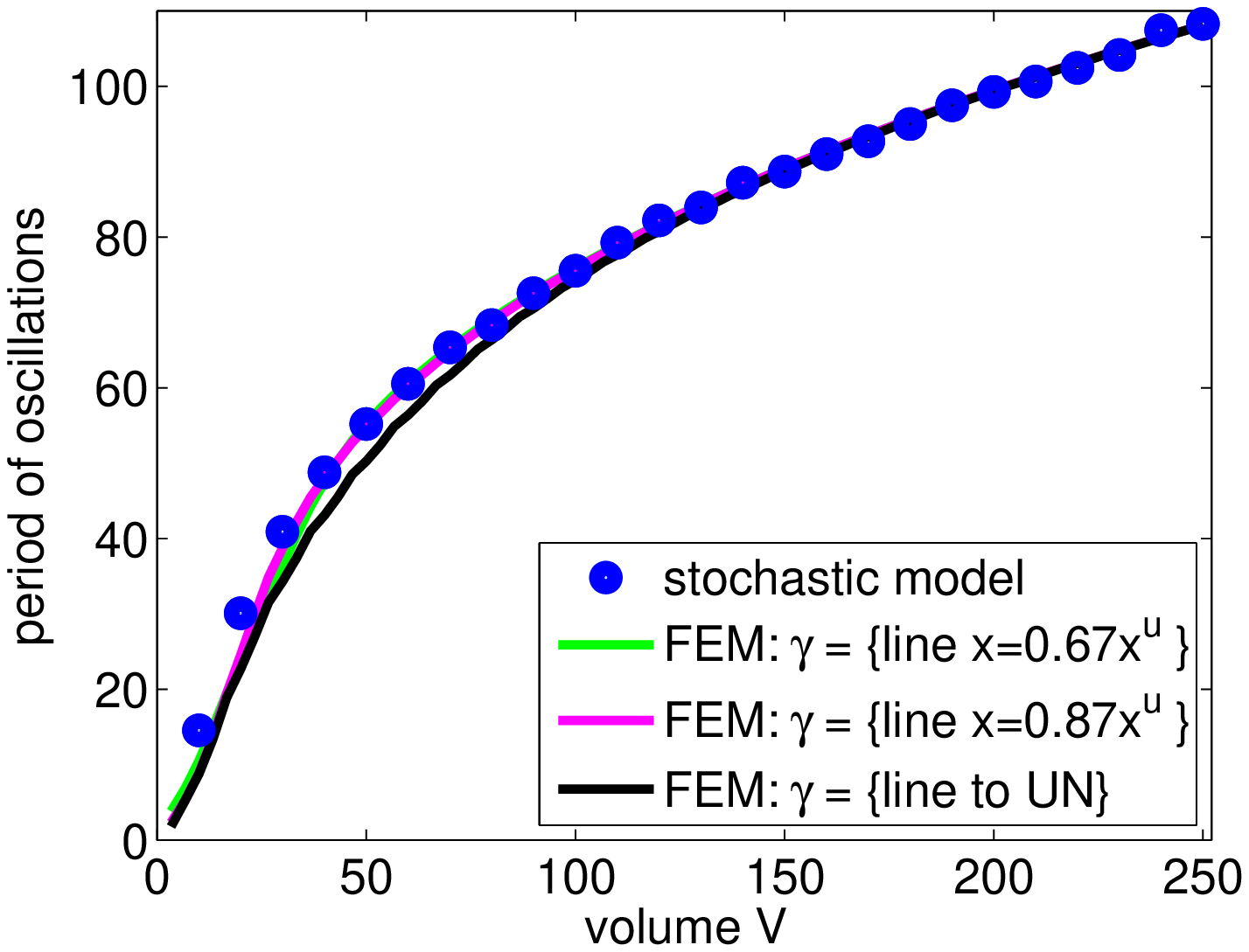}{2.03 in}{7mm}
\caption{{\rm (a)} 
{\it The dependence of the bifurcation value of the parameter $k_{1d}$ 
on the volume $V$ for the ODE system $(\ref{detSNIPERode})$ (solid line)
and for the ODE system $(\ref{eq1})$--$(\ref{eq2})$ (dashed line).
The other parameter values are given according to
$(\ref{sniperparam})$.}
{\rm (b)} 
{\it The period of oscillation as a function
of the volume $V$. The results obtained by the Gillespie
SSA (blue circles) are compared with the results 
obtained by $(\ref{apposc})$ for the lines $\gamma$ 
(scaled by $V$) which are shown in Figure 
}{\rm \ref{figsniperperiod2}(a)}.
}
\label{figsniperperiod3}
\end{figure}

In Figure~\ref{figsniperperiod2}(b), we present estimates of
the period of oscillation computed by (\ref{apposc}) as 
a function of the parameter $k_{1d}$. The results obtained
by the Gillespie SSA (blue circles) and by the ODEs 
(\ref{eq1})--(\ref{eq2}) (red line) were already presented
in Figure~\ref{figperosc}(a). The green, magenta and black
curves correspond to the results computed by (\ref{apposc}) 
for the lines $\gamma$ of the same colour 
in Figure~\ref{figsniperperiod2}(a). We also present
results computed by $\tau(x_f,y_f)$ where $[x_f,y_f]$ is
the point where $P_s(x,y)$ achieves its maximum. The dotted
line denotes the bifurcation values of the ODE
systems (\ref{eq1})--(\ref{eq2}) and (\ref{detSNIPERode}).
These values differ for finite values of the volume $V$. 
In Figure~\ref{figsniperperiod3}(a), we show the dependence
of the bifurcation value of $k_{1d}$ on the volume $V$
for both ODE systems. The ODE system (\ref{eq1})--(\ref{eq2})
is independent of $V$, and so its bifurcation value.
Fixing the value of $k_{1d}$ at the bifurcation value
$K_d \doteq 12.2$ of the ODE model (\ref{eq1})--(\ref{eq2}),
the period of oscillation of the ODE system (\ref{eq1})--(\ref{eq2})
is infinity. The ODE system (\ref{detSNIPERode}) does not
have a limit cycle at all as $K_d \doteq 12.2$ is below its
bifurcation value which we denote $K$. However, we saw in 
Figure~\ref{figperosc}(b) that the stochastic system has oscillatory 
solutions with a 
finite period for $k_{1d}=K_d.$ The period of oscillation
as a function of the volume $V$ can be also computed by 
(\ref{apposc}). The results are shown in Figure
\ref{figsniperperiod3}(b). We use the same lines $\gamma$
as in Figure~\ref{figsniperperiod2}(a). We have to take
into account that the system volume $V$ changes and 
that the phase plane axes (the number of species particles) 
scale linearly with $V$. Thus we define
the lines $\gamma$ as segments of $x=0.67 x^u$, 
$x=0.87 x^u$ and $x=x^u$ where $x^u$ is the $x$-coordinate
of the unstable node. This definition gives for $V=40$
the lines $\gamma$ plotted in Figure~\ref{figsniperperiod2}(a). 
We also have to scale the domain $\widetilde{S}$ with $V$:
we use $\widetilde{S} V/40$ instead of $\widetilde{S}$. 
Fortunately, we can simply
rescale the triangulation. Thus the volume dependence of
the computational domain does not causes any additional problems.
Let us note that the $y$ threshold for the lines
at $x=0.67 x^u$ and $x=0.87 x^u$ is defined as an intersection
with the stable direction at the saddle point of (\ref{detSNIPERode}). 
As shown 
in Figure~\ref{figsniperperiod3}(a), the saddle point
of the ODE system (\ref{detSNIPERode}) is always defined
for $k_{1d} \doteq K_d$ because its (volume dependent)
bifurcation value $K$ satisfies $K > K_d$.  Let us note 
that for $k_1$ greater than the bifurcation value 
of (\ref{detSNIPERode}), we 
use the lines $\gamma$ computed at the bifurcation point 
(since there is no saddle defined for such values of $k_{1d}$).
This definition was used in Figure~\ref{figsniperperiod2}(b)
for $k_{1d} > K$.

\section{Analytical results} \label{secanalFPres}
\begin{figure}
\picturesAB{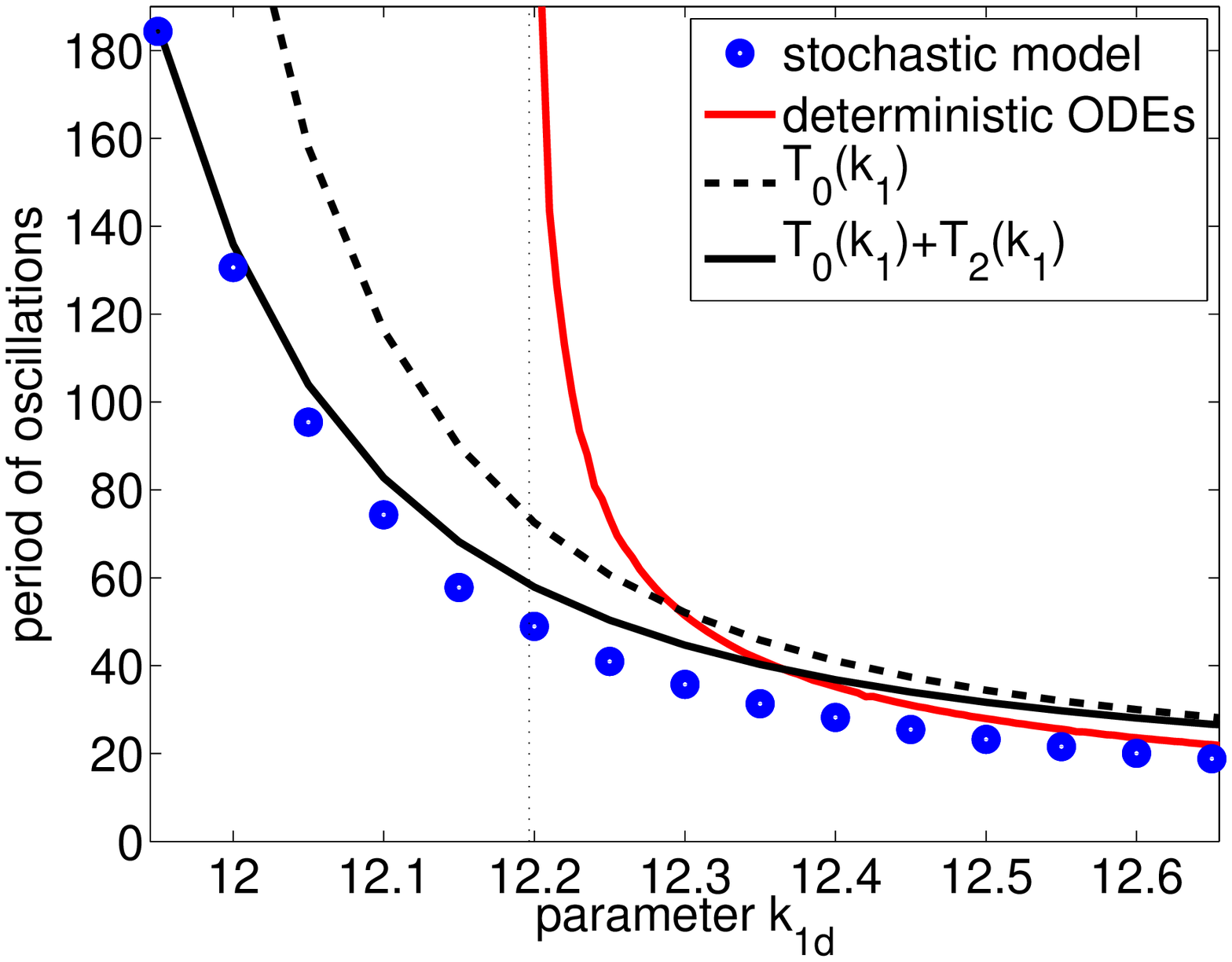}{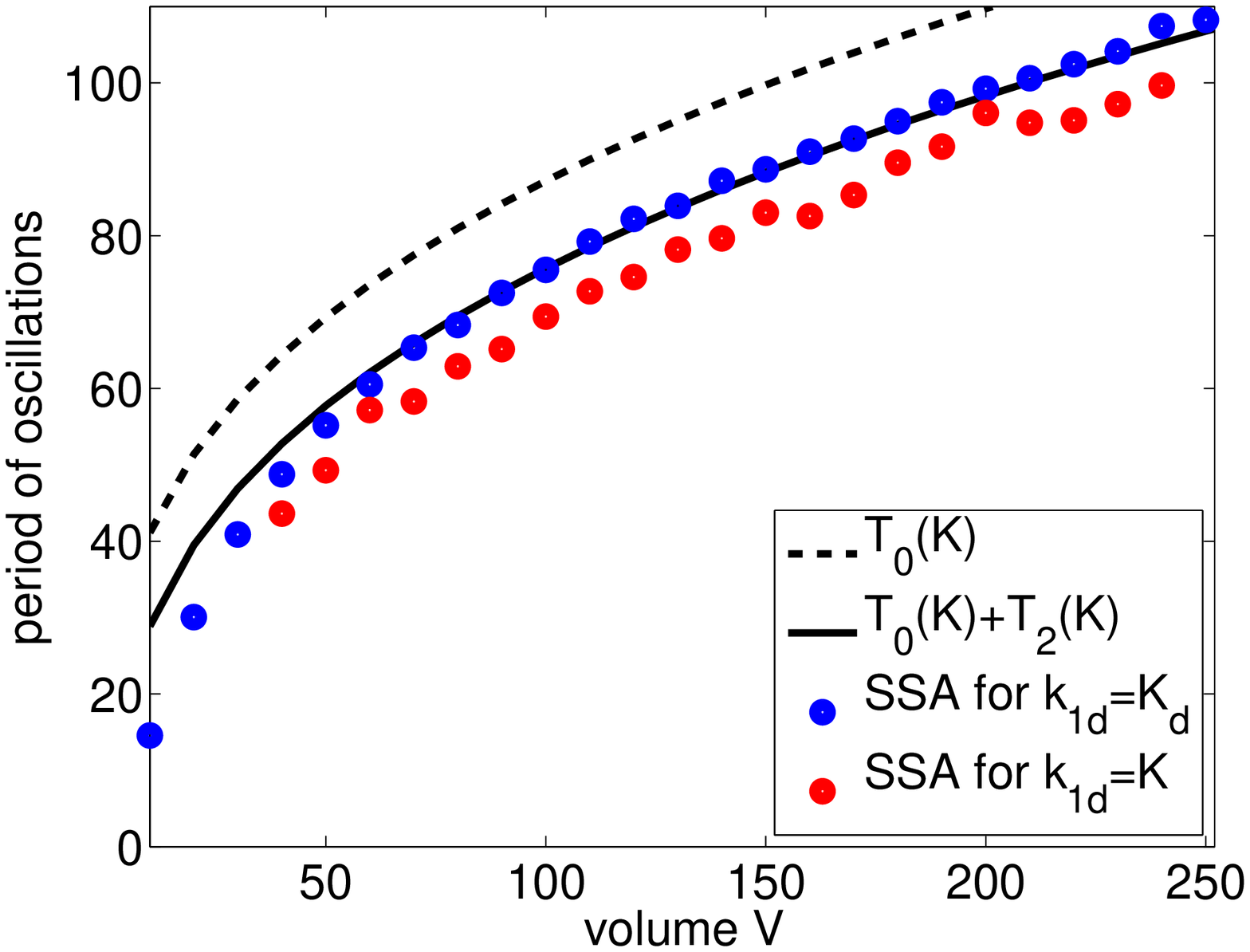}{2.05in}{6mm}
\caption{{\rm (a)} 
{\it Approximations of the mean period of oscillation $T_0(k_1)$
and $T_0(k_1) + T_2(k_1)$ given by $(\ref{T0approx})$ and
$(\ref{T2approx})$. We also replot the results of Figure
$\ref{figperosc}$(a) for comparison. Parameters are the same as in
Figure~$\ref{figperosc}$(a).}
{\rm (b)} 
{\it Approximations of the mean period of oscillation $T_0(K)$
and $T_0(K) + T_2(K)$ given by $(\ref{T0approx0})$ and
$(\ref{T2approx0})$. We also replot the results of Figure
$\ref{figperosc}$(b) for comparison. Parameters are given
by $(\ref{sniperparam})$.
}
}
\label{figsniperanalysis1}
\end{figure}

In this section, we derive analytical formulae for the dependence
of the period of oscillation of the chemical SNIPER
problem (\ref{model1})--(\ref{model2}) on the parameter
$k_{1d}$ and on the volume $V$, i.e. we obtain formulae
for the behaviour shown in Figure~\ref{figperosc}. 
In particular, we generalize the one-dimensional approach from 
Section \ref{secinner} to the chemical SNIPER problem.
As we mentioned before, 
we denote the bifurcation value of the parameter $k_{1d}$ 
of the ODE system (\ref{detSNIPERode}) by $K$. 
If $k_{1d} = K$, then the saddle and the stable node 
of the ODE system (\ref{detSNIPERode}) coincide at one 
point which we denote as $[x_{k},y_{k}].$ Let us define
$$
{\mathbf A}
=
\left(
\begin{matrix}
\displaystyle
\frac{\partial v_x}{\partial x}
&
\displaystyle
\frac{\partial v_x}{\partial y}
\\
\displaystyle
\frac{\partial v_y}{\partial x}
&
\displaystyle
\frac{\partial v_y}{\partial y}
\end{matrix}
\right)
(x_{k},y_{k},K),
$$
where the dependence on parameters $k_2,\dots,k_7$ is not indicated,
because their values are fixed and given by (\ref{sniperparam}).
The eigenvalues of ${\mathbf A}$ are $\lambda_{1} = 0$
and $\lambda_{2} =  - \lambda_0 < 0$. We denote the corresponding 
eigenvectors as ${\mathbf u}_1 = (u_{11},u_{21})$ (for the eignevalue
zero) and ${\mathbf u}_2 = (u_{12},u_{22})$ (for the eigenvalue 
$-\lambda_0$). The eigenvector ${\mathbf u}_1$ is in the direction
of the limit cycle.
There is an obvious separation of time scales. The dynamics
is much slower in the direction of ${\mathbf u}_1$ than in the
direction of ${\mathbf u}_2$. Thus we first transform the 
variables from the Cartesian coordinate system $x$ and $y$ to 
the directions of eigenvectors of ${\mathbf A}$ and then we 
analyse the transformed $\tau$-equation. To do it systematically, 
we use the following scaling (compare with (\ref{scalingone}))
$$
x = \frac{\overline{x}}{\varepsilon},
\quad
y = \frac{\overline{y}}{\varepsilon},
\quad
v_x =  \frac{\overline{v}_x}{\varepsilon^{2}},
\quad
v_y =  \frac{\overline{v}_y}{\varepsilon^{2}},
\quad
d_{x} = \frac{\overline{d}_x}{\varepsilon^{2}},
\quad
d_{xy} = \frac{\overline{d}_{xy}}{\varepsilon^{2}},
\quad
d_{y} = \frac{\overline{d}_y}{\varepsilon^{2}}.
$$
Then the $\tau$-equation (\ref{escapeequation}) reads as follows
\begin{equation}
\varepsilon \,
\overline{d}_x \, \frac{\partial^2 \tau}{\partial \overline{x}^2}
+ 
\varepsilon \, 
\overline{d}_{xy} \, 
\frac{\partial^2 \tau}{\partial \overline{x} \partial \overline{y}}
+ 
\varepsilon \, 
\overline{d}_y \, \frac{\partial^2 \tau}{\partial \overline{y}^2} 
+
\overline{v}_x \, 
\frac{\partial \tau}{\partial \overline{x}}
+
\overline{v}_y \,
\frac{\partial \tau}{\partial \overline{y}}
=
- \varepsilon.
\label{escapeequation2}
\end{equation}
For $k_{1d}$ close to $K$ and $[\overline{x},\overline{y}]$ 
close to $[\overline{x}_k,\overline{y}_k]$,
we use the local variables $\eta,$ $\xi$  and $\kappa$ which are 
related to $\overline{x}$ and $\overline{y}$ as follows
\begin{equation}
\left(
\begin{matrix}
\overline{x} \\
\overline{y}
\end{matrix}
\right)
=
\left(
\begin{matrix}
\overline{x}_k \\
\overline{y}_k
\end{matrix}
\right)
+
\left(
\begin{matrix}
\overline{u}_{11} & \overline{u}_{12} \\
\overline{u}_{21} & \overline{u}_{22}
\end{matrix}
\right)
\left(
\begin{matrix}
\varepsilon^{1/3} \, \eta \\
\varepsilon^{1/2} \, \xi
\end{matrix}
\right),
\qquad
k_{1d} V = k_1 = K + \varepsilon^{2/3} \kappa.
\label{etaxirotSNIPERbb}
\end{equation}
Then the line $\xi = 0$ corresponds to the direction of the limit cycle
and $\eta = 0$ to the stable direction. 
Let us denote $\por = [\overline{x}_k,\overline{y}_k,K]$ to shorten
the following formulae. Using the Taylor expansion at the point 
$\por = [\overline{x}_k,\overline{y}_k,K]$, we approximate 
\begin{eqnarray*}
\overline{d}_{x} (\overline{x},\overline{y},k_1) 
& \approx & 
\overline{d}_{x} (\por)
+
\varepsilon^{1/3} 
c_1(\overline{d}_x)
\, \eta
+
\varepsilon^{1/2} 
c_2(\overline{d}_x)
\, \xi
+
\varepsilon^{2/3} 
c_3(\overline{d}_x)
\, \eta^2
+
o(\varepsilon^{5/6}),
\\
\overline{d}_{xy} (\overline{x},\overline{y},k_1) 
& \approx & 
\overline{d}_{xy} (\por)
+
\varepsilon^{1/3} 
c_1(\overline{d}_{xy}) 
\, \eta
+
\varepsilon^{1/2} 
c_2(\overline{d}_{xy})
\, \xi
+
\varepsilon^{2/3} 
c_3(\overline{d}_{xy})
\, \eta^2
+
o(\varepsilon^{5/6}),
\\
\overline{d}_{y} (\overline{x},\overline{y},k_1) 
& \approx & 
\overline{d}_{y} (\por)
+
\varepsilon^{1/3} 
c_1(\overline{d}_{y})
\, \eta
+
\varepsilon^{1/2} 
c_2(\overline{d}_{y})
\, \xi
+
\varepsilon^{2/3} 
c_3(\overline{d}_{y})
\, \eta^2
+
o(\varepsilon^{5/6}),
\end{eqnarray*}
where $c_i(\overline{d}_{x}),$ $c_i(\overline{d}_{xy})$
and $c_i(\overline{d}_{y})$ are constants. 
To systematically derive the formulae for the 
mean period of oscillation, we need to take all the terms above
into account. However, we will show that the results
are actually independent of some coefficients in the expansion.
To save space, we explicitly specify only those coefficients
which actually appears in the final formulae. They are
given in Appendix \ref{appendixci}.
Using the fact that ${\mathbf u}_1 = (u_{11},u_{21})$
(resp. ${\mathbf u}_2 = (u_{12},u_{22})$) is an eigenvector
of the matrix ${\mathbf A}$ corresponding to the
eigenvalue $\lambda_{1} = 0$ (resp. $\lambda_{2} =  - \lambda_0 < 0$),
we obtain the expansion of $\overline{v}_x$ and 
$\overline{v}_y$ in the form
\begin{eqnarray*}
\overline{v}_{x} (\overline{x},\overline{y},k_1) 
& \approx & 
-
\varepsilon^{1/2} 
\lambda_0 \overline{u}_{12}
\, \xi
+
\varepsilon^{2/3} 
c_1(\overline{v}_x)
\, \eta^2
+
\varepsilon^{5/6} 
c_2(\overline{v}_x)
\, \eta \xi
\\
& + &
\varepsilon
(c_3(\overline{v}_x) \, \xi^2
+ 
c_4(\overline{v}_x) \, \eta^3)
+
\varepsilon^{7/6} 
c_5(\overline{v}_x)
\, \eta^2 \xi
+
o(\varepsilon^{4/3}),
\\
\overline{v}_{y} (\overline{x},\overline{y},k_1) 
& \approx & 
-
\varepsilon^{1/2} 
\lambda_0 \overline{u}_{22}
\, \xi
+
\varepsilon^{2/3} 
(
c_1(\overline{v}_y) \, \eta^2
+
\kappa
)
+
\varepsilon^{5/6} 
\xi c_2(\overline{v}_y)
\, \eta
\\
& + &
\varepsilon
(
c_3(\overline{v}_y) \, \xi^2 
+ 
c_4(\overline{v}_y)) \, \eta^3
)
+
\varepsilon^{7/6} 
c_5(\overline{v}_y)
\, \eta^2 \xi
+
o(\varepsilon^{4/3}),
\end{eqnarray*}
where the constants $c_i(\overline{v}_x)$ and $c_i(\overline{v}_y)$
are given in Appendix \ref{appendixci}. Using (\ref{etaxirotSNIPERbb}), 
the derivatives transform as follows
\begin{eqnarray*}
\frac{\partial}{\partial \overline{x}}
& = &
\frac{\partial}{\partial \eta}
\frac{\partial \eta}{\partial \overline{x}}
+
\frac{\partial}{\partial \xi}
\frac{\partial \xi}{\partial \overline{x}}
=
\varepsilon^{-1/3}
\frac{\overline{u}_{22}}{\det \overline{U}}
\frac{\partial}{\partial \eta}
-
\varepsilon^{-1/2}
\frac{\overline{u}_{21}}{\det \overline{U}}
\frac{\partial}{\partial \xi},
\\
\frac{\partial}{\partial \overline{y}}
& = &
\frac{\partial}{\partial \eta}
\frac{\partial \eta}{\partial \overline{y}}
+
\frac{\partial}{\partial \xi}
\frac{\partial \xi}{\partial \overline{y}}
=
-
\varepsilon^{-1/3}
\frac{\overline{u}_{12}}{\det \overline{U}}
\frac{\partial}{\partial \eta}
+
\varepsilon^{-1/2}
\frac{\overline{u}_{11}}{\det \overline{U}}
\frac{\partial}{\partial \xi},
\end{eqnarray*}
where 
$
\det \overline{U} 
= 
\overline{u}_{11} \overline{u}_{22} - \overline{u}_{12} \overline{u}_{21}$. 
Substituting our approximations to (\ref{escapeequation2})
and using the scaling $\tau = \varepsilon^{2/3}\overline{\tau}$, 
we obtain
\begin{eqnarray}
0
& = &
\left(
- 
\lambda_0 \, \xi
\frac{\partial \overline{\tau}}{\partial \xi}
+
\sigma_1
\frac{\partial^2 \overline{\tau}}{\partial \xi^2}
\right)
+
\varepsilon^{1/6}
\left(
\sigma_2 \, \eta^2
\frac{\partial \overline{\tau}}{\partial \xi}
+
\sigma_3 \, \kappa
\frac{\partial \overline{\tau}}{\partial \xi}
+
\sigma_4
\frac{\partial^2 \overline{\tau}}{\partial \xi \partial \eta}
\right)
\nonumber
\\
& + &
\varepsilon^{1/3}
\left(
\sigma_5 \, \eta^2 
\frac{\partial \overline{\tau}}{\partial \eta}
+ 
\sigma_6 \, \kappa
\frac{\partial \overline{\tau}}{\partial \eta}
+ 
\sigma_7 \, \xi \eta
\frac{\partial \overline{\tau}}{\partial \xi}
+
\sigma_8 
\frac{\partial^2 \overline{\tau}}{\partial \eta^2}
+
\sigma_9 \, \eta
\frac{\partial^2 \overline{\tau}}{\partial \xi^2}
+
1
\right)
\nonumber
\\
& + &
\varepsilon^{1/2}
\left(
\sigma_{10} \, \xi \eta
\frac{\partial \overline{\tau}}{\partial \eta}
+ 
\sigma_{11} \, \xi^2
\frac{\partial \overline{\tau}}{\partial \xi}
+ 
\sigma_{12} \, \eta^3
\frac{\partial \overline{\tau}}{\partial \xi}
+
\sigma_{13} \, \xi
\frac{\partial^2 \overline{\tau}}{\partial \xi^2}
+
\sigma_{14} \, \eta
\frac{\partial^2 \overline{\tau}}{\partial \xi 
\partial \eta}
\right)
\nonumber
\\
& + &
\varepsilon^{2/3}
\left(
\sigma_{15} \, \eta^2
\frac{\partial^2 \overline{\tau}}{\partial \xi^2}
+
\sigma_{16} \, \xi
\frac{\partial^2 \overline{\tau}}{\partial \xi \partial \eta}
+
\sigma_{17} \, \eta
\frac{\partial^2 \overline{\tau}}{\partial \eta^2}
+
\sigma_{18} \, \xi^2 
\frac{\partial \overline{\tau}}{\partial \eta}
\nonumber
\right.
\\
&&
\qquad \; \;
+ 
\left.
\sigma_{19} \, \eta^3 
\frac{\partial \overline{\tau}}{\partial \eta}
+
\sigma_{20} \, \eta^2 \xi 
\frac{\partial \overline{\tau}}{\partial \xi}
\right)
+ 
o(\varepsilon^{5/6}),
\label{etaxieq}
\end{eqnarray}
where the coefficients $\sigma_{i}$ are given in the
Appendix \ref{appendixci}. We assume that $\overline{\tau} \to 0$
as $\eta \to \infty$ and that $\overline{\tau}$ is bounded as 
$\eta \to -\infty$ and $\xi \to \pm \infty$. We expand 
$\overline{\tau}$ as
$$
\overline{\tau}
\sim
\tau_0
+
\varepsilon^{1/6} \tau_1
+
\varepsilon^{1/3} \tau_2
+
\varepsilon^{1/2} \tau_3
+
\varepsilon^{2/3} \tau_4
+
o(\varepsilon^{5/6}).
$$
Substituting into (\ref{etaxieq}), we obtain the following
equation, at $O(\varepsilon^{0})$,
\begin{equation}
\lambda_0 \, \xi
\frac{\partial \tau_0}{\partial \xi}
-
\sigma_1
\frac{\partial^2 \tau_0}{\partial \xi^2}
= 
0.
\label{ordereps0}
\end{equation}
Integrating over $\xi$, we obtain
$$
\frac{\partial \tau_0}{\partial \xi}
=
C(\eta) 
\exp \left[ 
\frac{\lambda_0 \xi^2}{2\sigma_1}
\right],
$$
where $C$ is a function of $\eta$. Since $\tau$ is assumed
to be bounded as $\xi \to \pm \infty$, we must have $C(\eta) \equiv 0$.
Consequently, we obtain
\begin{equation}
\tau_0 \equiv \tau_0(\eta),
\label{tau0xi}
\end{equation}
i.e. $\tau_0$ is not a function of $\xi$. Using (\ref{tau0xi})
in (\ref{etaxieq}), we obtain the following
equation, at $O(\varepsilon^{1/6})$,
\begin{equation}
\lambda_0 \, \xi
\frac{\partial \tau_1}{\partial \xi}
-
\sigma_1
\frac{\partial^2 \tau_1}{\partial \xi^2}
=
0,
\label{ordereps1o6}
\end{equation}
which similarly implies
\begin{equation}
\tau_1 \equiv \tau_1(\eta).
\label{tau1xi}
\end{equation}
Using (\ref{tau0xi}) and (\ref{tau1xi}) in (\ref{etaxieq}),
we have, at $O(\varepsilon^{1/3})$,
\begin{equation} 
\lambda_0 \, \xi
\frac{\partial \tau_2}{\partial \xi}
-
\sigma_1
\frac{\partial^2 \tau_2}{\partial \xi^2}
= 
\sigma_5 \, \eta^2 
\frac{\partial \tau_{0}}{\partial \eta}
+ 
\sigma_6 \, \kappa
\frac{\partial \tau_{0}}{\partial \eta}
+
\sigma_8 
\frac{\partial^2 \tau_{0}}{\partial \eta^2}
+
1.
\label{pomtau2eq}
\end{equation}
Thus
$$
-
\sigma_1
\frac{\partial}{\partial \xi}
\left(
\exp \left[
- 
\frac{\lambda_0 \xi^2}{2 \sigma_1}
\right] 
\frac{\partial \tau_2}{\partial \xi}
\right)
= 
\exp \left[
- 
\frac{\lambda_0 \xi^2}{2\sigma_1}
\right] 
\left(
\left(
\sigma_5 \eta^2 
+ 
\sigma_6 \kappa 
\right)
\frac{\partial \tau_0}{\partial \eta}
+
\sigma_8 
\frac{\partial^2 \tau_0}{\partial \eta^2}
+
1
\right).
$$
Integrating over $\xi$ in $(-\infty,\infty)$,
we obtain the solvability condition
\begin{equation}
\left(
\sigma_5 \eta^2 
+ 
\sigma_6 \kappa 
\right)
\frac{\partial \tau_0}{\partial \eta}
+
\sigma_8 
\frac{\partial^2 \tau_0}{\partial \eta^2}
=
- 1.
\label{tau0eq}
\end{equation}
This equation is the SNIPER equivalent of equation
(\ref{equationforovertauetayfyu}) for the chemical switch, 
and can be solved to find the leading-order period
of oscillation. Before we do so, we proceed to higher-order
with the expansion to determine the correction term to
$\tau_0.$
Using (\ref{tau0eq}) in (\ref{pomtau2eq}) gives
$$
\lambda_0 \, \xi
\frac{\partial \tau_2}{\partial \xi}
-
\sigma_1
\frac{\partial^2 \tau_2}{\partial \xi^2}
= 
0.
$$
Thus, again,
\begin{equation}
\tau_2 \equiv \tau_2(\eta).
\label{tau2xi}
\end{equation}
Using (\ref{tau0xi}), (\ref{tau1xi}) and (\ref{tau2xi}) 
in (\ref{etaxieq}), we have, at $O(\varepsilon^{1/2})$,
\begin{equation}
\lambda_0 \, \xi
\frac{\partial \tau_3}{\partial \xi}
+
\sigma_1
\frac{\partial^2 \tau_3}{\partial \xi^2}
=
\sigma_5 \, \eta^2 
\frac{\partial \tau_1}{\partial \eta}
+ 
\sigma_6 \, \kappa
\frac{\partial \tau_1}{\partial \eta}
+
\sigma_8 
\frac{\partial^2 \tau_1}{\partial \eta^2}
+ 
\sigma_{10} \, \xi \eta
\frac{\partial \tau_0}{\partial \eta}
\label{ordereps1o2}
\end{equation}
which is equivalent to
\begin{eqnarray*}
&-&
\sigma_1
\frac{\partial}{\partial \xi}
\left(
\exp \left[
- 
\frac{\lambda_0 \xi^2}{2 \sigma_1}
\right] 
\frac{\partial \tau_3}{\partial \xi}
\right)
\qquad\qquad\qquad\qquad
\\
&=& 
\exp \left[
- 
\frac{\lambda_0 \xi^2}{2\sigma_1}
\right] 
\left(
\left(
\sigma_5 \, \eta^2 
+ 
\sigma_6 \, \kappa
\right)
\frac{\partial \tau_1}{\partial \eta}
+
\sigma_8 
\frac{\partial^2 \tau_1}{\partial \eta^2}
+ 
\sigma_{10} \, \xi \eta
\frac{\partial \tau_0}{\partial \eta}
\right).
\end{eqnarray*}
Integrating over $\xi$ in $(-\infty,\infty)$,
we obtain the solvability condition
$$
\left(
\sigma_5 \, \eta^2 
+ 
\sigma_6 \, \kappa
\right)
\frac{\partial \tau_1}{\partial \eta}
+
\sigma_8 
\frac{\partial^2 \tau_1}{\partial \eta^2}
=
0
$$
which implies
\begin{equation}
\tau_1 \equiv 0.
\label{tau1is0}
\end{equation}
Consequently, (\ref{ordereps1o2}) yields 
$$
-
\sigma_1
\frac{\partial}{\partial \xi}
\left(
\exp \left[
- 
\frac{\lambda_0 \xi^2}{2 \sigma_1}
\right] 
\frac{\partial \tau_3}{\partial \xi}
\right)
= 
\exp \left[
- 
\frac{\lambda_0 \xi^2}{2\sigma_1}
\right] 
\left(
\sigma_{10} \, \xi \eta
\frac{\partial \tau_0}{\partial \eta}
\right).
$$
Integrating over $\xi$, we get
\begin{equation}
\frac{\partial \tau_3}{\partial \xi}
= 
\frac{\sigma_{10} \, \eta}{\lambda_0}
\frac{\partial \tau_0}{\partial \eta}.
\label{tau3xider}
\end{equation}
Using (\ref{tau0xi}), (\ref{tau1is0}) and (\ref{tau2xi}) 
in (\ref{etaxieq}), we have, at $O(\varepsilon^{2/3})$,
\begin{eqnarray}
\lambda_0 \, \xi
\frac{\partial \tau_4}{\partial \xi}
-
\sigma_1
\frac{\partial^2 \tau_4}{\partial \xi^2}
& = &
\sigma_2 \, \eta^2
\frac{\partial \tau_3}{\partial \xi}
+
\sigma_3 \, \kappa
\frac{\partial \tau_3}{\partial \xi}
+
\sigma_4
\frac{\partial^2 \tau_3}{\partial \xi \partial \eta}
\nonumber
\\
& + &
\sigma_5 \, \eta^2 
\frac{\partial \tau_2}{\partial \eta}
+ 
\sigma_6 \, \kappa
\frac{\partial \tau_2}{\partial \eta}
+
\sigma_8 
\frac{\partial^2 \tau_2}{\partial \eta^2}
\nonumber
\\
& + &
\sigma_{17} \, \eta
\frac{\partial^2 \tau_0}{\partial \eta^2}
+
\sigma_{18} \, \xi^2 
\frac{\partial \tau_0}{\partial \eta}
+ 
\sigma_{19} \, \eta^3 
\frac{\partial \tau_0}{\partial \eta}.
\qquad \quad
\label{ordereps2o3}
\end{eqnarray}
Using (\ref{tau3xider}) on the right hand side to eliminate the term
$\partial \tau_3/\partial \xi$ and
(\ref{tau0eq}) to eliminate 
$\partial^2 \tau_0/\partial \eta^2$, we get
$$
-
\sigma_1
\frac{\partial}{\partial \xi}
\left(
\exp \left[
- 
\frac{\lambda_0 \xi^2}{2 \sigma_1}
\right] 
\frac{\partial \tau_4}{\partial \xi}
\right)
= 
\exp \left[
- 
\frac{\lambda_0 \xi^2}{2 \sigma_1}
\right]
\times
\Bigg(
\left(
\sigma_5 \, \eta^2 
+
\sigma_6 \, \kappa
\right)
\frac{\partial \tau_2}{\partial \eta}
+
\sigma_8 
\frac{\partial^2 \tau_2}{\partial \eta^2}
$$
$$
-
\omega_0 \, \eta
+
\left(
\omega_1
+
\omega_2 \, \kappa \, \eta
+ 
\omega_3
\, \eta^3 
+
\sigma_{18} \, \xi^2 
-
\frac{\sigma_{18} \, \sigma_1}{\lambda_0}
\right)
\frac{\partial \tau_0}{\partial \eta}
\Bigg),
$$
where 
\begin{equation}
\omega_0
=
\frac{\sigma_4 \, \sigma_{10}}{\lambda_0 \sigma_8}
+ 
\frac{\sigma_{17}}{\sigma_8},
\qquad
\omega_1
=
\frac{\sigma_4 \, \sigma_{10} + \sigma_{18} \, \sigma_1}{\lambda_0},
\label{defom0om1}
\end{equation}
$$
\omega_2
=
\frac{\sigma_3 \,\sigma_{10}}{\lambda_0}
-
\frac{\sigma_4 \, \sigma_6 \, \sigma_{10}}{\lambda_0 \sigma_8}
-
\frac{\sigma_6 \sigma_{17}}{\sigma_8},
\qquad
\omega_3
=
\frac{\sigma_2 \,\sigma_{10}}{\lambda_0}
+
\sigma_{19} 
-
\frac{\sigma_4 \, \sigma_5 \, \sigma_{10}}{\lambda_0 \sigma_8}
-
\frac{\sigma_5 \, \sigma_{17}}{\sigma_8}.
$$
Integrating over $\xi$ in $(-\infty,\infty)$,
we obtain the solvability condition
\begin{equation}
\left(
\sigma_5 \, \eta^2 
+
\sigma_6 \, \kappa
\right)
\frac{\partial \tau_2}{\partial \eta}
+
\sigma_8 
\frac{\partial^2 \tau_2}{\partial \eta^2}
= 
\omega_0 \, \eta
-
\left(
\omega_1
+
\omega_2 \, \kappa \, \eta
+ 
\omega_3 \, \eta^3 
\right)
\frac{\partial \tau_0}{\partial \eta}.
\label{tau2eq}
\end{equation}
This is the equation which determines $\tau_2$, the
first non-zero correction term to $\tau_0$.
Let us now solve (\ref{tau0eq}) and (\ref{tau2eq})
for $\tau_0$ and $\tau_2$. Since (\ref{tau0eq}) is
of the same form as the equation (\ref{equationforovertauetayfyu}), 
we can use the same technique as in Section \ref{secinner} 
to analyse it. We find (compare with (\ref{limemi}))
\begin{equation}
\lim_{\eta \to -\infty}
\tau_0(\eta) 
=
\sqrt{\pi} 
\; 3^{1/6} 
\left(
\frac{2}{\sigma_5^2 \, \sigma_8}
\right)^{1/3}
\,
H \! \left(
-
\left(
\frac{12}{\sigma_5 \, \sigma_8^2}
\right)^{\!1/3}
\sigma_6 
\,
\kappa
\right), 
\label{tau0lim}
\end{equation}
where the function $H: \er \to (0,\infty)$ is given by (\ref{defH}).
Moreover, we have the following identity (compare with
(\ref{pomicb}))
\begin{equation}
\exp
\left[
\frac{\sigma_5 \eta^3 + 3 \sigma_6 \kappa \eta}{3 \sigma_8} 
\right]
\frac{\partial \tau_0}{\partial \eta}
= 
- 
\frac{1}{\sigma_8}
\int_{-\infty}^{\eta}
\exp
\left[
\frac{\sigma_5 z^3 + 3 \sigma_6 \kappa z}{3 \sigma_8} 
\right]
\dz.
\label{pomicb2D}
\end{equation}
Multiplying equation (\ref{tau2eq}) by $\exp \left[ 
(\sigma_5 \eta^3 + 3 \sigma_6 \kappa \eta)/(3 \sigma_8)
\right],$ using (\ref{pomicb2D}), integrating over $\eta$
and using integration by parts, we obtain
\begin{eqnarray*}
\frac{\partial \tau_2}{\partial \eta}
&=& 
\int_{-\infty}^\eta 
\left(
\frac{\omega_0}{\sigma_8} 
\, y
+
\frac{\omega_1}{\sigma_8^2} \, (\eta - y)
+
\frac{\omega_2}{2 \sigma_8^2} \, \kappa \, (\eta^2 - y^2)
+ 
\frac{\omega_3}{4 \sigma_8^2} \, (\eta^4 - y^4) 
\right) 
\\
&& 
\qquad \quad
\times 
\exp
\left[
\frac{\sigma_5 (y^3-\eta^3) + 3 \sigma_6 \kappa (y-\eta)}{3 \sigma_8} 
\right]
\dy.
\end{eqnarray*}
Integrating over $\eta$ in $(-\infty,\infty)$, we get
\begin{eqnarray}
\lim_{\eta \to -\infty} \tau_2(\eta)
&=& 
-
\frac{\omega_0}{\sigma_8} 
\int_{-\infty}^\infty
\int_{-\infty}^z 
y
\exp
\left[
\frac{\sigma_5 (y^3-z^3) + 3 \sigma_6 \kappa (y-z)}{3 \sigma_8} 
\right]
\dy
\, \dz
\nonumber
\\
& + &
\frac{\omega_1}{\sigma_8^2}
\int_{-\infty}^\infty
\int_{-\infty}^z 
(y - z)
\exp
\left[
\frac{\sigma_5 (y^3-z^3) + 3 \sigma_6 \kappa (y-z)}{3 \sigma_8} 
\right]
\dy
\, \dz
\nonumber
\\
& + &
\frac{\omega_2}{2 \sigma_8^2} \, \kappa 
\int_{-\infty}^\infty
\int_{-\infty}^z 
(y^2 - z^2)
\exp
\left[
\frac{\sigma_5 (y^3-z^3) + 3 \sigma_6 \kappa (y-z)}{3 \sigma_8} 
\right]
\dy
\, \dz
\nonumber
\\
& + &
\frac{\omega_3}{4 \sigma_8^2}  
\int_{-\infty}^\infty
\int_{-\infty}^z 
(y^4 - z^4)
\exp
\left[
\frac{\sigma_5 (y^3-z^3) + 3 \sigma_6 \kappa (y-z)}{3 \sigma_8} 
\right]
\dy
\, \dz. \qquad\quad
\label{dertau2}
\end{eqnarray}
Using the substitution
\begin{equation}
y = \left(\frac{3 \sigma_8}{2 \sigma_5} \right)^{1/3} \, (u-v),
\qquad
z = \left(\frac{3 \sigma_8}{2 \sigma_5} \right)^{1/3} \, (u+v),
\label{subyzuv}
\end{equation}
we simplify the integrals on the right hand side of (\ref{dertau2})
using the same technique as in (\ref{limemi}). In all
cases, we are able to integrate over $u$ variable explicitly
because the $u$-integral is the Gaussian integral. 
The last two integrals are zero
because we integrate odd functions in $u$ over $(-\infty,\infty)$.
Thus (\ref{dertau2}) becomes
\begin{equation}
\lim_{\eta \to -\infty} \tau_2(\eta)
= 
\left(
\omega_0 - \frac{2 \omega_1}{\sigma_8}
\right)
\frac{\sqrt{3 \pi}}{\sigma_5} 
\int_{0}^\infty
\sqrt{v}
\exp
\left[- v^3 
-  
\left(\frac{12}{\sigma_5 \,\sigma_8^2} \right)^{1/3} 
\sigma_6
\, \kappa \, v
\right]
\dv.
\label{dertau2b}
\end{equation}
Let us define the function $G: \er \to (0,\infty)$ by
\begin{equation}
G(z)
=
\int_{0}^\infty
\sqrt{v}
\exp \left[ - v^3 + z v \right] \, \dv.
\label{defG}
\end{equation}
Using (\ref{defom0om1}) and the definition (\ref{defG}), the
formula (\ref{dertau2b}) implies
\begin{equation}
\lim_{\eta \to -\infty} \tau_2(\eta)
= 
\left(
\sigma_{17}
-
\frac{\sigma_4 \, \sigma_{10}}{\lambda_0}
-
\frac{2 \sigma_{18} \, \sigma_1}{\lambda_0}
\right)
\frac{\sqrt{3 \pi}}{\sigma_5 \sigma_8} 
\,\, G \! \left(
- 
\left(\frac{12}{\sigma_5 \,\sigma_8^2} \right)^{1/3} 
\sigma_6
\, \kappa 
\right).
\label{tau2limit}
\end{equation}
The limits (\ref{tau0lim}) and (\ref{tau2limit}) give us the
desired estimates of the period of oscillation. Transforming
back to the original variables, the zero-order approximation
to the mean period of oscillation is given by
\begin{equation}
T_0(k_1)
=
\sqrt{\pi} 
\; 3^{1/6} 
\left(
\frac{2}{\sigma_5^2 \, \sigma_8}
\right)^{1/3}
H \! \left(
-
\left(
\frac{12}{\sigma_5 \, \sigma_8^2}
\right)^{\!1/3}
\sigma_6 
\,
(k_1 - K)
\right)
\label{T0approx}
\end{equation}
where the constants $\sigma_5$, $\sigma_6$ and $\sigma_8$ are given
in Appendix \ref{appendixci}. More precisely, the formulae for
the constants $\sigma_5$, $\sigma_6$ and $\sigma_8$ are obtained
by dropping the overbars in their definitions in Appendix \ref{appendixci}.
They only depend on the derivatives of the coefficients 
(\ref{begcovx})--(\ref{endcodxy}) and on the eigenvectors
${\mathbf u}_1$ and ${\mathbf u}_2$. The approximation $T_0(k_1)$
is plotted in Figure~\ref{figsniperanalysis1}(a) as the black dashed 
line. 
Transforming (\ref{tau2limit}) into original variables,
we can obtained an improved (second-order) approximation of the
period of oscillation as $T_0(k_1) + T_2(k_1)$ where
\begin{equation}
T_2(k_1)
= 
\left(
\sigma_{17}
-
\frac{\sigma_4 \, \sigma_{10}}{\lambda_0}
-
\frac{2 \sigma_{18} \, \sigma_1}{\lambda_0}
\right)
\frac{\sqrt{3 \pi}}{\sigma_5 \sigma_8} 
\,\, G \! \left(
- 
\left(\frac{12}{\sigma_5 \,\sigma_8^2} \right)^{1/3} 
\sigma_6
\, (k_1 - K) 
\right)
\label{T2approx}
\end{equation}
and the constants $\sigma_i$, $i=1,4,5,6,8,10,17,18$, are given
in Appendix \ref{appendixci}. Again, the overbars have to be dropped.
The approximation $T_0(k_1)+T_2(k_1)$
is plotted in Figure~\ref{figsniperanalysis1}(a) as the black 
solid line. 

In Figure~\ref{figperosc}(b), we studied the period of oscillation as 
a function of the volume $V$ for $k_{1d} = K_d$. To approximate
this dependence, we put $k_{1d} = K$ into formulae (\ref{T0approx})
and (\ref{T2approx}). Since $G(0)=\sqrt{\pi}/3$ and
$H(0)=\Gamma(1/6)/3$, we obtain
\begin{eqnarray}
T_0(K)
& = &
\sqrt{\pi} 
\; 3^{-5/6} 
\left(
\frac{2}{\sigma_5^2 \, \sigma_8}
\right)^{1/3}
\Gamma(1/6),
\label{T0approx0}
\\
T_2(K)
& = &
\left(
\sigma_{17}
-
\frac{\sigma_4 \, \sigma_{10}}{\lambda_0}
-
\frac{2 \sigma_{18} \, \sigma_1}{\lambda_0}
\right)
\frac{\pi}{\sigma_5 \sigma_8\sqrt{3}},
\label{T2approx0}
\end{eqnarray}
where $\Gamma$ is the Gamma function.
The values of $T_0(K)$ and $T_0(K)+T_2(K)$ are plotted
as functions of the volume $V$ in Figure~\ref{figsniperanalysis1}(b). 
We see that $T_0(K)+T_2(K)$ compare well with the results
obtained by the stochastic simulation for $k_{1d}=K_d$. To be
more precise, we should compare it with stochastic results
obtained for $k_{1d} = K$ where $K$ depends on the volume
as shown in Figure~\ref{figsniperperiod3}(a). Such data are plotted
in Figure~\ref{figsniperanalysis1}(b) too. 

\begin{figure}
\picturesAB{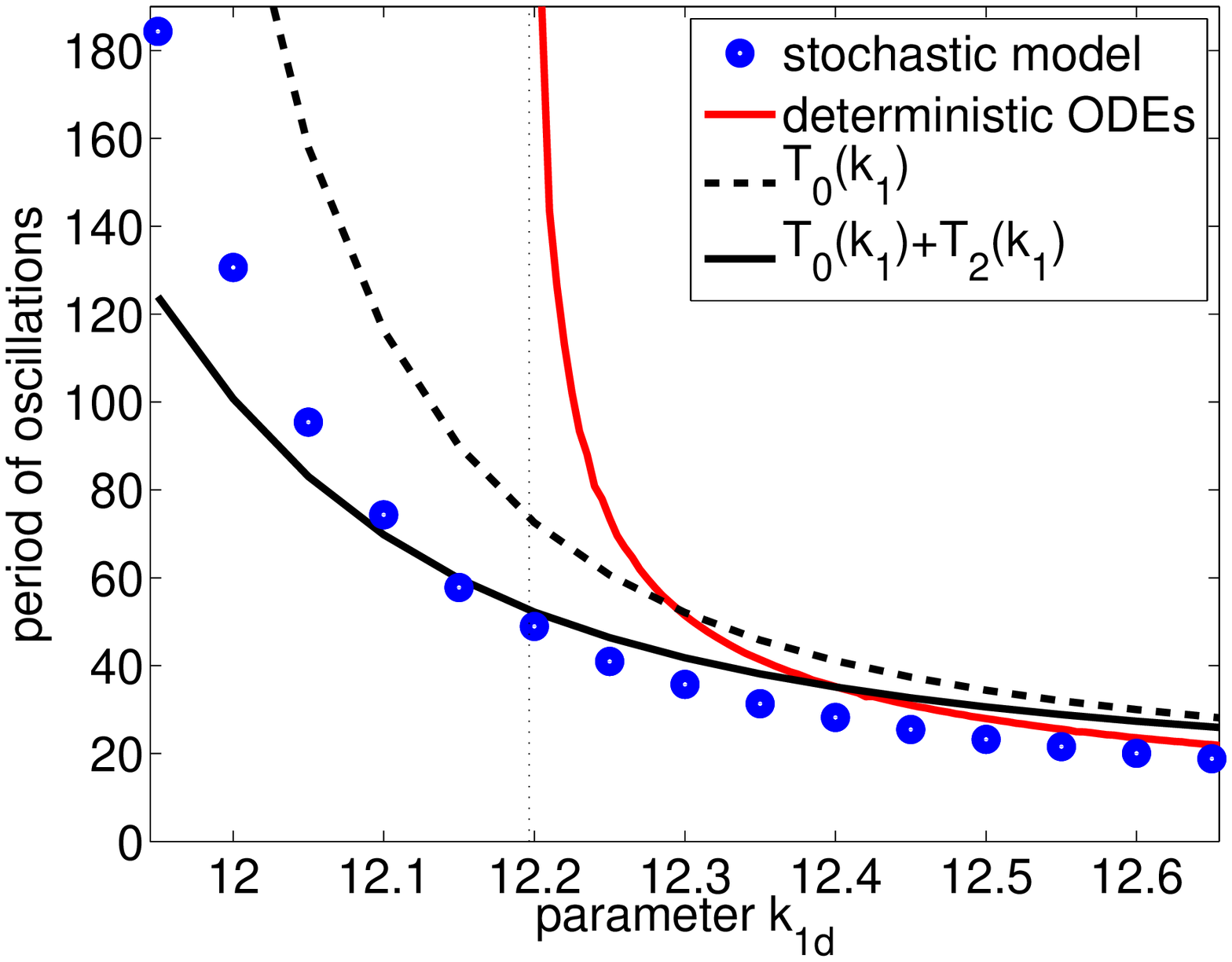}{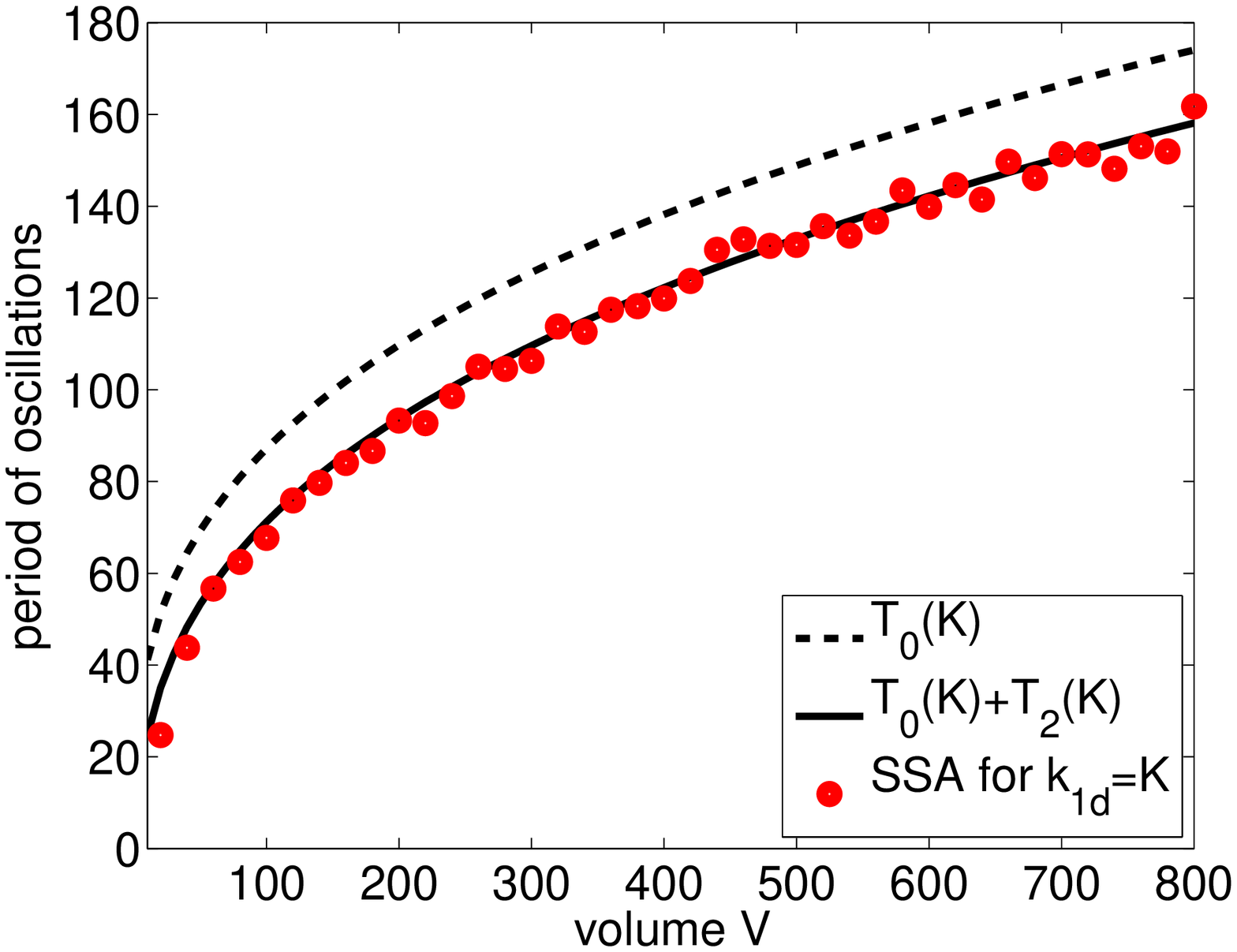}{2.05in}{6mm}
\caption{{\rm (a)} 
{\it Approximations of the mean period of oscillation $T_0(k_1)$
and $T_0(k_1) + T_2(k_1)$ given by $(\ref{T0approx})$ and
$(\ref{T2approx})$ with $\sigma_{17}=0$. 
We also replot the results of Figure
$\ref{figperosc}$(a) for comparison. Parameters are the same as in
Figure~$\ref{figperosc}$(a).}
{\rm (b)} 
{\it Approximations of the mean period of oscillation $T_0(K)$
and $T_0(K) + T_2(K)$ given by $(\ref{T0approx0})$ and
$(\ref{T2approx0})$ with $\sigma_{17}=0$. We also plot the results obtained 
by the stochastic simulation for $k_{1d} = K$. The other parameters are 
given by $(\ref{sniperparam})$.
}
}
\label{figsniperanalysis2}
\end{figure}

Another way to approximate the period of oscillation is
to approximate $\overline{d}_{x} (\overline{x},\overline{y},k_1)$
(resp. $\overline{d}_{xy} (\overline{x},\overline{y},k_1)$ 
and $\overline{d}_{y} (\overline{x},\overline{y},k_1)$)
by $\overline{d}_{x} (\por)$ (resp. $\overline{d}_{xy} (\por)$
and $\overline{d}_{y} (\por)$) only in the equation 
(\ref{escapeequation2}). If the approximation of
$\overline{v}_{x} (\overline{x},\overline{y},k_1)$
and $\overline{v}_{y} (\overline{x},\overline{y},k_1)$
is the same as before, the resulting formulae are
equal to (\ref{T0approx})--(\ref{T2approx0}) with 
 $\sigma_{17}=0$. The comparison of the results
for $\sigma_{17}=0$ is shown in Figure~\ref{figsniperanalysis2}.
We see that we have an excellent comparison with the results
of the stochastic simulation for $k_{1d}=K$ as a function
of the volume $V$ - panel Figure~\ref{figsniperanalysis2}(b).

\section{Conclusion}
\label{secconclusion}

Bifurcation diagrams of ODEs computed by standard numerical continuation
methods constitute a systematic way to study and summarize the 
dependence of the behaviour of the ODE system on its parameters. 
An exploration of such a dependence could be in principle done by a
direct integration in time, but it would be much more computationally 
intensive 
than the numerical continuation. In a similar way, the exploration
of the dependence of the behaviour of a stochastic chemical model 
on its parameters can in principle be done by the Gillespie SSA; yet, 
it is more efficient to study it by solving the underlying stationary
Fokker-Planck equation numerically or by analysing it. For example,
computation of the mean period of oscillation by the Gillespie SSA
is more computationally intensive for large values of the system
size (volume $V$) because the large values of $V$ correspond 
to chemical systems with many molecules. On the other hand, 
the computational intensity
of the FEM solution of the Fokker-Planck equation is independent
of $V,$ making it more suitable for studying the dependence of
the mean period of oscillation on the volume $V$. The asymptotic
formulae (\ref{T0approx}) and (\ref{T2approx}) provide further 
insights into the behaviour of the mean period of oscillation.
At the bifurcation point $k_{1d} = K$ of (\ref{detSNIPERode}),
the formula (\ref{T0approx}) simplifies to (\ref{T0approx0}). 
Inspecting (\ref{T0approx0}), we conclude that the mean period 
of oscillation asymptotes to infinity as $V^{1/3}$ for 
$V \to \infty.$ If $V$ is finite, the stochastic model of the chemical 
system (\ref{model1})--(\ref{model2}) possesses oscillatory 
solutions for any value of $k_{1d}$ which is close to $K$
or close to $K_d$ where $k_{1d} = K_d$ is the bifurcation point 
of the classical deterministic ODE model (\ref{eq1})--(\ref{eq2}). 
On the other hand, the trajectories of (\ref{eq1})--(\ref{eq2})
only oscillate for $k_{1d} > K_d$. The period of oscillation 
asymptotes 
to infinity as $(k_{1d} - K_d)^{-1/2}$ for $k_{1d} \to K_d^+.$

In this paper, we used an illustrative example of the chemical 
system with two chemical species for which the deterministic 
mean-field description is undergoing a SNIPER bifurcation.
The system was simple enough that we could directly compute many 
realisations of the Gillespie SSA. Averaging over many 
realisations, we estimated important characteristics of the 
system, for example, the mean period of oscillation. These estimates 
were used for the visual comparison with the results obtained
by the asymptotic analysis of the chemical
Fokker-Planck equation and by solving this equation numerically. 
Another advantage of the illustrative chemical system was that
the chemical Fokker-Planck equation was two-dimensional, which
simplified its asymptotic analysis and numerical solution.
If we have a system of $N$ chemical species, the resulting 
chemical Fokker-Planck equation will be $N$-dimensional. We are currently
investigating the advantages and disadvantages of this approach
for $N$ larger than 2. Clearly, a suitable numerical method 
for solving the higher-dimensional PDEs must be applied.
The analysis presented here can also be generalized to 
higher-dimensional cases. All we need to find are the eigenvectors 
of the Jacobian matrix at the saddle. The biggest contribution to 
the mean exit time or the mean period of oscillation will be 
in the direction of the eigenvector corresponding to the 
saddle-node connection. Thus the approach presented 
is potentially equally applicable to the $N$-dimensional case and lead
to the estimates of the dependence of the important system
characteristics (e.g. the mean period of oscillation)
on the model parameters. Alternative methods to obtain
these estimates include: accelerating the Gillespie SSA by
using approximate SSAs \cite{Gillespie:2001:AAS}, or estimating the 
low-dimensional
effective Fokker-Planck equation by using short bursts of
appropriately initialized stochastic simulations \cite{Erban:2006:GRN}. 
Both approaches use SSAs. In 
contrast, the methods presented in this paper were based on
the numerical solution and asymptotic analysis of PDEs,
i.e. no stochastic simulation and no generators of random numbers
were needed.

\appendix

\section{Chemical Fokker-Planck equation for $N$-dimensional
systems} \label{appendixCFPE}
Let us consider a well-stirred mixture of $N \ge 1$ molecular
species that chemically interact through $M \ge 1$ chemical
reactions $R_j$, $j=1, \dots, M$. The state of this system is described
by $\boldX = [X_1, \dots, X_N]$ where $X_i \equiv X_i(t)$
is the number of molecules of the $i$-th chemical species,
$i=1, \dots, N.$ Let $\alpha_j(\boldx)$ be the propensity
function of the chemical reaction $R_j$, $j=1, \dots, M$, i.e. 
$\alpha_j(\boldx) \dt$ is the probability that, given $\boldX(t)=\boldx$,
one $R_j$ reaction will occur in the next infinitesimal time
interval $[t,t+\dt)$. Let $\nu_{ji}$ be the change in the
number of $X_i$ produced by one $R_j$ reaction. Then the
chemical system can be described by the chemical master
equation
$$
\frac{\partial}{\partial t}
p(\boldx,t)
=
\sum_{j=1}^M
\Big[
\alpha_j(\boldx - \boldnu_j)
p(\boldx - \boldnu_j,t)
-
\alpha_j(\boldx)
p(\boldx,t)
\Big],
$$
where $p(\boldx,t)$ is the probability that $\boldX(t)=\boldx$
and $\boldnu_j = [\nu_{j1},\dots,\nu_{jN}]$.
Considering the integer valued vector $\boldx$ as a real
variable, Gillespie \cite{Gillespie:2000:CLE} derived 
the chemical Langevin equation by approximating Poisson
random variables by normal random variables. This approximation
is possible if many reaction events happen 
before the propensity functions change significantly their
values, see \cite{Gillespie:2000:CLE} for details. The chemical
Langevin equation can be written in the form
$$
\dX_i = \left( \sum_{j=1}^M \nu_{ji} \alpha_j(\boldX(t)) \right) \dt
+ \sum_{j=1}^M \nu_{ji} \alpha_j^{1/2}(\boldX(t)) \dW_j
$$
which corresponds to the {\it chemical Fokker-Planck equation}
\begin{eqnarray}
\frac{\partial}{\partial t}
p(\boldx,t)
& = &
\sum_{i=1}^N
\frac{\partial}{\partial x_i}
\left[ - \left( \sum_{j=1}^M \nu_{ji} \alpha_j(x) \right) p(\boldx,t) \right]
+ 
\frac{1}{2}
\frac{\partial^2}{\partial x_i^2}
\left[ 
\left( \sum_{j=1}^M \nu^2_{ji} \alpha_j(x) \right) 
p(\boldx,t)
\right]
\nonumber
\\
& &
+ 
\sum_{k<i} 
\frac{\partial^2}{\partial x_i x_k}
\left[ 
\left( \sum_{j=1}^M \nu_{ji} \nu_{jk} \alpha_j(x) \right) 
p(\boldx,t)
\right]
\label{FPEgengen}.
\end{eqnarray}

\section{Equation for the exit time $\tau$} \label{appendixmtime}
Let $p(x^\prime,y^\prime,t; x,y,0)$ be the probability that
$X(t) = x^\prime$, $Y(t) = y^\prime$ given that
$X(0) = x$ and $Y(0) = y$. It satisfies the
backward Kolmogorov equation
\begin{equation}
\frac{\partial p}{\partial t}
=
d_x \frac{\partial^2 p}{\partial x^2}
+ 
d_{xy} \frac{\partial^2 p}{\partial x \partial y}
+ 
d_y \frac{\partial^2 p}{\partial y^2} 
+
v_x 
\frac{\partial p}{\partial x}
+
v_y
\frac{\partial p}{\partial y}.
\label{backKolm2D}
\end{equation}
Let $h(x,y,t)$ be the probability that $[X(t),Y(t)] \in \Omega$
at time $t$ given that it started at $[X(0),Y(0)] = [x,y].$ 
Then
$$
h(x,y,t)
=
\int_{\Omega}
p(x^\prime,y^\prime,t; x,y,0)
\,
\dx^\prime
\dy^\prime.
$$
Integrating (\ref{backKolm2D}), we obtain
\begin{equation}
\frac{\partial h}{\partial t}
=
d_x \frac{\partial^2 h}{\partial x^2}
+ 
d_{xy} \frac{\partial^2 h}{\partial x \partial y}
+ 
d_y \frac{\partial^2 h}{\partial y^2} 
+
v_x 
\frac{\partial h}{\partial x}
+
v_y
\frac{\partial h}{\partial y}.
\label{equationforh2D}
\end{equation}
The mean exit time $\tau$ to leave $\Omega$,
given that initially $[X(0),Y(0)]=[x,y]$, can be computed 
as follows \cite{Fox:1986:FAS}
$$
\tau(x,y)
=
- \int_0^\infty t \, \frac{\partial h}{\partial t} (x,y,t) \dt
=
\int_0^\infty h(x,y,t) \dt.
$$
Integrating equation (\ref{equationforh2D}) over $t$, we obtain
the following elliptic problem:
\begin{equation}
d_x \frac{\partial^2 \tau}{\partial x^2}
+ 
d_{xy} \frac{\partial^2 \tau}{\partial x \partial y}
+ 
d_y \frac{\partial^2 \tau}{\partial y^2} 
+
v_x 
\frac{\partial \tau}{\partial x}
+
v_y
\frac{\partial \tau}{\partial y}
=
- 1,
\quad
\mbox{for} \; [x,y] \in \Omega.
\label{tau2Dequation}
\end{equation}

\section{Formulae for coefficients $c_i$ and $\omega_i$} \label{appendixci}
The coefficients $c_i(\overline{d}_x)$, $c_i(\overline{d}_{xy})$,
$c_i(\overline{d}_{y})$, $c_i(\overline{v}_x)$ and $c_i(\overline{v}_y)$
can be computed by the Taylor expansion. Below, we provide the
expressions for those coefficients which actually appear in the final
formulae for the mean period of oscillation:
\begin{eqnarray*}
&&
c_1(\overline{v}_x) 
= 
\displaystyle
\frac{\partial^2 \overline{v}_{x}}{\partial \overline{x}^2} (\por) 
\frac{\overline{u}_{11}^2}{2},
\qquad
c_2(\overline{v}_x) 
= 
\displaystyle
\frac{\partial^2 \overline{v}_{x}}{\partial \overline{x}^2} (\por) 
\overline{u}_{11} \overline{u}_{12},
\qquad
c_3(\overline{v}_x) 
=
\displaystyle
\frac{\partial^2 \overline{v}_{x}}{\partial \overline{x}^2} (\por) 
\frac{\overline{u}_{12}^2}{2},
\\
&&
c_1(\overline{v}_y)
=  
\displaystyle
\frac{\partial^2 \overline{v}_{y}}{\partial \overline{x}^2} (\por) 
\frac{\overline{u}_{11}^2}{2}
+ 
\displaystyle
\frac{\partial^2 \overline{v}_y}{\partial \overline{x} \partial \overline{y}} 
(\por)
\overline{u}_{11}\overline{u}_{21},
\\
&&
c_2(\overline{v}_y)
=  
\displaystyle
\frac{\partial^2 \overline{v}_{y}}{\partial \overline{x}^2} (\por) 
\overline{u}_{11} \overline{u}_{12}
+
\displaystyle
\frac{\partial^2 \overline{v}_y}{\partial \overline{x} \partial \overline{y}} 
(\por)
(\overline{u}_{11}\overline{u}_{22} + \overline{u}_{12}) \overline{u}_{21} ), 
\\
&&
c_3(\overline{v}_y)
 =  
\displaystyle
\frac{\partial^2 \overline{v}_{y}}{\partial \overline{x}^2} (\por) 
\frac{\overline{u}_{12}^2}{2}
+
\displaystyle
\frac{\partial^2 \overline{v}_y}{\partial \overline{x} \partial \overline{y}} 
(\por)
 \overline{u}_{12}
 \overline{u}_{22}, 
\qquad
c_1(\overline{d}_x)
= 
\frac{\partial \overline{d}_{x}}{\partial \overline{x}} 
(\por) \, \overline{u}_{11}
+
\frac{\partial \overline{d}_{x}}{\partial \overline{y}} 
(\por) \, \overline{u}_{21},
\\
&&
c_1(\overline{d}_{xy})
=
\frac{\partial \overline{d}_{xy}}{\partial \overline{x}} 
(\por) \, \overline{u}_{11}
+
\frac{\partial \overline{d}_{xy}}{\partial \overline{y}} 
(\por) \, \overline{u}_{21},
\qquad
c_1(\overline{d}_y)
=
\frac{\partial \overline{d}_{y}}{\partial \overline{x}} 
(\por) \, \overline{u}_{11}
+
\frac{\partial \overline{d}_{y}}{\partial \overline{y}} 
(\por) \, \overline{u}_{21},
\end{eqnarray*}
where we used the formulae (\ref{begcovx})--(\ref{midcovy}) 
to simplify the Taylor expansion of 
$\overline{v}_{x} (\overline{x},\overline{y},k_1)$ 
and $\overline{v}_{y} (\overline{x},\overline{y},k_1)$.
Note that some of the second derivatives of $\overline{v}_{x}$
and $\overline{v}_{y}$ are zero which makes the resulting formulae
shorter. However, even if they were nonzero, the same analysis
could be carried through. The $\sigma$ coefficients are given by
\begin{eqnarray*}
\sigma_1
& = &
(\det \overline{U})^{-2}
\left(
\overline{d}_{x} (\por) 
\overline{u}_{21}^2
-
\overline{d}_{xy} (\por) 
\overline{u}_{21} \overline{u}_{11}
+ 
\overline{d}_y (\por) 
\overline{u}_{11}^2
\right),
\\
\sigma_4
& = &
(\det \overline{U})^{-2}
\left(
-
\overline{d}_{x} (\por) 
2 u_{21} u_{22}
+
\overline{d}_{xy} (\por) 
\left(
\overline{u}_{11} \overline{u}_{22}
+ 
\overline{u}_{12} \overline{u}_{21}
\right)
- 
\overline{d}_y (\por) 
2 u_{11}u_{12}
\right),
\\
\sigma_5
& = &
(\det \overline{U})^{-1}
\left(
c_1(\overline{v}_{x}) \overline{u}_{22}
-
c_1(\overline{v}_{y}) \overline{u}_{12}
\right),
\\
\sigma_6
& = &
-
(\det \overline{U})^{-1} \overline{u}_{12},
\\
\sigma_8
& = &
(\det \overline{U})^{-2}
\left(
\overline{d}_{x} (\por) 
\overline{u}_{22}^2
-
\overline{d}_{xy} (\por) 
\overline{u}_{12} \overline{u}_{22}
+ 
\overline{d}_y (\por) 
\overline{u}_{12}^2
\right),
\\
\sigma_{10}
& = &
(\det \overline{U})^{-1}
\left(
c_2(\overline{v}_{x}) \overline{u}_{22}
-
c_2(\overline{v}_{y}) \overline{u}_{12}
\right),
\\
\sigma_{17}
& = &
(\det \overline{U})^{-2}
\left(
c_1(\overline{d}_{x}) (\por) 
\overline{u}_{22}^2
-
c_1(\overline{d}_{xy}) (\por) 
\overline{u}_{12} \overline{u}_{22}
+ 
c_1(\overline{d}_y) (\por) 
\overline{u}_{12}^2
\right),
\\
\sigma_{18}
& = &
(\det \overline{U})^{-1}
\left(
c_3(\overline{v}_{x}) \overline{u}_{22}
-
c_3(\overline{v}_{y}) \overline{u}_{12}
\right).
\end{eqnarray*}
To get the coefficients $\sigma_i$ in the final formulae
(\ref{T0approx})--(\ref{T2approx0}), we simply drop the
overbars in the expressions above.

\medskip

\noindent
{\bf Acknowledgements.} This publication is based on work supported 
by St. John's College, Oxford; Linacre College, Oxford; 
Somerville College, Oxford and by Award No. KUK-C1-013-04, made by 
King Abdullah University of Science and Technology (RE).
IGK was supported by the National Science Foundation and the US DOE.
TV acknowledges the support of the Czech Science Foundation,
project No.~102/07/0496, of the Grant Agency of the Academy 
of Sciences, project No.~IAA100760702, and of the Czech Academy 
of Sciences, institutional research plan No.~AV0Z10190503.
The authors also benefited from helpful discussions with Bill Baumann and 
John Tyson.

\bibliographystyle{amsplain}
%\bibliography{bibrad}

\providecommand{\bysame}{\leavevmode\hbox to3em{\hrulefill}\thinspace}
\providecommand{\MR}{\relax\ifhmode\unskip\space\fi MR }
\providecommand{\MRhref}[2]{%
  \href{http://www.ams.org/mathscinet-getitem?mr=#1}{#2}
}
\providecommand{\href}[2]{#2}

\end{document}